\documentclass[prd,aps,eqsecnum,floatfix,nofootinbib,preprint,tightenlines]{revtex4-2}

\usepackage{latexsym}
\usepackage{graphicx}
\usepackage{amsmath}
\usepackage[dvipsnames]{xcolor}

\usepackage{hyperref}

\def\bibi{\bibitem}
\def\floatcaption#1#2{ \caption{#2 \label{#1}} }
\def\ttl#1{{\it #1}}

\let\br=\u                      

\def\a{\alpha}
\def\b{\beta}
\def\c{\chi}
\def\d{\delta}
\def\e{\epsilon}                
\def\f{\phi}                    
\def\g{\gamma}

\def\j{\psi}
\def\k{\kappa}
\def\l{\lambda}
\def\m{\mu}
\def\n{\nu}

\def\p{\pi}                     
\def\r{\rho}                    
\def\s{\sigma}                  
\def\t{\tau}

\def\x{\xi}
\def\z{\zeta}
\def\D{\Delta}

\def\G{\Gamma}

\def\L{\Lambda}
\def\O{\Omega}
\def\P{\Pi}

\def\S{\Sigma}

\def\X{\Xi}

\def\cf{{\cal F}}

\def\cl{{\cal L}}
\def\cm{{\cal M}}

\def\co{{\cal O}}

\def\cu{{\cal U}}
\def\cv{{\cal V}}

\def\bo{\raisebox{-.4ex}{\large$\Box$}}                 
\def\cbo{{\,\raise-.15ex\Sc [\,}}                       

\def\sl#1{\rlap{\hbox{$\mskip 1 mu /$}}#1}      

\def\ddt#1{{\buildrel {\hbox{\LARGE .\kern-2pt.}} \over {#1}}}

\def\ie{\mbox{\it i.e.}}
\def\eg{\mbox{\it e.g.}}
\def\etc{\mbox{\it etc.}}

\def\tr{{\rm tr}\,}
\def\Tr{{\rm Tr}\,}
\def\hc{{\rm h.c.\,}}

\def\half{{1\over 2}}

\def\bj{\overline\psi}
\def\tj{\tilde\psi}
\def\tc{\tilde\c}
\def\tA{\tilde{A}}
\def\tbj{\tilde{\overline\psi}}
\def\tbc{\tilde{\overline\c}}
\def\bd{\overline\d}

\def\bc{\overline\chi}
\def\bff{\overline\phi}

\def\hf{\hat{\f}}
\def\hbf{\hat{\bff}}

\def\bu{\overline{u}}
\def\bdd{\overline{d}}
\def\bd{\overline{\d}}
\def\bq{\overline{q}}
\def\br{\overline{r}}
\def\bg{\bar\gamma}

\def\bp{\bar{p}}
\def\bc{\bar{\c}}
\def\hp{\hat{\p}}
\def\hT{\hat{T}}
\def\hS{\hat{S}}
\def\hI{\hat{I}}
\def\hP{\hat{P}}
\def\hC{\hat{C}}
\def\hQ{\hat{Q}}
\def\tX{\tilde{X}}
\def\hX{\hat{\X}}
\def\hchi{\hat{\c}}

\def\tp{\tilde{p}}
\def\tq{\tilde{q}}
\def\tk{\tilde{k}}

\def\textit#1{{\it \!\!\! #1 \!\!}}

\def\dlrm{{\overleftrightarrow{\partial}_{\!\!\!\m}}}

\begin{document}

\begin{center}
{\large{\bf Staggered Fermions}}\\[8mm]
Maarten Golterman\\[8 mm]
{\it Department of Physics and Astronomy, San Francisco State University,\\
San Francisco, CA 94132, USA}\\[10mm]
\end{center}

\begin{quotation}
  These notes are based on a series of lectures on staggered fermions given
  at the Centre de Physique Th\'eorique, Luminy, in Marseille, France, January 17-25, 2024.
\end{quotation}

\section{\label{intro} Introduction}
These lecture notes present the lattice formulation of vectorlike gauge theories 
such as QED and QCD employing staggered fermions, which are extensively used in 
numerical simulations of QCD.   While many aspects
of the use of staggered fermions are discussed, these notes are not intended as
a comprehensive review.   Several topics and alternative formulations are only briefly mentioned,
or not
covered at all (see Sec.~\ref{reading} for pointers to additional reading).    Likewise, while important references on staggered 
fermions and related topics have been included in the bibliography, the list
of references is not complete.   Staggered fermions were first introduced in a 
hamiltonian formalism \cite{TB,LSU}, but here we will only be concerned with 
the euclidean theory.   These notes complement the review of Ref.~\cite{MILC}.

\section{\label{action} Staggered fermion action}
In this section we begin with the naive fermion lattice action, from which we derive the 
staggered fermion action through spin diagonalization.   
We then derive the Feynman rules, show that the 
classical continuum limit reproduces the expected gauge theory, and we list and
discuss the symmetries of staggered fermions.   We end with the
formulation of the staggered theory on the so-called ``taste basis."

\subsection{\label{naivesec} Naive fermions}
The naive fermion action is given by
\begin{equation}
\label{naive}
S=\half\sum_{x\m}\left(\bj(x)\g_\m U_\m(x)\j(x+\m)-\bj(x+\m)\g_\m U^\dagger_\m(x)\j(x)\right)
+m\sum_x\bj(x)\j(x)\ .
\end{equation}
Points on the hypercubic lattice are labeled by $x$, and $x+\m$ is the point one lattice
spacing away from $x$ in the $+\m$ direction.   Most of the time we will set the lattice spacing
$a=1$; if not, this point will be denoted as $x+a\m$.   $\j(x)$ is a Dirac spinor transforming in the fundamental irreducible representation
(irrep) of $SU(3)$, gauged by the link variables $U_\m(x)$.  We are in four-dimensional euclidean space,
and the Dirac matrices $\g_\m$ are hermitian.  The parameter $m$ is the bare mass; later on,
we will generalize this to a mass matrix $M$ when we consider $N_f>1$ flavors of staggered
fermions.   For now, we take $N_f=1$.

Let us obtain the free fermion propagator in momentum space.   We define the Fourier
transform of $\j$ and $\bj$ by
\begin{equation}
\label{FT}
\j(x)=\int_p e^{ipx}\tj(p)\ ,\qquad \bj(x)=\int_p e^{-ipx}\tbj(p)\ ,
\end{equation}
where
\begin{equation}
\label{intp}
\int_p\equiv \int_{-\p}^\p\frac{d^4p}{(2\p)^4}\ .
\end{equation}
In momentum space, the free action, obtained by setting $U_\m(x)=1$, is
\begin{equation}
\label{freemom}
S_{\rm free}=\int_p\sum_\m\tbj(p)\,i\g_\m\sin(p_\m)\tj(p)+m\int_p\tbj(p)\tj(p)\ .
\end{equation}
We see that the inverse propagator (at $m=0$) not only has a zero at $p=0$, but
at all momenta
\begin{equation}
\label{piA}
p=\p_A\ ,
\end{equation}
where $\p_A$ has components $0$ or $\p$.   There are 16 such vectors, and the
index $A$ thus runs from 1 to 16.   For
$p_\m=\tp_\m+\p_A$, 
\begin{equation}
\sin(p_\m)=\sin(\tp_\m+\p_{A\m})=S^A_\m\sin(\tp_\m)\ ,
\end{equation}
which defines the signs $S^A_\m=e^{i\p^A_\m}$.
Therefore, the inverse propagator near $p=\p_A$ for small $\tp$ equals
\begin{equation}
\label{invprop}
S^{-1}(p)=\sum_\m i\g_\m S^A_\m\tp_\m+m\ ,
\end{equation}
showing that the continuum limit of the naive fermion action contains 16 relativistic
fermions.  These are the so-called fermion ``species doublers.''  In particular, the matrices
\begin{equation}
\label{gA}
\g^A_\m=\g_\m S^A_\m
\end{equation} 
satisfy the Dirac algebra $\{\g_\m^A,\g_\n^A\}=2\d_{\m\n}$, and are unitarily equivalent to 
the original set.   For the relativistic fermion
near $p=\p_A$ we can define
\begin{equation}
\label{g5}
\g_5^A=\g^A_1\g^A_2\g^A_3\g^A_4=S_1^AS_2^AS_3^AS_4^A\g_1\g_2\g_3\g_4=\pm\g_5\ ,
\end{equation}
with the plus (minus) sign if the number of components of $\p_A$ equal to $\p$ is even (odd).
This implies that half of the doublers have axial charge $+1$, and half of them have axial
charge $-1$.   The multiplet formed by the 16 species is thus anomaly free.   This explains
how the naive fermion action can regulate a Dirac fermion while maintaining exact chiral
symmetry (for $m=0$) \cite{KS}.   

We note that the action~(\ref{naive}) is invariant under hypercubic rotations in the $\m\n$ plane with 
the matrix
\begin{equation}
\label{Rhyper}
R_\half=e^{\half\omega\g_\m\g_\n}\ ,\qquad \m<\n
\end{equation}
with $\omega=\p/2$ acting on the Dirac spinor $\j$.\footnote{Note that the generators
$\half\g_\m\g_\n$, $\m<\n$, are anti-hermitian.}

\subsection{\label{spindiagsec} Spin diagonalization}
We now carry out the unitary basis transformation
\begin{equation}
\label{spindiag}
\j(x)=\g_1^{x_1}\g_2^{x_2}\g_3^{x_3}\g_4^{x_4}\c(x)\ ,\qquad \bj(x)=\bc(x)\g_4^{x_4}\g_3^{x_3}\g_2^{x_2}\g_1^{x_1}\ .
\end{equation}
This brings the action~(\ref{naive}) into the form
\begin{equation}
\label{stag}
S=\half\sum_{x\m}\eta_\m(x)\left(\bc(x)U_\m(x)\c(x+\m)-\bc(x+\m)U^\dagger_\m(x)\c(x)\right)
+m\sum_x\bc(x)\c(x)\ ,
\end{equation}
with the phase factors $\eta_\m(x)$ defined by
\begin{equation}
\label{etas}
\eta_\m(x)=(-1)^{x_1+\dots+x_{\m-1}}\ .
\end{equation}
We can make different choices in Eq.~(\ref{spindiag}) by reordering the $\g$ matrices, and this
will lead to different results for the phase factors $\eta_\m(x)$.   However, all essential
properties of these phase factors that we will use below are independent of this choice.

In Eq.~(\ref{stag}) the spin matrices $\g_\m$ have been diagonalized.   We now change the 
theory by dropping the spin index on the fields $\c$ and $\bc$.  This defines the staggered
fermion action,
with the ``one-component'' fields\footnote{Of course, these fields still carry
a color index and possibly a flavor index, so ``one-component'' here only refers to spin.}
 $\c$ and $\bc$
(for a discussion of the euclidean action and earlier references, see Ref.~\cite{KawS}).

The reduction from the naive to the staggered theory reduces the number of degrees of
freedom by a factor 4, so one might expect the staggered theory to only have 4 species
doublers in the continuum limit.   However, as this reduction was obtained by spin
diagonalization, it is not obvious that this is the correct interpretation of the staggered
theory.   In order to demonstrate this, one first should show that the action~(\ref{stag})
has the correct classical continuum limit, and then that quantum corrections do not 
change this conclusion.   If the theory is asymptotically free, so that the continuum limit
is taken at the gaussian fixed point, the latter question can be studied in weak-coupling
perturbation theory (WCPT), and we will do so in the next chapter.

In order to see that the staggered action~(\ref{stag}) may have a sensible continuum
limit, we will define several operators on the field $\c$.   First, we define
\begin{equation}
\label{Tmu}
T_{\pm\m}:\ \c(x)\to\eta_\m(x)\c(x\pm\m)\ ,
\end{equation}
from which it follows that
\begin{equation}
\label{twoTs}
T_\m T_\n:\ \c(x)\to \eta_\m(x)\eta_\n(x+\m)\c(x+\m+\n)=\varepsilon_{\m\n}(-1)^{x_1+\dots+x_{\m-1}+x_1+\dots+x_{\n-1}}\c(x+\m+\n)\ ,
\end{equation}
with
\begin{equation}
\label{epsmunu}
\varepsilon_{\m\n}=\Biggl\{\begin{array}{cc}+1\ , & \m\ge\n\\-1\ , & \m<\n\end{array}\ .
\end{equation}
From this, it follows that
\begin{equation}
\label{DiracT}
\{T_\m,T_\n\}:\ \c(x)\to 2\d_{\m\n}\c(x+2\m)\ .
\end{equation}
We see that the square of $T_\m$ is a normal translation over two lattice spacings in the $\m$ direction.
Moreover the $T_\m$ themselves satisfy the Dirac algebra modulo these translations by $2\m$.
The irreps of the group generated by the $T_\m$ live in momentum space, as $T_\m$ involves a
translation.
If we define $\hT_\m$ as the translation~(\ref{Tmu}) modding out the translations over $2\m$, 
the $\hT_\m$ generate a 32-element finite group isomorphic with that generated by the Dirac matrices
$\g_\m$.   We will denote this group as $\G_4$.\footnote{Not to be confused with the matrix $\G_4$
that will be introduced later on.  The context should help avoid confusion.}

This group has 16 one-dimensional irreps (obtained by mapping the matrices
$\g_\m$ onto $\pm 1$ in all possible
ways), and one four-dimensional irrep generated by the $\g_\m$.    Since the 16 fields $\c(x)$ within a hypercube form a
16-dimensional representation of $\G_4$, this representation is reducible (but it will be irreducible 
under the full lattice symmetry group, see Sec.~\ref{symmetries}).   We thus expect that a basis
transformation exists such that the $\hT_\m$ can be written as
\begin{equation}
\label{gammamu}
\hT_\m=\g_\m\times 1_4\ ,
\end{equation}
with $1_4$ the $4\times 4$ unit matrix.

We expect similar transformations to exist of the form $1_4\times \x_\n$,
with $\x_\n$ also satisfying the Dirac algebra.   Indeed, we can define {\em shifts}
\cite{vdDS,GS}
\begin{equation}
\label{shift}
S_{\pm\m}:\ \c\to\z_\m(x)\c(x\pm\m)\ ,
\end{equation}
and require the new phase factors $\z_\m(x)$ to be such that
\begin{equation}
\label{STcomm}
[S_\m,T_\n]=0\ .
\end{equation}  
This implies that
\begin{equation}
\label{etaxi}
\z_\m(x)\eta_\n(x+\m)=\eta_\n(x)\z_\m(x+\n)\ ,
\end{equation}
which can be rewritten as
\begin{equation}
\label{zetacond}
\z_\m(x+\n)=\eta_\n(x)\eta_\n(x+\m)\z_\m(x)=\varepsilon_{\m\n}\z_\m(x)\ ,
\end{equation}
and thus
\begin{equation}
\label{zeta}
\z_\m(x)=(-1)^{x_{\m+1}+\dots+x_4}\ .
\end{equation}
With $S_\m$ acting on the gauge fields as
\begin{equation}
\label{SU}
S_\m:\ U_\n(x)\to U_\n(x+\m)\ ,
\end{equation}
$S_\m$ is a symmetry of the action~(\ref{stag}), and
\begin{equation}
\label{DiracS}
\{S_\m,S_\n\}:\ \c(x)\to 2\d_{\m\n}\c(x+2\m)\ .
\end{equation}
We note the similarity to Eq.~(\ref{DiracT}).
We will refer to this symmetry as
shift symmetry.   It is a crucial symmetry for staggered fermions, as we will see below.
In contrast, the $T_\m$ are not a symmetry of the theory, but they are helpful 
for the interpretation of the theory:  as we will see in more detail below, they 
``generate'' the spin matrices $\g_\m$ needed to interpret the continuum limit as
a theory of four Dirac fermions.

It is instructive to recast the operators $T_\m$ and $S_\m$ in momentum space.
We first define a subset of the vectors $\p_A$:
\begin{eqnarray}
\label{pieta}
\p_{\eta_1}&=&(0,0,0,0)\ ,\\
\p_{\eta_2}&=&(\p,0,0,0)\ ,\nonumber\\
\p_{\eta_3}&=&(\p,\p,0,0)\ ,\nonumber\\
\p_{\eta_4}&=&(\p,\p,\p,0)\ ,\nonumber
\end{eqnarray}
in terms of which 
\begin{equation}
\label{etarew}
\eta_\m(x)=e^{i\p_{\eta_\m}x}\ .
\end{equation}
We then have
\begin{equation}
\eta_\m(x)\c(x+\m)=\int_p e^{ipx+i\p_{\eta_\m}x+ip_\m}\tc(p)=\int_p e^{ipx}e^{ip_\m}\tc(p+\p_{\eta_\m})\ ,
\end{equation}
where we used that $(\p_{\eta_\m})_\m=0$ and that the
exponential has a periodicity $2\p$ in each direction because $x_\m$ is integer on the lattice.   We now define the full Brillouin zone (BZ) as
\begin{equation}
\label{BZ}
-\frac{\p}{2}<p_\m\le\frac{3\p}{2}\ ,
\end{equation}
and the {\em reduced} BZ as
\begin{equation}
\label{rBZ}
-\frac{\p}{2}<\tp_\m\le\frac{\p}{2}\ .
\end{equation}
We write $p=\tp+\p_A$ and define
\begin{equation}
\label{phi}
\f_A(\tp)=\c(\tp+\p_A)\ .
\end{equation}
It follows that
\begin{equation}
\label{Tmumom}
T_\m:\ \f_A(\tp)=\c(\tp+\p_A)\to S^A_\m e^{i\tp_\m}\c(\tp+\p_A+\p_{\eta_\m})=S_\m^A(\hp_{\eta_\m})_{AB}e^{i\tp_\m}\f_B(\tp)\ ,
\end{equation}
with
\begin{equation}
\label{pihat}
(\hp_{\eta_\m})_{AB}=\bd(\p_{\eta_\m}+\p_A+\p_B)\ ,
\end{equation}
and $\bd(\p_A)=1$ for $\p_A=0$ (mod $2\p$) and zero otherwise.
Defining also
\begin{equation}
\label{Smu}
(S_\m)_{AB}=S_\m^A\d_{AB}
\end{equation}
and
\begin{equation}
\label{Gamma}
\G_\m=S_\m\hp_{\eta_\m}\ ,
\end{equation}
Eq.~(\ref{Tmumom}) can be rewritten as
\begin{equation}
\label{Tmumom2}
T_\m:\ \f(\tp)\to e^{i\tp_\m}\G_\m\f(\tp)\ .
\end{equation}
From Eq.~(\ref{DiracT}) it then follows that
\begin{equation}
\label{GDirac}
\{\G_\m,\G_\n\}=2\d_{\m\n}\ ,
\end{equation}
\ie, the $16\times 16$ matrices $\G_\m$ obey the Dirac algebra.

Analogously, we can define
\begin{eqnarray}
\label{pizeta}
\p_{\z_1}&=&(0,\p,\p,\p)\ ,\\
\p_{\z_2}&=&(0,0,\p,\p)\ ,\nonumber\\
\p_{\z_3}&=&(0,0,0,\p)\ ,\nonumber\\
\p_{\z_4}&=&(0,0,0,0)\ ,\nonumber
\end{eqnarray}
and
\begin{eqnarray}
\label{pihatXi}
(\hp_{\z_\m})_{AB}&=&\bd(\p_{\z_\m}+\p_A+\p_B)\ ,\\
\X_\m&=&S_\m\hp_{\z_\m}\ ,\nonumber
\end{eqnarray}
to rewrite Eq.~(\ref{shift}) in momentum space as
\begin{equation}
\label{Smom}
S_\m:\ \f(\tp)\to e^{i\tp_\m}\X_\m\f(\tp)\ ,
\end{equation}
with $\X_\m$ another set of $16\times 16$ matrices.
From Eqs.~(\ref{STcomm}) and~(\ref{DiracS}), it follows that
\begin{equation}
\label{Xirules}
\{\X_\m,\X_\n\}=2\d_{\m\n}\ ,\qquad [\G_\m,\X_\n]=0\ .
\end{equation}
As in Eq.~(\ref{gammamu}), we can find a basis transformation such that
\begin{equation}
\label{basistrtogxi}
\G_\m\X_\n\to\g_\m\times\x_\n\ ,
\end{equation}
with $\g_\m$ and $\x_\n$ both sets of $4\times 4$ matrices satisfying the
Dirac algebra.   We note that
\begin{eqnarray}
\label{rels}
&&[S_\m,\hp_{\eta_\m}]=[S_\m,\hp_{\z_\m}]=0\ ,\\
&&S_\m=S_\m^T\ ,\quad \hp_{\eta_\m}=\hp_{\eta_\m}^T\ ,\quad \hp_{\z_\m}=\hp_{\z_\m}^T
\quad\Rightarrow\quad \G_\m=\G_\m^T\ ,\quad\X_\m=\X_\m^T\ .\nonumber
\end{eqnarray}
For an explicit representation, see Ref.~\cite{GS}.

\subsection{\label{Feynman} Feynman rules and classical continuum limit}
Using Eq.~(\ref{etarew}), the free staggered action can be written in momentum space
as
\begin{equation}
\label{stagmomfree}
S_{\rm free}=\half\int_p\int_q\tbc(q)\left(\sum_\m\d(p-q+\p_{\eta_\m})(e^{ip_\m}-e^{-iq_\m})+m\d(p-q)\right)\tc(p)\ ,
\end{equation}
from which we obtain the inverse propagator (using $(\p_{\eta_\m})_\m=0$ again)
\begin{equation}
\label{invG}
G^{-1}(p,-q)=\sum_\m\d(p-q+\p_{\eta_\m})i\sin(p_\m)+m\d(p-q)\ ,
\end{equation}
where $\d$ is the periodic delta function with period $2\p$ in each direction.
We now set
\begin{equation}
\label{redBZ}
p=\tp+\p_A\ ,\qquad q=\tq+\p_B\ ,
\end{equation}
to obtain
\begin{eqnarray}
\label{Ginv2}
G^{-1}(\tp+\p_A,-\tq-\p_B)&=&\sum_\m\d(\tp-\tq+\p_A+\p_B+\p_{\eta_\m})\,i\sin(p_\m)\\
&&\hspace{1cm}+m\d(\tp-\tq+\p_A+\p_B)\nonumber\\
&=&\d(\tp-\tq)\left(iS^A_\m(\hp_{\eta_\m})_{AB}\sin(\tp_\m)+m\d_{AB}\right)\nonumber\\
&\equiv& \d(\tp-\tq)S^{-1}_{AB}(\tp)\ .\nonumber
\end{eqnarray}
The second line follows, because, from Eq.~(\ref{rBZ}), $-\frac{\p}{2}<\tp_\m-\tq_\m<\frac{\p}{2}$, with
strict inequalities.   We thus find for the free propagator
\begin{equation}
\label{Sprop}
S(\tp)=\left(\sum_\m i\G_\m\sin(\tp_\m)+m\right)^{-1}\ ,
\end{equation}
and the free action becomes
\begin{equation}
\label{Sfreephi}
S_{\rm free}=\int_{\tp}\sum_{AB}\bff_A(\tp)\left(\sum_\m i(\G_\m)_{AB}\sin(\tp_\m)+m\d_{AB}\right)\f_B(\tp)\ .
\end{equation}
This action is invariant under a group $U(4)$ generated by the $\X_\m$ (all products of these
matrices precisely give the generators of the group $U(4)$), because this group commutes with $\G_\m$.
Free staggered fermions contain 4 degenerate species of Dirac spinors in the continuum limit.   We
will refer to these 4 species as ``tastes,'' and we find that the free theory has an exact
$U(4)$ taste symmetry on the lattice.   In the interacting theory, the product of this $U(4)$ and lattice
translations is broken to the discrete symmetries $S_\m$, {\it cf.} Eqs.~(\ref{shift}),~(\ref{SU}) and~(\ref{Smom}).

To obtain the one-gluon vertex, we expand
\begin{equation}
\label{link}
U_\m(x)=e^{igA_\m(x)}=1+igA_\m(x)-\half\, g^2A_\m^2(x)+\dots\ ,
\end{equation}
and Fourier transform
\begin{equation}
\label{FTA}
A_\m(x)=\int_k e^{ikx}\tA_\m(k)\ .
\end{equation}
The one-gluon part of the action is thus
\begin{equation}
\label{S1gluon}
S_{\rm 1-gluon}=ig\int_p\int_q\int_k\sum_\m\tbc(q)\d(p+k-q+\p_{\eta_\mu})\half(e^{ip_\m}+e^{-iq_\m})\tA_\m(k)\tc(p)\ ,
\end{equation}
from which we read off the one-gluon vertex
\begin{equation}
\label{1gvertex}
V^m_\m(p,-q,k)=-ig\,\d(p+k-q+\p_{\eta_\mu})\,\half(e^{ip_\m}+e^{-iq_\m})\,t_m\ ,
\end{equation}
where $\tA_\m(k)=\sum_m\tA_\m^m(k)t_m$, with $t_m$ the gauge group generators.

We can take the classical continuum limit by setting $p=\tp+\p_A$, $q=\tq+\p_B$ with
$\tp$, $\tq$ and $k$ small.   Using
\begin{equation}
\label{cos}
\half(e^{ip_\m}+e^{-iq_\m})=\half(S_\m^A e^{i\tp_\m}+S_\m^Be^{-i\tq_\m})\approx\half(S_\m^A
+S_\m^B)
\end{equation}
and
\begin{equation}
\label{pqksmall}
\d(\tp-\tq+k+\p_A+\p_B+\p_{\eta_\m})=\d(\tp-\tq+k)(\hp_{\eta_\m})_{AB}
\end{equation}
for $\tp$, $\tq$ and $k$ small, the one-gluon vertex becomes
\begin{equation}
\label{1glclass}
V^m_\m(\tp+\p_A,-\tq-\p_B,k)\approx -ig\,\d(\tp-\tq+k)(\G_\m)_{AB}t_m\ ,
\end{equation}
which is the correct classical continuum limit.   In particular, the classical 
continuum limit is invariant under $U(4)_{\rm taste}$.

The two-gluon vertex can be read off using the $O(g^2)$ part of the expansion~(\ref{link}),
with $\ell$ the momentum of the second gluon
\begin{equation}
\label{2gvertex}
V^{m_1m_2}_{\m_1\m_2}(p,-q,k,\ell)=\half\, g^2\sum_\m\d(p+k+\ell-q+\p_{\eta_\mu})\,\half(e^{ip_\m}-e^{-iq_\m})
\,\d_{\m\m_1}\d_{\m\m_2}\,\half\{t_{m_1},t_{m_2}\}\ .
\end{equation}
Because of the minus sign between the momentum-dependent phase factors, it is straightforward
to show that this vertex vanishes in the classical continuum limit.

We end this subsection with the observation that if we take $k_\m=\p$ in Eq.~(\ref{1gvertex}), and
use
\begin{equation}
\label{kmupi}
\half(e^{ip_\m}+e^{-iq_\m})=\half(e^{ip_\m}-e^{-ip_\m})=i\sin(p_\m)=iS_\m^A\sin(\tp_\m)
\end{equation}
for this choice of $k$, we find that the one-gluon vertex for a gluon with momentum $k_\m=\p$ is 
suppressed for small $\tp_\m$.   We will return to this point in Sec.~\ref{improvement}.
For the complete Feynman rules, we refer to Ref.~\cite{GS}.

\subsection{\label{symmetries} More symmetries}
We have already identified one symmetry of the staggered action, the shifts defined by
Eqs.~(\ref{shift}) and~(\ref{SU}).
There are more symmetries of course, which we will discuss in this subsection.   For 
more detail, see Ref.~\cite{GS}.

\subsubsection{\label{rot} Hypercubic rotations}
The action is invariant under hypercubic rotations, which are rotations over an angle
$\p/2$ in each plane.   Clearly, the action~(\ref{Sfreephi}) is invariant under 
rotations $R_{\r\s}$ in the $\r\s$ plane over an angle $\p/2$ which takes $\tp_\r\to\tp_\s$
and $\tp_\s\to-\tp_\r$, and
\begin{equation}
\label{rot1}
\f(\tp)\to e^{\frac{\p}{4}\G_\r\G_\s}\f(R_{\r\s}^{-1}\tp)\qquad(\mbox{tentative})\ .
\end{equation}
However, this is in the free theory, which has more symmetry than the interacting theory.
Noting that the operators $T_\m$ and $S_\m$ are expected to transform similarly under
hypercubic rotations, we replace Eq.~(\ref{rot1}) by
\begin{equation}
\label{rot2}
\f(\tp)\to e^{\frac{\p}{4}\G_\r\G_\s}e^{\frac{\p}{4}\X_\r\X_\s}\f(R_{\r\s}^{-1}\tp)\ .
\end{equation}
Of course, the gauge fields have to be rotated accordingly, and Eq.~(\ref{rot2}) can be
translated back to position space.   The form of rotations on the original staggered
fields $\c(x)$ and $\bc(x)$ as well as the links $U_\m(x)$ is given in Ref.~\cite{GS}.
Equation~(\ref{rot2}), along with the transformation on the gauge fields, constitutes
the hypercubic rotational symmetry of the staggered action.   These rotations form
the discrete hypercubic subgroup of a group $SO(4)$, which is the diagonal
subgroup of $SO(4)_{\rm spin}\times SO(4)_{\rm taste}$, with $SO(4)_{\rm taste}$
the unique $SO(4)$ subgroup of $SU(4)_{\rm taste}$.

The field $\f$ transforms in a bosonic irrep of Eq.~(\ref{rot2}).   However, we can define
a different group of lattice symmetries under which it transforms as a fermion field.
Since we have shift and rotational symmetry, we can define a composite symmetry
\begin{equation}
\label{RS}
\tilde{R}_{\r\s}=R_{\r\s}S_\r\ ,
\end{equation}
which can be visualized as rotations around the center of the hypercube based at
the origin.   Using that
\begin{equation}
\label{RonS}
R_{\r\s}^{-1}S_\r R_{\r\s}=S_\s\ ,\qquad R_{\r\s}^{-1}S_\s R_{\r\s}=-S_\r\ ,
\end{equation}
one can prove that, modulo lattice translations over an even number of lattice
spacings in any direction,
\begin{equation}
\label{Rtilde4}
(\tilde{R}_{\r\s})^4 = -1\ ,
\end{equation}
\ie, that a rotation over $2\p$ exists which is equal to $-1$, as one would expect
for a fermionic theory.  In contrast, the fourth power of $R_{\r\s}$ equals
$+1$. In the construction of the symmetry~(\ref{RS}) shift symmetry
plays a key role.     In momentum space, we find that
\begin{equation}
\label{tRmom}
\tilde{R}_{\r\s}=e^{i\tp_\r}e^{\frac{\p}{4}\G_\r\G_\s}\frac{1}{\sqrt{2}}(\X_\r-\X_\s)
=-ie^{i\tp_\r}e^{\frac{\p}{4}\G_\r\G_\s}e^{i\frac{\p}{2}\frac{1}{\sqrt{2}}(\X_\r-\X_\s)}\ ,
\end{equation}
using
\begin{equation}
\label{XrmXs}
\left(\frac{1}{\sqrt{2}}(\X_\r-\X_\s)\right)^2=1\ .
\end{equation}

\subsubsection{\label{reversal} Axis reversal}
Axis reversal $I_\r$ in the $\r$ direction takes $x_\r\to -x_\r$, or $\tp_\r\to-\tp_\r$ in momentum
space.   Again, we
expect symmetry under the interchange $\G_\m\leftrightarrow\X_\m$, and in momentum
space,
\begin{equation}
\label{axisrev}
\f(\tp)\to\G_\r\G_5\X_\r\X_5\f(I_\r\tp)\ ,
\end{equation}
where
\begin{equation}
\label{G5X5}
\G_5=\G_1\G_2\G_3\G_4\ ,\qquad \X_5=\X_1\X_2\X_3\X_4\ .
\end{equation}
This form follows because we need the matrix acting on $\f$ to commute with $\G_\m$
for $\m\ne\r$ and anticommute with $\G_\r$, because $\sin(I_\r\tp_\r)=-\sin(\tp_\r)$
({\it cf.} Eq.~(\ref{Sfreephi})).  In position space, we have that \cite{GS}
\begin{equation}
\label{Irhopos}
I_\r:\quad\c(x)\to(-1)^{x_\r}\c(I_\r x)\ .
\end{equation}
We can define a parity transformation,
\begin{equation}
\label{Is}
I_s=I_1I_2I_3\ ,
\end{equation}
but we note that this involves a taste transformation:
\begin{equation}
\label{Isphi}
I_s:\ \f(\tp)\to\G_4\X_4\f(I_s\tp)\ .
\end{equation}
Thus, ``tasteless'' parity $P$ can be defined by combining this with $S_4$,
\begin{equation}
\label{P}
P:\ \f(\tp)\to e^{i\tp_4}\G_4\f(I_s\tp)\ ,
\end{equation} 
or, in position space,
\begin{equation}
\label{Ppos}
P:\ \c(x)\to (-1)^{x_1+x_2+x_3}\c(I_s x+4)=\eta_4(x)\c(I_s x+4)\ .
\end{equation}
Of course, the gauge field also transforms non-trivially under axis reversal \cite{GS}.

\subsubsection{\label{FN} Fermion number}
Because the action is bilinear in $\c$ and $\bc$, it is invariant under
\begin{equation}
\label{fermnum}
\c\to e^{i\a}\c\ ,\qquad\bc\to\bc e^{-i\a}\ .
\end{equation}

\subsubsection{\label{U1eps} $U(1)_\e$ symmetry}
For $m=0$, the action is invariant under \cite{KS}
\begin{equation}
\label{U1epssymm}
\c(x)\to e^{i\b\e(x)}\c(x)\ ,\qquad \bc(x)\to\bc(x) e^{i\b\e(x)}\ ,
\end{equation}
in which
\begin{equation}
\label{eps}
\e(x)=(-1)^{x_1+x_2+x_3+x_4}=e^{i\p_\e x}\ ,\qquad \p_\e=(\p,\p,\p,\p)\ .
\end{equation}
Since this is only a symmetry of the action for $m=0$, we can interpret this symmetry as a 
chiral symmetry.   (As always, this interpretation depends on the form of the mass term.
For instance, for a one-link mass term, $U(1)_\e$ is a vectorlike symmetry \cite{GS}!)

Using that 
\begin{equation}
\label{epsTS}
T_4T_3T_2T_1S_{-1}S_{-2}S_{-3}S_{-4}:\ \c(x)\to\e(x)\c(x)\ ,
\end{equation}
and using the fact that $T_{\pm\m}$ ($S_{\pm\m}$) generates a $\G_\m$ ($\X_\m$), we find that
\begin{equation}
\label{hatpieps}
\hp_\e=\G_5\X_5\ ,
\end{equation}
and thus that in momentum space $U(1)_\e$ symmetry takes the form
\begin{equation}
\label{U1epsmom}
\f(\tp)\to e^{i\b\G_5\X_5}\f(\tp)\ ,\qquad \bff(\tp)\to\bff(\tp)e^{i\b\G_5\X_5}\ .
\end{equation}
Indeed, this looks like an non-singlet axial transformation.   

\subsubsection{\label{charge} Charge conjugation}
The staggered action is invariant under charge conjugation $C_0$:
\begin{eqnarray}
\label{cconj}
&&\c(x)\to i\e(x)\bc^T(x)\ ,\qquad \bc(x)\to i\e(x)\c^T(x)\ ,\\
&&U_\m(x)\to U^*_\m(x)\ ,\nonumber
\end{eqnarray}
where the transpose acts on the color (and possible flavor) indices.   The factors
$i\e(x)$ are needed to make both the mass term and the kinetic term invariant
under this symmetry.   This implies that charge conjugation involves a taste
transformation; in momentum space it looks like
\begin{equation}
\label{cconjmom}
\f(\tp)\to i\G_5\X_5\bff^T(-\tp)\ ,\qquad \bff(\tp)\to i\G_5\X_5\f^T(-\tp)\ .
\end{equation}

The hope is, of course, that the lattice symmetries are sufficient to restore the
full $SO(4)_{\rm rot}\,\times\,SU(4)_{\rm taste}$ in the continuum limit without any
fine tuning of counter terms.
We have noted already that the hypercubic rotations~(\ref{rot2}) are a discrete subgroup of
the diagonal $SO(4)_{\rm diag}$ of $SO(4)_{\rm rot}\,\times\,SO(4)_{\rm taste}$, with
$SO(4)_{\rm taste}\,\subset\,SU(4)_{\rm taste}$.   Likewise, the group
$\G_4$ generated by shifts (modulo even translations) is a discrete subgroup
of $SU(4)_{\rm taste}$.   Finally, $U(1)_\e$ provides one exact (non-singlet, 
in taste space) chiral symmetry.   At the classical level, we have seen that
indeed the continuum limit coincides with QCD.   We will revisit this question 
at the one-loop level in the next chapter, and in terms of the Symanzik
effective theory in Sec.~\ref{SET}.

\subsection{\label{taste} Taste basis}
Another basis that can sometimes be useful has been introduced a long time
ago, the so-called taste basis \cite{Gliozzi,Saclay}.   We will review the construction
here for the free theory.   

We define a unitary transformation
\begin{equation}
\label{tasteb}
\j_{\a a}(y)=\frac{1}{2^{3/2}}\sum_A(\g_A)_{\a a}\c(2y+A)\ ,\qquad
\bj_{a\a}(y)=\frac{1}{2^{3/2}}\sum_A\bc(2y+A)(\g_A)^\dagger_{a\a}\ .
\end{equation}
Here $A$ runs over the 16 vectors with components $0$ or $1$.  
The $\j$ fields live on a ``coarse'' lattice, labeled by lattice sites $y$,
and 
\begin{equation}
\label{gammaA}
\g_A = \g_1^{A_1}\g_2^{A_2}\g_3^{A_3}\g_4^{A_4}\ .
\end{equation}
The index $\a$ will be interpreted as a spin index, and the index $a$
as a taste index below.   In order to write the (free) one-component
staggered fermion action in terms of the fields $\j$ and $\bj$, we need
to invert Eq.~(\ref{gammaA}).  Using 
\begin{equation}
\label{trace}
\tr(\g_B^\dagger\g_A)=4\d_{AB}\ ,
\end{equation}
we find that
\begin{equation}
\label{tastebinv}
\c(2y+A)=\frac{1}{2^{5/2}}\tr(\g_A^\dagger\j(y))\ ,\qquad
\bc(2y+A)=\frac{1}{2^{5/2}}\tr(\g_A\bj(y))\ .
\end{equation}
We will also need the completeness relation
\begin{equation}
\label{compl}
\sum_A(\g_A)_{\a a}(\g_A)^\dagger_{b\b}=4\d_{\a\b}\d_{ab}\ .
\end{equation}
Using this, it is straightforward to show that
\begin{equation}
\label{massterm}
m\sum_x\bc(x)\c(x)=\frac{1}{8}m\sum_y\bj(y)\j(y)\ .
\end{equation}
The kinetic term is a little more involved:
\begin{eqnarray}
\label{kinterm}
&&\half\sum_{x\m}\eta_\m(x)(\bc(x)\c(x+\m)-\bc(x+\m)\c(x))\\
&=&\half\sum_{yA\mu}\eta_\m(A)(\bc(2y+A)\c(2y+A+\m)-\bc(2y+A+\m)\c(2y+A))\nonumber\\
&=&\half\sum_{yA\mu,A_\m=0}\eta_\m(A)(\bc(2y+A)\c(2y+A+\m)-\bc(2y+A+\m)\c(2y+A))\nonumber\\
&&+\half\sum_{yA\mu,A_\m=1}\eta_\m(A)(\bc(2y+A)\c(2y+A+\m)-\bc(2y+A+\m)\c(2y+A))\nonumber\\
&=&\frac{1}{2^6}\sum_{yA\mu,A_\m=0}\eta_\m(A)(\tr(\g_A\bj(y))\tr(\g_{A+\m}^\dagger\j(y))-
\tr(\g_{A+\m}\bj(y))\tr(\g_A^\dagger\j(y)))\nonumber\\
&&+\frac{1}{2^6}\sum_{yA\mu,A_\m=0}\eta_\m(A)(\tr(\g_{A+\m}\bj(y))\tr(\g_A^\dagger\j(y+\m))-
\tr(\g_A\bj(y+\m))\tr(\g_{A+\m}^\dagger\j(y)))\ .\nonumber
\end{eqnarray}
We now use that
\begin{eqnarray}
\label{rel1}
\sum_{A,A_\m=0}\eta_\m(A)(\g_{A+\m})_{\a a}(\g_A^\dagger)_{b\b}&=&
\sum_A\eta_\m(A)(\g_{A+\m})_{\a a}(\g_A^\dagger)_{b\b}\half((1+(-1)^{A_\m})\\
&&\hspace{-4cm}=\half\sum_A(\g_\m)_{\a\g}(\g_A)_{\g a}(\g^\dagger_A)_{b\b}
+\half\sum_A(-1)^{A_\m}\eta_\m(A)\z_\m(A)(\g_A)_{\a c}(\g_\m)_{ca}(\g^\dagger_A)_{b\b}\nonumber\\
&&\hspace{-4cm}=\half\sum_A(\g_\m)_{\a\g}(\g_A)_{\g a}(\g^\dagger_A)_{b\b}
+\half\sum_A\e(A)(\g_A)_{\a c}(\g_\m)_{ca}(\g^\dagger_A)_{b\b}\nonumber\\
&&\hspace{-4cm}=\half\sum_A(\g_\m)_{\a\g}(\g_A)_{\g a}(\g^\dagger_A)_{b\b}
+\half\sum_A(\g_5)_{\a\g}(\g_A)_{\g d}(\g_5)_{dc}(\g_\m)_{ca}(\g^\dagger_A)_{b\b}\nonumber\\
&&\hspace{-4cm}=2(\g_\m)_{\a\b}\d_{ab}+2(\g_5)_{\a\b}(\g_5\g_\m)_{ba}\ ,
\nonumber
\end{eqnarray}
and, likewise,
\begin{equation}
\label{rel2}
\sum_{A,A_\m=0}\eta_\m(A)(\g_A)_{\a a}(\g_{A+\m}^\dagger)_{b\b}
=2(\g_\m)_{\a\b}\d_{ab}-2(\g_5)_{\a\b}(\g_5\g_\m)_{ba}\ .
\end{equation}
This allows us to write Eq.~(\ref{kinterm}) as
\begin{eqnarray}
\label{kinterm2}
&&\frac{1}{2^4}\half\sum_{y\m}\Biggl(\tr(\bj(y)\g_\m\j(y+\m)-\bj(y+\m)\g_\m\j(y))\\
&&\hspace{1.3cm}+\tr(\bj(y)\g_5\j(y+\m)\g_5\g_\m+\bj(y+\m)\g_5\j(x)\g_5\g_\m-2\bj(y)\g_5\j(y)\g_5\g_\m)
\Biggr)\ .
\nonumber
\end{eqnarray}
We see that in this form, the free action is that of a naive fermion, with the addition of a 
spin ($\g_5$) and taste ($\g_5\g_\m$) non-diagonal Wilson-like mass term, which removes
the doublers.   The fermion field $\j$ is thus undoubled, and shows the explicit
four-fold degeneracy of staggered fermions in the continuum limit.   

One can also translate the shift symmetry to the new basis.   One obtains, in the free
theory, that 
\begin{equation}
\label{shifttasteb}
S_\m:\ \j(y)\to\half(\j(x)+\j(x+\m))\g_\m+\half\g_5\g_\m(\j(x)-\j(x+\m))\g_5\approx \j(y)\g_\m\ ,
\end{equation}
where the approximate equality on the right applies in the continuum limit.   We can identify a taste
matrix $\x_\m=\g_\m^T$ acting on the taste index of the field $\j$.

It is tempting to gauge Eq.~(\ref{kinterm2}) on the coarse lattice, instead of starting from 
the one-component staggered action.   However, it is clear that doing this breaks
shift symmetry,\footnote{Gauging also Eq.~(\ref{shifttasteb}) does not make it into a
symmetry of the gauged version of Eq.~(\ref{kinterm2}).} and we would thus end up 
with a different interacting theory, with 
fewer lattice symmetries.   We will see in the next chapter that this would undo
a key property of staggered fermions, which is that the mass renormalizes
multiplicatively.

\section{\label{oneloop} One loop calculations}
In this section, we calculate the one-loop vacuum polarization, and demonstrate that
it satisfies the Ward--Takahashi identity (WTI), which follows from the gauge invariance of
the lattice theory.   We also consider the one-loop fermion
self energy, and show how staggered symmetries limit possible counter terms.   These 
examples illustrate the conjecture that lattice QCD (or QED) with staggered fermions 
has a well-defined relativistic continuum limit, with no need for additional counter terms
in the lattice action.

\subsection{\label{vacpolcalc} Vacuum polarization}
In Fig.~\ref{vacpol} we show the two diagrams contributing at one loop to the vacuum 
polarization.   For simplicity, we consider the case of QED with gauge coupling $g$.
Using the Feynman rules of Sec.~\ref{Feynman}, the tadpole diagram is equal to
\begin{equation}
\label{tadpole}
\P^{\rm tadpole}_{\m\n}(k,\ell)=-g^2\,\d_{\m\n}\int_{pq}\d(q-p+k-\ell+\p_{\eta_\m})\,\half\left(
e^{i(q+k-\ell)_\m}-e^{-iq_\m}\right)G(p,-q)\ .
\end{equation}
Setting $p=\tp+\p_A$ and $q=\tq+\p_B$ this becomes
\begin{equation}
\label{tadpole2}
-g^2\d_{\m\n}e^{\frac{i}{2}(k-\ell)_\m}\!\!\int_{\tp\tq}\sum_{AB}\d(\tq-\tp+k-\ell+\p_A+\p_B+\p_{\eta_\m})
i\sin(\tq_\m+\mbox{$\small{\half}$}(k_\m-\ell_\m))S_\m^B G(\tp+\p_A,-\tq-\p_B)\ .
\end{equation}
We now use that ({\it cf.} Eq.~(\ref{Ginv2}))
\begin{equation}
\label{G2}
G(\tp+\p_A,-\tq-\p_B)=\d(\tp-\tq)S_{AB}(\tp)\ ,
\end{equation}
which sets the combination $\tq-\tp$ in $\d(\tq-\tp+k-\ell+\p_A+\p_B+\p_{\eta_\m})$ equal to
zero.
Then, we use that $k$ and $\ell$, being external momenta, are taken small, so that we
can factorize 
\begin{equation}
\label{fact}
\d(k-\ell+\p_A+\p_B+\p_{\eta_\m})=\d(k-\ell)(\hp_{\eta_\m})_{AB}\ ,
\end{equation}
and use Eq.~(\ref{Gamma}) to arrive at
\begin{eqnarray}
\label{tadpole3}
\P^{\rm tadpole}_{\m\n}(k)&=&-g^2\d_{\m\n}\int_{\tq}\tr\left(\G_\m\frac{-i\sum_\k \G_\k\sin(\tq_\k)+m}{\sum_\k\sin^2(\tq_\k)+m^2}\right)i\sin(\tq_\m)\\
&=&-16g^2\d_{\m\n}\int_{\tq}\frac{\sin^2(\tq_\m)}{\sum_\k\sin^2(\tq_\k)+m^2}\ ,\nonumber
\end{eqnarray}
where we omitted $\d(k-\ell)$.   We see that $\P^{\rm tadpole}_{\m\n}(k)$ does not actually 
depend on $k$, as one would expect for a tadpole diagram.

\begin{figure}[!t]
\vspace*{1ex}
\begin{center}
\begin{picture}(50,100)(5,0)
  \put(-150,0){\includegraphics*[width=2in]{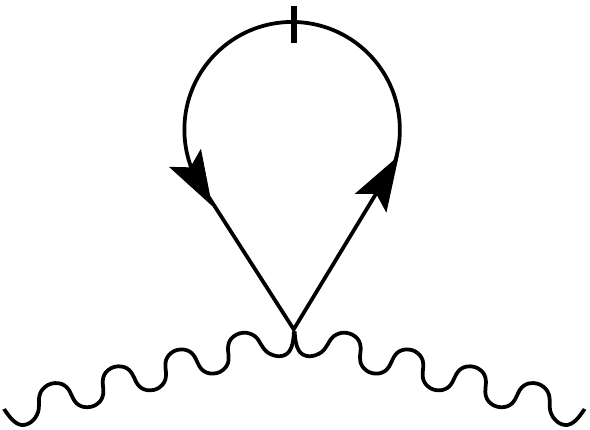}}
  \put(51,8){\includegraphics*[width=2.5in]{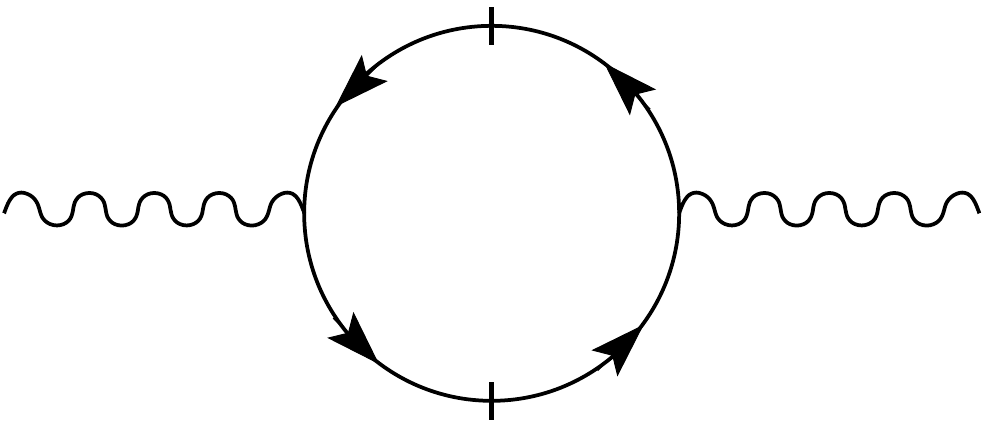}}
  \put(-160,0){$\m$}
  \put(0,0){$\n$}
  \put(-120,-5){$\to k$}
  \put(-60,-5){$\to \ell$}
  \put(-45,65){$p$}
  \put(-118,65){$q$}
  \put(40,43){$\m$}
  \put(235,43){$\n$}
  \put(105,75){$q'$}
  \put(172,75){$p'$}
  \put(105,15){$p$}
  \put(172,15){$q$}
  \put(70,30){$\to k$}
  \put(190,30){$\to \ell$}
\end{picture}
\vspace*{5ex}
\floatcaption{vacpol}
             {The two diagrams contributing to the one-loop vacuum polarization.
The left panel shows the tadpole diagram, the right panel the sunset diagram.}
\end{center}
\end{figure}

The other diagram in Fig.~\ref{vacpol} is
\begin{eqnarray}
\label{sunset}
\P^{\rm sunset}_{\m\n}(k,\ell)&=&g^2\int_{pqp'q'}\d(q'-p+k+\p_{\eta_\m})\,\half\left(e^{i(q'+k)_\m}+e^{-iq'_\m}\right)G(p,-q)\\
&&\hspace{1.2cm}\times\d(q-p'-\ell+\p_{\eta_\n})\,\half\left(e^{i(q-\ell)_\n}+e^{-iq_\n}\right)G(p',-q')\ .\nonumber
\end{eqnarray}
Integrating over $p'$ and $q'$ and writing $p=\tp+\p_A$, $q=\tq+\p_B$ makes this equal to
\begin{eqnarray}
\label{sunset2}
\P^{\rm sunset}_{\m\n}(k,\ell)&=&g^2\int_{\tp\tq}\sum_{AB}\half S_\m^A\left(e^{i\tp_\m}+e^{-i(\tp-k)_\m}\right)\d(\tp-\tq)S_{AB}(\tp)\\
&&\times\half S_\n^B\left(e^{i(\tq-\ell)_\n}+e^{-i\tq_\n}\right)G(\tq-\ell+\p_{\eta_\n}+\p_B,-\tp+k+\p_{\eta_\m}+\p_A)\ ,\nonumber
\end{eqnarray}
where we used the inverse of Eq.~(\ref{Ginv2}).
Using that we can set $\tq=\tp$ because of the $\d(\tp-\tq)$ inside the integrand, we use Eqs.~(\ref{invG}) to write
\begin{eqnarray}
\label{GtoS}
&&G^{-1}(\tp-\ell+\p_{\eta_\n}+\p_B,-\tp+k+\p_{\eta_\m}+\p_A)=\\
&&\hspace{1cm}\sum_\k\d(-\ell+k+\p_{\eta_\k}+\p_{\eta_\n}+\p_B+\p_{\eta_\m}+\p_A)i S_\k^{\eta_\n}S_\k^B\sin(\tp_\k-\ell_\k)\nonumber\\
&&\hspace{1cm}+m\d(-\ell+k+\p_{\eta_\n}+\p_B+\p_{\eta_\m}+\p_A)\ .\nonumber
\end{eqnarray}
Now, using that $k$ and $\ell$ are small,
\begin{eqnarray}
\label{massdelta}
\d(-\ell+k+\p_{\eta_\n}+\p_B+\p_{\eta_\m}+\p_A)
&=&\d(k-\ell)\bd(\p_{\eta_\n}+\p_B+\p_{\eta_\m}+\p_A)\\
&=&\d(k-\ell)\sum_C\bd(\p_{\eta_\n}+\p_B+\p_C)\bd(\p_{\eta_\m}+\p_A+\p_C)\nonumber\\
&=&\d(k-\ell)(\hp_{\eta_\n}\hp_{\eta_\m})_{BA}\ ,\nonumber
\end{eqnarray}
and, likewise,
\begin{eqnarray}
\label{kindelta}
&&\d(-\ell+k+\p_{\eta_\k}+\p_{\eta_\n}+\p_B+\p_{\eta_\m}+\p_A)S_\k^{\eta_\n}S_\k^B=\\
&&\hspace{1cm}=\d(k-\ell)\bd(\p_{\eta_\n}+\p_B+\p_{\eta_\k}+\p_{\eta_\m}+\p_A)S_\k^{\eta_\n}S_\k^B\nonumber\\
&&\hspace{1cm}=\d(k-\ell)\sum_{CD}\bd(\p_{\eta_\n}+\p_B+\p_D)\bd(\p_{\eta_\k}+\p_D+\p_C)
\bd(\p_{\eta_\m}+\p_C+\p_A)S_\k^D\nonumber\\
&&\hspace{1cm}=\d(k-\ell)(\hp_{\eta_\n}\G_\k\hp_{\eta_\m})_{BA}\ .
\nonumber
\end{eqnarray}
We thus obtain
\begin{equation}
\label{GtoS2}
G(\tp-\ell+\p_{\eta_\n}+\p_B,-\tp+k+\p_{\eta_\m}+\p_A)=\d(k-\ell)(\hp_{\eta_\m}S(\tp-\ell)\hp_{\eta_\n})_{BA}\ .
\end{equation}
Combining with the factors $S_\n^B$ and $S_\m^A$ in Eq.~(\ref{sunset2}),
we arrive at
\begin{equation}
\label{Psunset}
\P^{\rm sunset}_{\m\n}(k,\ell)=g^2e^{\frac{i}{2}(k_\m-\ell_\n)}\d(k-\ell)\int_{\tp}
\tr\left(S(\tp)\G_\m S(\tp-k)\G_\n\right)\cos(\tp_\m-\mbox{$\small{\half}$} k_\m)\cos(\tp_\n-\mbox{$\small{\half}$} k_\n)\ .
\end{equation}
Using
\begin{equation}
\label{trGamma}
\tr(\G_\k\G_\n\G_\l\G_\m)=16(\d_{\k\n}\d_{\l\m}-\d_{\k\l}\d_{\m\n}+\d_{\k\m}\d_{\l\n})\ ,
\end{equation}
this equals 
\begin{eqnarray}
\label{Psunset2}
&&\P^{\rm sunset}_{\m\n}(k)=16g^2e^{\frac{i}{2}(k_\m-k_\n)}\\
&&\int_{\tp}
\frac{\d_{\m\n}(\sum_\k\sin(\tp_\k)\sin(\tp_\k-k_\k)+m^2)-\sin(\tp_\m)\sin(\tp_\n-k_\n)-\sin(\tp_\n)\sin(\tp_\m-k_\m)}{(\sum_\k\sin^2(\tp_\k)+m^2)(\sum_\l\sin^2(\tp_\l-k_\l)+m^2)}\nonumber\\
&&\phantom{\int_{\tp}}\times\cos(\tp_\m-\mbox{$\small{\half}$} k_\m)\cos(\tp_\n-\mbox{$\small{\half}$} k_\n)\ ,
\nonumber
\end{eqnarray}
where again we omitted $\d(k-\ell)$.   Formally, in the continuum we would have obtained 
one quarter of this result
(with $\sin(\tp_\m)\to\tp_\m$, $\cos(\tp_\m-\half k_\m)\to 1$, \etc), confirming that there are
four species doublers (tastes) in the staggered theory.

Defining
\begin{equation}
\label{Ptotal}
\P_{\m\n}(k)=\P^{\rm sunset}_{\m\n}(k)+\P^{\rm tadpole}_{\m\n}(k)\ ,
\end{equation}
we first show that this vanishes at $k=0$, thus demonstrating that the tadpole diagram
removes the quadratic divergence present in the sunset diagram.   At $k=0$,
\begin{equation}
\label{Pk0}
\P^{\rm sunset}_{\m\n}(0)=16g^2\int_{\tp}\left(-2\frac{\sin(\tp_\m)\sin(\tp_\n)\cos(\tp_\m)\cos(\tp_\n)}
{(\sum_\k\sin^2(\tp_\k)+m^2)^2}+\d_{\m\n}\frac{\cos^2(\tp_\m)}{\sum_\k\sin^2(\tp_\k)+m^2}\right)\ .
\end{equation}
Using
\begin{equation}
\label{partial}
-2\,\frac{\sin(\tp_\m)\cos(\tp_\m)}
{(\sum_\k\sin^2(\tp_\k)+m^2)^2}=\frac{\partial}{\partial\tp_\m}\frac{1}{\sum_\k\sin^2(\tp_\k)+m^2}
\end{equation}
and partially integrating the first term,
one can show that $\P^{\rm sunset}_{\m\n}(0)$ is equal to minus $\P^{\rm tadpole}_{\m\n}(0)$.
There are no boundary terms, because the integrand is periodic in every direction with period $\p$.

Next, we demonstrate that the WTI is satisfied, which  on the lattice takes the form
\begin{equation}
\label{WTI}
\sum_\m(1-e^{-ik_\m})\P_{\m\n}(k)=0\ .
\end{equation}
Using that
\begin{equation}
\label{WTIstep1}
(1-e^{-ik_\m})e^{\frac{i}{2}k_\m}\G_\m\cos(\tp_\m-\half k_\m)=S^{-1}(\tp)-S^{-1}(\tp-k)\ ,
\end{equation}
one can simplify
\begin{eqnarray}
\label{WTIstep2}
\sum_\m(1-e^{-ik_\m})\P^{\rm sunset}_{\m\n}(k)&\!\!=\!\!&
16g^2e^{-\frac{i}{2}k_\n}\int_{\tp}\tr(\G_\n (S(\tp-k)-S(\tp))\cos(\tp_\n-\mbox{$\small{\half}$} k_\n)\\
&&\hspace{-4cm}=-16ig^2e^{-\frac{i}{2}}\int_{\tp}\left(\frac{\sin(\tp_\n-k_\n)}{\sum_\k\sin^2(\tp_\k-k_\k)+m^2}
-\frac{\sin(\tp_\n)}{\sum_\k\sin^2(\tp_\k)+m^2}\right)\cos(\tp_\n-\mbox{$\small{\half}$} k_\n)\nonumber\\
&&\hspace{-4cm}=-16ig^2e^{-\frac{i}{2}k_\n}\int_{\tp}\frac{\sin(\tp_\n)}{\sum_\k\sin^2(\tp_\k)+m^2}
(\cos(\tp_\n+\mbox{$\small{\half}$}k_\n)-\cos(\tp_\n-\mbox{$\small{\half}$} k_\n)\nonumber\\
&&\hspace{-4cm}=16g^2(1-e^{-ik_\n})\int_{\tp}\frac{\sin^2(\tp_\n)}{\sum_\k\sin^2(\tp_\k)+m^2}
\ ,\nonumber
\end{eqnarray}
where we again used that the integrand is periodic in $\tp_\m$ with period $\p$.  This equals $-\sum_\m(1-e^{-ik_\m})\P^{\rm tadpole}_{\m\n}(k)$,
thus proving Eq.~(\ref{WTI}).

Finally, in this subsection, we will indicate how one may actually go about completing the calculation of the 
vacuuum polarization at one-loop.   We start from the expression
\begin{eqnarray}
\label{fullvacpol}
&&\P_{\m\n}(k)=16g^2e^{\frac{i}{2}(k_\m-k_\n)}\\
&&\int_{\tp}\Biggl(
\frac{\d_{\m\n}(\sum_\k\sin(\tp_\k)\sin(\tp_\k-k_\k)+m^2)-\sin(\tp_\m)\sin(\tp_\n-k_\n)-\sin(\tp_\n)\sin(\tp_\m-k_\m)}{(\sum_\k\sin^2(\tp_\k)+m^2)(\sum_\l\sin^2(\tp_\l-k_\l)+m^2)}\nonumber\\
&&\phantom{\int_{\tp}}\times\cos(\tp_\m-\mbox{$\small{\half}$} k_\m)\cos(\tp_\n-\mbox{$\small{\half}$} k_\n)\nonumber\\
&&\phantom{\int_{\tp}}-\frac{\d_{\m\n}(\sum_\k\sin^2(\tp_\k)+m^2)-2\sin(\tp_\m)\sin(\tp_\n)}{(\sum_\k\sin^2(\tp_\k)+m^2)^2}\cos(\tp_\m)\cos(\tp_\n)\Biggr)\ ,
\nonumber
\end{eqnarray}
where the last line, obtained by setting $k=0$ in the middle two lines, is equal to the tadpole contribution because $\P_{\m\n}(0)=0$.   This form
is useful because it shows explicitly that $\P_{\m\n}(k)$ is logarithmically divergent;
the last line explicitly subtracts the quadratic divergence present in $\P^{\rm sunset}_{\m\n}(k)$.
Moreover, by shifting $\tp-\half k\to\tp$, which is allowed in an integral that is only logarithmically
divergent, one can show that the integral is quadratic in $k$
in the continuum limit (modulo, of course, a logarithmic dependence on $k$).

Returning to Eq.~(\ref{fullvacpol}), we split the integral into two regions:
\begin{equation}
\label{intsplit}
\int_{\tp}=\int_{|\tp|<\d}+\int_{|\tp|>\d}\ ,
\end{equation}
where we take 
\begin{equation}
\label{delta}
|ak|\sim |am|\ll \d\ll 1\ ,
\end{equation}
and we take the limit $a\to 0$ first, followed by $\d\to 0$. (In Eq.~(\ref{delta}) we wrote $k$ and $m$
in physical units, temporarily restoring the lattice spacing $a$.)   We will refer to these
two regions as the ``inner'' and ``outer'' regions.

In the inner region, we may replace the integrand by its covariant form:
\begin{eqnarray}
\label{inner}
\P^{\rm inner}_{\m\n}(k)&=&16g^2\int_{|\tp|<\d}\Biggl(\frac{\d_{\m\n}(\tp(\tp-k)+m^2)-\tp_\m(\tp_\n-k_\n)-\tp_\n(\tp_\m-k_\m)}{(\tp^2+m^2)((\tp-k)^2+m^2)}\nonumber\\
&&\phantom{16g^2\int_{|\tp|<\d}\Biggl(}-\frac{\d_{\m\n}(\tp^2+m^2)-2\tp_\m\tp_\n}{(\tp^2+m^2)^2}\Biggr)\\
&=&\frac{1}{3\p^2}g^2(k_\m k_\n-\d_{\m\n}k^2)\log(\d^2)+\mbox{finite\ terms}\ .\nonumber
\end{eqnarray}
Since $\P_{\m\n}(k)$ does not depend on $\d$, the same logarithmic divergence (up to a sign)
also has to be present in the outer region integral.   We handle this by adding and subtracting
\begin{equation}
\label{addsub}
\frac{1}{3}g^2(k_\m k_\n-\d_{\m\n}k^2)\int_{|p|>\d}\frac{1}{(\sum_\m\sin^2(\half p_\m))^2}\ ,
\end{equation}
because, as we will show,
\begin{equation}
\label{addsub1}
I\equiv\lim_{\d\to 0}\left(\int_{|p|>\d}\frac{1}{(\sum_\m\sin^2(\half p_\m))^2}+\frac{1}{\p^2}\log(\d^2)\right)
\end{equation}
is finite and can be explicitly calculated.
This means that the sum of Eq.~(\ref{addsub}) and the inner region is finite, and thus that the
difference of the outer region integral and Eq.~(\ref{addsub}) is finite as well.   Since, in the
outer region, $|ak|\sim|am|\ll\d<|\tp|$, we can expand the subtracted outer-region integral
in powers of $ak_\m$ and $am$, keep those powers that survive in the continuum limit, 
and then take $\d$ to zero, leading to purely numerical integrals that can be 
computed numerically.   All non-analytic dependence on $k$ and $m$ resides in the
inner-region integral.

What remains to be done is to calculate Eq.~(\ref{addsub1}).   We begin rewriting this as
\begin{equation}
\label{addsub2}
\lim_{\d\to 0}\left(\lim_{m\to 0}\left(\int_p\frac{1}{(\sum_\m\sin^2(\half p_\m)+\frac{1}{4}m^2)^2}
-\int_{|p|<\d}\frac{16}{(p^2+m^2)^2}\right)+\frac{1}{\p^2}\log(\d^2)\right)\ .
\end{equation}
The integral over the components $p_\m$ of $p$ in the first integral runs from $-\p\to\p$.
We use
\begin{eqnarray}
\label{basic}
\int_p\frac{1}{\sum_\m\sin^2(\half p_\m)+\frac{1}{4}m^2}&=&
2\int_0^\infty dx\int_p e^{-\half x m^2-2x\sum_\m\sin^2(\half p_\m)}\\
&=&2\int_0^\infty dx \,e^{-\half x m^2}e^{-4x}I_0^4(x)\ ,\nonumber
\end{eqnarray}
where $I_0(x)$ is the modified Bessel function, and hence
\begin{eqnarray}
\label{basic2}
\int_p\frac{1}{(\sum_\m\sin^2(\half p_\m)+\frac{1}{4}m^2)^2}&=&
-4\,\frac{\partial}{\partial m^2}\int_p\frac{1}{\sum_\m\sin^2(\half p_\m)+\frac{1}{4}m^2}\\
&=&4\int_0^\infty dx \,x\,e^{-\half x m^2}e^{-4x}I_0^4(x)\ .\nonumber
\end{eqnarray}
Furthermore,
\begin{equation}
\label{basic3}
-16\int_{|p|<\d}\frac{1}{(p^2+m^2)^2}=-16\int_{|p|<1}\frac{1}{(p^2+\frac{m^2}{\d^2})^2}
=-\frac{1}{\p^2}\log\frac{\d^2}{m^2}+O\left(\frac{m^2}{\d^2}\right)\ ,
\end{equation}
showing that the double limit in Eq.~(\ref{addsub2}) is finite.   We then calculate the integral
on the left of Eq.~(\ref{basic3}) in a way similar to the calculation in Eq.~(\ref{basic2}),
obtaining
\begin{equation}
\label{basic4}
-16\int_{|p|<\d}\frac{1}{(p^2+m^2)^2}=-\frac{1}{\p^2}\int_0^\infty\frac{dx}{x}\,e^{-\half x\frac{m^2}{\d^2}}
\left(1-(1+\half x)e^{-\half x}\right)\ .
\end{equation}
We can now combine all terms in Eq.~(\ref{addsub2}), take first $m$ and then $\d$ to zero, 
obtaining
\begin{equation}
\label{result}
I=4\int_0^\infty dx\,x\left(e^{-4x}I_0^4(x)-\frac{1}{4\p^2 x^2}\left(1-(1+\half x)e^{-\half x}\right)\right)
=0.4855321\ .
\end{equation}

\subsection{\label{self} Fermion self energy}
The one-loop fermion self energy can be calculated similarly.   The steps are similar
to those followed in Sec.~\ref{vacpolcalc} (for details, see 
Ref.~\cite{GS}), and here we just give the result (in Feynman gauge):
\begin{eqnarray}
\label{selfenergy}
\S(\tp)&=&g^2\sum_\m\int_k\half(1+\cos(k_\m+2\tp_\m))\G_\m S(k+\tp)\G_\m D(k)\\
&&-g^2\sum_\m \G_\m\,i\sin(\tp_\m)\int_k D(k)\ ,\nonumber\\
D(k)&=&\frac{1}{4\sum_\m\sin^2(\half k_\m)}\ .\nonumber
\end{eqnarray}
The term on the second line comes from a gauge-field tadpole diagram.   This
result can be further evaluated by splitting into inner and outer regions, \etc, 
as we did for the vacuum polarization.
$\S(\tp)$ is a $16\times 16$ matrix, and thus a linear combination of products
of the matrices $\G_\m$ and $\X_\n$.   In the continuum limit, non-analytic 
dependence on $\tp$ and $m$ is the same as we would obtain in any other
regularization, and comes from the inner-region integral.

By dimensional analysis, the most general contact term one can find in the 
continuum limit has the form
\begin{equation}
\label{contact}
C\equiv\tp_\m X_\m+\frac{1}{a}\,Y_0+mY \ ,
\end{equation}
where again we restored the lattice spacing, and where  $X_\m$, $Y_0$ and $Y$ 
are $16\times 16$ matrices.
We will now see how lattice
symmetries restrict the form of $C$.   First, shift symmetry implies that $C$
should commute with all $\X_\n$, and therefore $X_\m$, $Y_0$ and $Y$
cannot contain any $\X_A$.

Then (hypercubic) rotational symmetry implies that $X_\m$ has to be a vector,
which implies that $X_\m\propto\G_\m$ or $X_\m\propto\G_\m\G_5$,
while both $Y_0$ and $Y$ have to be proportional to the $16\times 16$ unit matrix
 or $\G_5$.   Reflection symmetry $I_s$ of Eq.~(\ref{Isphi}) excludes $\G_5$ from appearing in 
any of these, and thus we find that
\begin{equation}
\label{contact2}
C=\a \,\tp_\m\G_\m+\b\,\frac{1}{a}+\g\, m\ ,
\end{equation}
with $\a$, $\b$ and $\g$ constants.
Finally, we use that for $m=0$ we have $U(1)_\e$ symmetry, which implies that,
for $m=0$, $C$ should anti-commute with $\G_5\X_5$.   This sets $\b=0$, and
we thus see that $m$ renormalizes multiplicatively, while the contact terms
take on the form expected in the continuum theory \cite{GS} (with coefficients
specific to the lattice regularization, of course).

We note that, in this argument, shift symmetry played a crucial role.   Suppose
we were to gauge the staggered theory on the taste basis, with action~(\ref{kinterm2})
plus~(\ref{massterm}) on the coarse lattice, losing shift symmetry in the process because
the gauge fields $U_\m(y)$ would now live on the coarse lattice.
We would expect a counter term of the form
\begin{equation}
\label{tastect}
\sum_\m\tr(\bj(y)\g_5\j(y)\g_5\g_\m)
\end{equation}
because of the form of the ``Wilson'' term in Eq.~(\ref{kinterm2}).   On our momentum
basis, this translates into a counter term $Y_0\propto\sum_\m\G_5\X_5\X_\m$.   
Indeed, this anti-commutes with $\G_5\X_5$, so it is {\it not} excluded by $U(1)_\e$
symmetry!   And, indeed, such a counter term appears at one loop in the theory on the
taste basis if it is gauged on the coarse lattice \cite{MW}.   One can verify that
the action on the taste basis is invariant under hypercubic rotations of the 
form~(\ref{tRmom}).  In particular, 
\begin{equation}
\label{rotcoarse}
\tilde{R}^{-1}_{\r\s}\left(\sum_\m\G_5\X_5\X_\m\right)\tilde{R}_{\r\s}=\sum_\m\G_5\X_5\X_\m\ ,
\end{equation}
showing that a counter term of the form~(\ref{tastect}) is consistent with rotational symmetry.

\section{\label{SET} Symanzik effective theory for staggered fermions}
The Symanzik effective theory (SET) is a continuum effective theory that reproduces 
correlation functions of the underlying lattice theory expanded in powers of the lattice spacing,
for momenta $\L_{\rm QCD}\ll |p|\ll 1/a$, where $p$ is the typical momentum of these
correlation functions.\footnote{In these notes we ignore anomalous dimensions.   However,
these could become important with high precision computations near the continuum limit \cite{HMS}.}   
In this chapter, we discuss the SET for QCD with $N_f$ flavors
of staggered fermions, for $N_f > 1$.    The SET is formulated in terms of continuum
quark fields $q_i$ ($i$ is the flavor index; we leave spin and color indices implicit) and gluons
$A_\m$, with field strength $G_{\m\n}$ \cite{SET,LLH}, because at these momenta
quarks and gluons are still the relevant degrees of freedom.

At leading order, the SET consists of all 4-dimensional operators consistent with the
lattice symmetries.  As we have seen in the previous chapter, the fermion mass 
renormalizes multiplicatively, and there are no dimension-3 operators.\footnote{In the 
previous chapter, we demonstrated this only to one loop, but the symmetry arguments
we used generalize to higher loops.}   As we have seen, a
crucial role in arriving at this conclusion is played by shift and $U(1)_\e$ symmetries.  
Dimensional analysis, gauge invariance, and staggered symmetries imply that the 
dimension-4 operators in the SET are just those defining continuum QCD.

Before we go on to discuss operators with dimension $d$ larger than 4 (which are thus
multiplied by $a^{d-4}$, with $a$ the lattice spacing), we make a general 
observation on how shift symmetry works in the SET, using the fact that this is a 
continuum theory, which, by assumption, is invariant under continuous translation
symmetry \cite{BGS}.   In momentum space, a shift symmetry takes the form
\begin{equation}
\label{Smuagain}
\f(p)\to S_\m\f(p)=e^{iap_\m}\X_\m\f(p)\ ,
\end{equation}
where $-\p/(2a)<p\le \p/(2a)$ is the physical momentum of the field $\f$ on which $S_\m$ acts, and we made the lattice spacing dependence explicit.   The SET is 
invariant under shift symmetry because the underlying lattice theory is.   At the same time, it is also invariant under 
continuous translations of the form
\begin{equation}
\label{transl}
\f(p)\to e^{ipr}\f(p)\ ,
\end{equation}
where $r$ is the vector over which we translate the field.   Combining these two symmetries,
choosing $r_\m=-a$, $r_{\n\ne\m}=0$, we find that the
SET is invariant under
\begin{equation}
\label{discrtaste}
\f(p)\to\X_\m\f(p)\ .
\end{equation}
This is convenient, as this symmetry does not mix operators of different dimensionality
in the SET.   Below we will make extensive use of Eq.~(\ref{discrtaste}), and we will still
refer to this symmetry as ``shift'' symmetry.   For additional arguments
supporting this claim, diagramatically or using the taste basis, see Ref.~\cite{BGS}.

In the discussion below, we will assume that a basis transformation has been carried out 
such that
\begin{equation}
\label{basistr}
\G_\m\to \g_\m 1_4\ ,\qquad \X_\m\to 1_4 \x_\m\ ,
\end{equation}
where we omit the direct-product sign (\ie, $\g_\m 1_4=\g_\m\times 1_4$, \etc),
and we will work with the spin matrices $\g_\m$ and the taste matrices $\x_\m$, both 
in the 4-dimensional irrep of the Dirac algebra in four dimensions.   We will also
drop the index $4$ on the $4\times 4$ unit matrix; in fact, we will mostly not write the
unit matrix at all, when no confusion is possible.

\subsection{\label{dim5} Dimension-5 operators}
We begin with operators of dimension 5, which would be of order $a$ in the
Symanzik expansion \cite{LS}.
These operators can at most contain one bilinear in $q_i$ and $\bq_i$.
They cannot contain any non-trivial taste matrix, because of the symmetry~(\ref{discrtaste}).
Bilinears of the form
\begin{equation}
\label{d51}
\bq_i\s_{\m\n}G_{\m\n}q_i\ ,\qquad \bq_i D_\m D_\m q_i
\end{equation}
(where $D_\m$ is the color-covariant derivative) are ruled out immediately by
$U(1)_{\e}$ symmetry.   In order to consider other dimension-5 operators,
we first define
\begin{eqnarray}
\label{qLqR}
q_{Li} \!\!\! &=& \!\!\! \half(1-\g_5\x_5)q_i\ ,\qquad q_{Ri} = \half(1+\g_5\x_5)q_i\ ,\\
\bq_{Li} \!\!\! &=& \!\!\! \bq_i\half(1+\g_5\x_5)\ ,\qquad \bq_{Ri} = \bq_i\half(1-\g_5\x_5)\ .
\nonumber
\end{eqnarray}
The lattice theory has a $U(N_f)_L\times U(N_f)_R$ symmetry,\footnote{Note that this symmetry
is smaller than the $U(4N_f)_L\times U(4N_f)_R$ symmetry of the continuum limit.} under which these fields transform 
as
\begin{eqnarray}
\label{UNeps}
&q_L\to V_L q_L\ ,\qquad &q_R\to V_R q_R\ ,\\
&\bq_L\to \bq_L V_L^\dagger\ ,\qquad &\bq_R\to\bq_RV_R^\dagger\ ,\nonumber
\end{eqnarray}
with 
\begin{equation}
\label{VLVR}
V_L=e^{\frac{i}{2}\a^a T^a(1-\g_5\x_5)}\ ,\qquad V_R=e^{\frac{i}{2}\b^a T^a(1+\g_5\x_5)}\ ,
\end{equation}
in which $T^a$ are the $U(N_f)$ generators.   This symmetry is a generalization of
$U(1)\times U(1)_\e$ symmetry for one flavor to the case of $N_f$ flavors.   Note that other
symmetries, such as shift and hypercubic symmetry, do not enlarge with the presence
of more flavors, because the gauge field transforms non-trivially under these symmetries.

Mass terms are not invariant under this symmetry, but can be made invariant by
making the $N_f\times N_f$ mass matrix $\cm$ into a {\it spurion} field (\ie, a non-dynamical
field) that transforms as
\begin{equation}
\label{Mspurion}
\cm\to V_R\cm V_L^\dagger\ ,
\end{equation}
so that the 4-dimensional staggered mass term
\begin{equation}
\label{stagmass}
\bc_R(x)\cm\c_L(x)+\bc_L(x)\cm^\dagger\c_R(x)
\end{equation}
is invariant (with the symmetry $U(N_f)_L\times U(N_f)_R$ acting on the staggered
fields $\c$ and $\bc$ as in Eq.~(\ref{VLVR}) with $\g_5\x_5$ replaced by $\e(x)$).   The idea
is to use the spurion form of $\cm$ to see what terms can show up in the SET, and then,
once we have found all such terms, set $\cm$ equal to the physical mass matrix $M$.
 The
dimension-4 mass term in the SET,
\begin{equation}
\label{SETmass}
\bq_R\cm q_L+\bq_L\cm^\dagger q_R
\end{equation}
is also invariant.   We are now ready to return to dimension-5 quark bilinears containing
one or more masses.   Using the transformations~(\ref{UNeps}) and~(\ref{Mspurion}), 
the symmetry $U(N_f)_L\times U(N_f)_R$ rules out terms of the form
\begin{equation}
\label{d52}
\bq M\sl{\!D}q\ ,\qquad \bq M^2 q\ ,\qquad \tr(M)G_{\m\n}G_{\m\n}\ .
\end{equation}
This exhausts all possible dimension-5 contributions to the SET, and we thus conclude
that dependence on the lattice spacing starts at order $a^2$, \ie, at dimension 6.
We note that no use of equations of motion or field redefinitions was made to rule out all dimension-5
operators, and our conclusion thus holds for off-shell correlation functions as well.

We end this subsection with a comment on the taste basis of Sec.~\ref{taste}.  The action~(\ref{kinterm2})
has a Wilson term, which is, in fact, a dimension-5 operator.   However, this action is invariant under
shift symmetry in the form~(\ref{shifttasteb}), but not under the symmetry $\j\to\j\g_\m$, which 
is the continuum form equivalent to Eq.~(\ref{discrtaste}).    As it is the latter symmetry
which is relevant for restricting the form of the SET, there is thus no contradiction.

\subsection{\label{dim6} Dimension-6 operators}
In this section, we will discuss the order-$a^2$ part of the SET, \ie, all dimension-6
operators that contribute, with the focus on operators that break $U(4)_{\rm taste}$
(which is, of course, only a symmetry of the continuum limit).   These operators were 
constructed in Refs.~\cite{LS,YL} (see also Ref.~\cite{AB}).

There are no taste-breaking bilinears.   This follows from the fact that any bilinear has
to be invariant under the discrete group $\G_4$, generated by the $\x_\m$.   (There 
exist taste-invariant bilinear operators, such as $\bq\,\sl{\!D}^3q$, see the end of this subsection.)

Many 4-quark operators can be constructed, and if we require them to be 
invariant under $SU(N_f)$, they take the form\footnote{They can be color-mixed or -unmixed.}
\begin{equation}
\label{4qop}
(\bq_i Xq_i)(\bq_j Yq_j)\ ,
\end{equation}
in which sums over $i$ and $j$ are implied, and 
where $X=\g_A\x_B$ and $Y=\g_C\x_D$, with
\begin{eqnarray}
\label{gAxB}
\g_A\ ,\g_C&\in&\{1,\ \g_\m,\ i\g_\m\g_\n,\ i\g_5\g_\m,\ \g_5\}\ ,\\
\x_B\ ,\x_D&\in&\{1,\ \x_\m,\ i\x_\m\x_\n,\ i\x_5\x_\m,\ \x_5\}\ ,\nonumber
\end{eqnarray}
in which $\m<\n$.  Since no mass matrix can appear in these operators, 
$U(N_f)_L\times U(N_f)_R$ symmetry implies that both $\g_A\x_B$ and
$\g_C\x_D$ have to anti-commute with $\g_5\x_5$, which, in turn, implies
that
\begin{eqnarray}
\label{odd}
&\g_A\in\{1\ \mbox{(S)},\ i\g_\m\g_\n\ \mbox{(T)},\ \g_5\ \mbox{(P)}\}\quad&\Rightarrow\quad
\x_B\in\{\x_\m\ \mbox{(V)},\ i\x_5\x_\m\ \mbox{(A)}\}\ ,\\
&\hspace{-1cm}\g_A\in\{\g_\m\ \mbox{(V)},\ i\g_5\g_\m\ \mbox{(A)}\}\quad&\Rightarrow\quad
\x_B\in\{1\ \mbox{(S)},\ i\x_\m\x_\n\ \mbox{(T)},\ \x_5\ \mbox{(P)}\}\ .\nonumber
\end{eqnarray}
Here we denoted the various tensor structures with S (scalar), T (tensor), P (pseudo-scalar),
V (vector) and A (axial-vector), for later convenience.    Both bilinears thus have to be ``odd,''
which means that the spin-taste matrix has to have an odd number of Lorentz indices.  
Translating this back to lattice operators implies that the corresponding lattice bilinears
have to be odd-link operators, since each $\g_\m$ or $\x_\m$ is generated by a shift over
one lattice spacing in the $\m$ direction.

Next, $\G_4$ symmetry implies that
$\x_B=\x_D$, because, if $\x_B$ were not equal to $\x_D$, we could find a $\x_X$
such that $\x_B\x_X=-\x_X\x_B$ but $\x_D\x_X=+\x_X\x_D$, and the operator would 
not be invariant under $q\to \x_Xq$, $\bq\to\bq \x_X$.   Our operators thus have the form
\begin{equation}
\label{4qop2}
(\bq_i \g_A\x_B q_i)(\bq_j \g_C\x_Bq_j)\ ,
\end{equation}
with the constraint~(\ref{odd}).   A similar argument sets $\g_A=\g_C$.   Rotations
over 180 degrees take the form ({\it cf.} Eq.~(\ref{rot2}))
\begin{equation}
\label{R180}
q\to\g_\r\g_\s\x_\r\x_\s q\ ,\quad \r\ne \s\ ,
\end{equation}
and if we combine these with shifts in the $\r$ and $\s$ directions, it follows that the SET
has to be invariant under 
\begin{equation}
\label{R1802}
q(x)\to\g_\r\g_\s q(R^{-1}_{\r\s}(\p)x) ,\quad \r\ne \s\ .
\end{equation}
Likewise, we can combine $I_s$ of Eq.~(\ref{Isphi}) with a shift in the 4th direction, 
resulting in a SET symmetry
\begin{equation}
\label{Par}
q(x)\to\g_4 q(I_sx) \ .
\end{equation}
Together, these imply that $\g_A$ has to be equal to $\g_C$ (with
Eq.~(\ref{Par}) excluding $\g_C=\g_A\g_5$).   Our 4-fermion
operators thus take the form
\begin{equation}
\label{fffinal}
(\bq_i \g_A\x_B q_i)(\bq_j \g_A\x_Bq_j)\ ,
\end{equation}
with the proviso that $\g_A\x_B$ has to be odd.   It remains to consider 
rotations over 90 degrees.   In order to make sure our 4-fermion operators 
are invariant under these, we need to sum over Lorentz indices, to make
all operators into scalars under 90-degree rotations.   There will be two types of operators \cite{LS}:
those operators which are invariant not only under hypercubic rotations,
but also under $SO(4)$, which we will refer to as ``type A," and those which
are not invariant under $SO(4)$, but only under hypercubic rotations, ``type B.''
To show how this works, it is easiest to give some examples.   Examples
of type-A operators are
\begin{eqnarray}
\label{typeA}
&&\sum_\m(\bg_i\x_\m q_i)(\bq_j\x_\m q_j)\ ,\\
&&\sum_{\m,\n\ne\r}(\bq_i\g_\m\,i\x_\n\x_\r q_i)(\bq_j\g_\m\,i\x_\n\x_\r q_j)\ .\nonumber
\end{eqnarray}
These operators are both invariant under $SO(4)$.   The first of these two operators
we refer to as $S\times V$, since $\g_A$ is scalar and $\x_B$ is vector; the 
second we refer to as $V\times T$, because $\g_A$ is vector and $\x_B$ is tensor.
An example of a type-B operator is
\begin{equation}
\label{typeB}
\sum_{\m\ne\n}(\bq_i\g_\m\,i\x_\m\x_\n q_i)(\bq_j\g_\m\,i\x_\m\x_\n q_j)\ .
\end{equation}
Because the index $\m$ is repeated four times, this operator is invariant 
under hypercubic rotations, but not under $SO(4)$.   We may refer to this operator
as $V_\m\times T_\m$, where the repeated $\m$ reminds us of the four times
repeated index $\m$ in Eq.~(\ref{typeB}).   This type can only occur if either $\g_A$
or $\x_B$ is tensor (and thus the other is vector or axial-vector).

In fact, the type-B operator of our example can be taken to be
\begin{equation}
\label{typeB2}
\sum_{\m\ne\n}\left((\bq_i\g_\m\,i\x_\m\x_\n q_i)(\bq_j\g_\m\,i\x_\m\x_\n q_j)
-(\bq_i\g_\m\,i\x_\m\x_\n\x_5 q_i)(\bq_j\g_\m\,i\x_\m\x_\n\x_5 q_j)\right)\ ,
\end{equation}
because the orthogonal combination
\begin{eqnarray}
\label{typeB3}
&&\sum_{\m\ne\n}\left((\bq_i\g_\m\,i\x_\m\x_\n q_i)(\bq_j\g_\m\,i\x_\m\x_\n q_j)
+(\bq_i\g_\m\,i\x_\m\x_\n\x_5 q_i)(\bq_j\g_\m\,i\x_\m\x_\n\x_5 q_j)\right)\\
&=&\sum_{\m\ne\n}\left((\bq_i\g_\m\,i\x_\m\x_\n q_i)(\bq_j\g_\m\,i\x_\m\x_\n q_j)
+(\bq_i\g_\m\,i\x_\r\x_\s q_i)(\bq_j\g_\m\,i\x_\r\x_\s q_j)_{\r\s\ne\m\n}\right)\nonumber\\
&=&\sum_{\m,\r\ne\s}(\bq_i\g_\m\,i\x_\r\x_\s q_i)(\bq_j\g_\m\,i\x_\r\x_\s q_j)\ ,\nonumber
\end{eqnarray}
which is type A.   We see that the requirement that the 4-quark operators are
invariant under the staggered symmetry group leads to a huge reduction in the
number of 4-quark operators.   With $\g_A$, $\x_B$, $\g_C$ and $\x_D$
unrestricted, there are $16^4=65536$ choices!

For $N_f=1$ there is a subtlety, based on the fact that the flavor symmetry group is
just $U(1)\times U(1)_\e$.   It is straightforward to verify that, \eg,
\begin{equation}
\label{SSPP}
(\bq q)(\bq q)-(\bq\g_5\x_5 q)(\bq\g_5\x_5 q)
\end{equation}
is invariant under this group.   Using our shorthand notation, we can label this 
operator as $S\times S-P\times P$.
The fermion bilinears in this combination are even,
and this operator thus does not fit into the general pattern we established above.
However, it can be made to fit, by a Fierz transformation \cite{LS}, as we will now show.

A Fierz transformation is based on identities of the form
\begin{equation}
\label{Fierz}
(\g_A)_{\a\b}(\g_A)_{\g\d}=\sum_B c(A,B)(\g_B)_{\a\d}(\g_B)_{\g\b}\ ,
\end{equation}
with numerical coefficients $c(A,B)$ that depend on $A$ and $B$.   For instance,
\begin{equation}
\label{Fierzex}
1_{\a\b}1_{\g\d}=\frac{1}{4}\sum_B (\g_B)_{\a\d}(\g_B)_{\g\b}\ .
\end{equation}
Similar identities hold, of course, for the $\x_A$.   Using the abbreviations 
$S$, $V$, $T$, $A$ and $P$ introduced above, we can use Eq.~(\ref{Fierzex}) for 
both the $\g$ and $\x$ matrices 
to write
\begin{eqnarray}
\label{SSPP2}
S\times S&=&\frac{1}{16}(S+V+T+A+P)\times (S+V+T+A+P)\ ,\\
P\times P&=&\frac{1}{16}(S-V+T-A+P)\times (S-V+T-A+P)\ .\nonumber
\end{eqnarray}
Subtracting these two, we find that
\begin{eqnarray}
\label{SSPP3}
&&S\times S-P\times P=\frac{1}{8}\Bigl(S\times V+V\times S+T\times V+V\times T+P\times V
+V\times P\nonumber\\
&&\hspace{2cm}+S\times A+A\times S+T\times A+A\times T+P\times A
+A\times P\Bigr)\ ,
\end{eqnarray}
which allows us to write the operator~(\ref{SSPP}) in the general form we derived above
for arbitrary $N_f$.  Of course, Fierzing interchanges color-mixed and color-unmixed operators,
but both types occur in the general analysis as well.

There are also dimension-6 operators with just one quark bilinear.   Staggered and flavor 
symmetries restrict these to the list\footnote{The operator $\sum_\m\bq_i(D_\m^2\,\sl{\!D}-\,\sl{\!D}D_\m^2)q_i$ has negative charge parity \cite{LS}.}
\begin{eqnarray}
\label{d6bil}
&&\bq_i\,\sl{\!D}^3q_i\ ,\quad\sum_\m\bq_i(D_\m^2\,\sl{\!D}+\,\sl{\!D}D_\m^2)q_i\ ,\quad
\sum_\m\bq_iD_\m\,\sl{\!D}D_\m q_i\ ,\quad\sum_\m\bq_i\g_\m D_\m^3 q_i\ ,\nonumber\\
&&\bq_i M\,\sl{\!D}^2q_i\ ,\quad \bq_i MD_\m^2q_i\ ,\quad \bq_i M^2\,\sl{\!D}q_i\ ,\quad
\bq_i M^3q_i\ .
\end{eqnarray} 
In particular, the fermion bilinear cannot contain any taste matrix, because of shift symmetry.
The operators containing the mass matrix $M$ can be constructed using Eqs.~(\ref{UNeps})
and~(\ref{Mspurion}).   For example, the last operator is obtained from
\begin{equation}
\label{last}
\bq_{Ri}\cm\cm^\dagger\cm q_{Li}+\bq_{Li}\cm^\dagger\cm\cm^\dagger q_{Ri}\ ,
\end{equation}
and setting $\cm=\cm^\dagger=M$.
Finally, there are purely gluonic dimension-6 operators, shared with other discretizations of
QCD, which we do not list here.

\subsection{\label{dimlg6} Operators with dimension larger than 6}
We briefly discuss what happens beyond dimension 6, \ie, at order $a^3$ and
higher. 
\subsubsection{\label{d7} Dimension 7}
Let us start with dimension-7 operators.   We again restrict ourselves to 
operators containing quark fields (no purely gluonic operators with odd dimension
exist).   Dimension-7 operators, if they exist, can contain one or two quark
bilinears.

First, consider operators with only one quark bilinear.   Starting from the list
in Eq.~(\ref{d6bil}), we need to add one mass dimension, which can be done by
adding an extra covariant derivative $D_\m$, or an extra mass matrix $M$.
However, if we insert a $D_\m$, Lorentz indices still have to balance, and the
only way to do that is to remove or add a $\g_\m$ inside the bilinear.   But this
would change the properties of this bilinear under $U(N_f)_L\times U(N_f)_R$,
so this does not work.   An analogous argument excludes adding a mass matrix $M$
(using again a spurion-based argument).

Essentially the same argument applies to operators with two quark bilinears.  
Inserting a $D_\m$ leads to an unmatched Lorentz index, or a change in the
symmetry properties of one fermion bilinear, making it inconsistent with
$U(N_f)_L\times U(N_f)_R$.   A similar argument applies with inserting
a mass matrix $M$.   If we replace one of the two bilinears
$\bq_i Xq_i$ with $\bq_i MYq_i$, we need
to choose $Y$ such that $\bq_i MYq_i$ is invariant under $U(N_f)_L\times U(N_f)_R$
(after promoting $M$ to the spurions $\cm$ and $\cm^\dagger$).
But this implies that $Y\ne X$, and thus shift and/or lattice rotation/reflection symmetries would be
broken.   We conclude that no dimension-7 operators exist that are part of the SET.

\subsubsection{\label{d8} Dimension 8}
In contrast, many operators of dimension 8 exist, and they are easily constructed from
dimension-6 operators, \eg, by inserting $D_\m^2$ or $\sl{\!D}^2$, or by
multiplying with $\tr(M^2)$.    However, all continuum symmetries are already broken
down to the staggered symmetry group at the level of dimension-6 operators (because
of type-B operators!), so no {\it new} symmetry breaking occurs at 
order $a^4$ \cite{SvdW}.\footnote{In particular, no new taste spurions occur, because
the number of quark bilinears is still restriced to two.   See next section.}

\subsubsection{\label{d9} Odd dimension $\ge 9$}
At dimension 9, we should consider operators with three quark bilinears.   Because of
dimensional reasons, these cannot contain any derivatives or masses.   Each of the
bilinears needs to be invariant under $U(N_f)_L\times U(N_f)_R$, which implies
that each bilinear has to be odd---it has an odd number of Lorentz indices.   But this
is impossible, because all Lorentz indices need to be contracted because of 
hypercubic invariance.   In general, any operator, at any dimension, has to have an
even number of Lorentz indices.

What about dimension-9 operators with only one quark bilinear?   Either the bilinear
is odd, which means it has 1 or 3 Lorentz indices, or, if it contains a mass matrix $M$,
it carries an even number of Lorentz indices, because of $U(N_f)_L\times U(N_f)_R$
symmetry.   (If there are two mass matrix insertions, it has to be odd again; if there are
three mass matrices, it has to be even again, \etc)   

The rest of the operator has to be made out of $k$ appearances of $D_\m$ and/or
$\ell$ appearances of $G_{\k\l}$.   This part of the operator thus has dimension
$d=k+2\ell$, and it has $k+2\ell$~mod~2 Lorentz indices.   If the quark bilinear has
dimension 3 (no $M$), the complete operator thus has dimension $d=3+k+2\ell$, and
it has $1+k+2\ell$~mod~2 Lorentz indices.   For odd $d$, we thus need $k+2\ell$ even,
but that implies that the number of Lorentz indices is odd.    Likewise, if the bilinear
contains one factor of $M$, we need $d=4+k+2\ell$ to be odd, and thus that
$k+2\ell$ is odd.   But now the quark bilinear needs to be even, so again the number
of Lorentz indices is odd.   This argument generalizes to operators with more 
insertions of $M$.

A very similar argument takes care of dimension-9 operators with two quark 
bilinears.   The general pattern is that the dimension of the gluonic part to be 
added is always such that the number of Lorentz indices is odd.   The conclusion
appears to be that no operators with odd dimension, and thus an odd power
of the lattice spacing, can appear in the SET.

\subsection{\label{tastebrsec} What causes taste breaking?}
Finally, in this chapter, let us have a closer look at what causes taste-breaking operators
to appear in the SET.   As they first appear at dimension 6 through 4-quark operators,
we expect them to come from quark-quark scattering, as depicted in Fig.~\ref{tastebr}.
Indeed, when the gluon momentum exchanged between the two quarks has momentum
$ak=\p_E\ne 0$, with $\p_E$ one of the momentum vectors with components equal to
0 or $\p$, this short-distance gluon exchange generates taste-breaking 4-quark operators, as we will
show below.   Note that in this figure, we restored the lattice spacing $a$.

\begin{figure}[!t]
\vspace*{1ex}
\begin{center}
\begin{picture}(50,100)(5,0)
  \put(-60,0){\includegraphics*[width=2.5in]{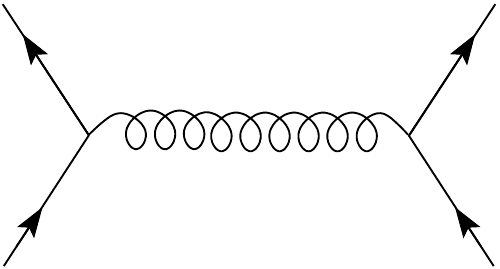}}
  \put(10,25){$ak=\p_E$}
  \put(-50,95){$a\tilde{p}'+\p_C$}
  \put(70,95){$a\tilde{q}'+\p_D$}
  \put(-50,-5){$a\tilde{p}+\p_A$}
  \put(75,-5){$a\tilde{q}+\p_B$}
\end{picture}
\vspace*{5ex}
\floatcaption{tastebr}
             {Exchange of a gluon with momentum equal to $\p_E$ between two
staggered quarks, in momentum space.}
\end{center}
\end{figure}

In Feynman gauge, the gluon propagator $D(k)$ in Eq.~(\ref{selfenergy}), for such momenta, is equal to
\begin{equation}
\label{gluon}
D_{\m\n}(k=\p_E/a)=\frac{\d_{\m\n}}{\sum_\m\frac{4}{a^2}\sin^2(\half ak_\m)}\Bigg|_{k=\p_E/a}=\frac{a^2}{4n}\,\d_{\m\n}\ ,
\end{equation}
where $n$ is the number of components of $\p_E$ equal to $\p$.  Using Eq.~(\ref{1gvertex}), and 
working in Feynman gauge,
the diagram is thus equal to (retaining the explicit lattice spacing $a$ only in the prefactor and omitting color generators)
\begin{eqnarray}
\label{quarkscat}
&&-g^2\frac{a^2}{4n}\sum_\m\d(\tp+\p_A-\tp'+\p_C+\p_E+\p_{\eta_\m})\half(S_\m^A e^{i\tp_\m}+S_\m^C e^{-i\tp'_\m})\\
&&\phantom{-g^2\frac{a^2}{4n}\sum_\m}\times
\d(\tq+\p_B-\tq'+\p_D+\p_E+\p_{\eta_\m})\half(S_\m^B e^{i\tq_\m}+S_\m^D e^{-i\tq'_\m})\nonumber\\
&=&-g^2\frac{a^2}{4n}\sum_\m\d(\tp-\tp')\d(\tq-\tq')\half ( S_\m^Ae^{i\tp_\m}+S_\m^C e^{-i\tp_\m})
\half ( S_\m^Be^{i\tq_\m}+S_\m^D e^{-i\tq_\m})\nonumber\\
&&\phantom{-g^2\frac{a^2}{4n}\sum_\m}\times\bd(\p_A+\p_C+\p_E+\p_{\eta_\m})\bd(\p_B+\p_D+\p_E+\p_{\eta_\m})\nonumber\\
&\approx&-g^2\frac{a^2}{4n}\frac{1}{4}\sum_\m\d(\tp-\tp')\d(\tq-\tq') \Biggl( (\hp_E\G_\m)_{CA}(\hp_E\G_\m)_{DB}+(\G_\m\hp_E)_{CA}(\G_\m\hp_E)_{DB}\nonumber\\
&&\phantom{-g^2\frac{a^2}{4n}\frac{1}{4}\sum_\m}
+ (\hp_E\G_\m)_{CA}(\G_\m\hp_E)_{DB}+
(\G_\m\hp_E)_{CA}(\hp_E\G_\m)_{DB}\Biggr)\ ,
\nonumber
\end{eqnarray}
where in the last expression we approximated $e^{i\tp_\m}\approx 1$, \etc, and
we used
\begin{equation}
\label{pEGmu}
\bd(\p_A+\p_C+\p_E+\p_{\eta_\m})=\sum_F\bd(\p_E+\p_C+\p_F)\bd(\p_{\eta_\m}+\p_F+\p_A)
=(\hp_E\hp_{\eta_\m})_{CA}=(\hp_{\eta_\m}\hp_E)_{CA}
\end{equation}
and likewise for $\bd(\p_B+\p_D+\p_E+\p_{\eta_\m})$.   (We recall that $(\p_{\eta_\m})_\m=0$.)

We now use that 
\begin{equation}
\label{piE}
\hp_E\in\{\G_\m\X_\m,\ \G_\m\G_\n\X_\m\X_\n,\ \G_\m\G_5\X_\m\X_5,\ \G_5\X_5\}\ .
\end{equation}
This is easy to see, because in these combinations, the factors $S_\m$ in Eqs.~(\ref{Gamma})
and~(\ref{pihatXi}) cancel, so $\G_\m\X_\m=\hp_{\eta_\m}\hp_{\z_\m}$, \etc, and hence
the matrices in Eq.~(\ref{piE}) generate ``pure'' $\hp$ matrices, a total of 15 of them.
Equation~(\ref{quarkscat}) thus generates the 4-quark operators with spin-taste structure
\begin{eqnarray}
\label{gen4q}
\hp_E=&\G_\m\X_\m:\qquad &S\times V\ ,\\
&\G_\n\X_\n|_{\n\ne\m}:\qquad &T_\n\times V_\n\ ,\nonumber\\
&\G_\m\G_\n\X_\m\X_\n:\qquad & V_\n\times T_\n\ ,\nonumber\\
&\G_\r\G_\s\X_\r\X_\s|_{\r,\s\ne\m}:\qquad & A_\n\times T_{\r\s}\ (\n\ne\r,\s)\ ,\nonumber\\
&\G_\m\G_5\X_\m\X_5:\qquad & P\times A\ ,\nonumber\\
&\G_\n\G_5\X_\n\X_5|_{\n\ne \m}:\qquad & T_{\r\s}\times A_\n\ (\n\ne\r,\s)\ ,\nonumber\\
&\G_5\X_5:\qquad &A\times P\ .
\nonumber
\end{eqnarray}
As expected, these are all odd operators, because the $U(N_f)_L\times U(N_f)_R$ 
symmetry is respected by the kinetic term in the staggered action from which the
Feynman rule used above derives.   We note that we do not obtain all 4-quark operators
from this diagram,   but other processes, 
for instance with more than one gluon exchanged, can occur.   
The scale of the coupling at the quark-gluon vertex in Fig.~\ref{tastebr} is that of the
exchanged gluon, and we thus expect taste breaking to be suppressed by a factor of
order $\a_s(\p/a)$ (with $\a_s$ the strong coupling).   
This already follows from the fact that the free theory has exact $U(4)$ taste symmetry, so internal gluons
are needed to produce taste-breaking operators.

\section{\label{SChPT} Staggered chiral perturbation theory}
The continuum limit of QCD with $N_f$ staggered quarks has $4N_f$ quarks in the
continuum limit, with flavor symmetry group $U(1)\times SU(4N_f)_L\times SU(4N_f)_R$.
It follows that this theory is expected to have $(4N_f)^2-1$ Nambu--Goldstone bosons (NGBs).
Of course, the number of NGBs is too large, and this is why, in practical numerical
computations, the 4th root of the staggered determinant is taken, in order to reduce the 
number of quark degrees of freedom by a factor of 4.\footnote{The staggered tastes can also
be interpreted as physical flavors, and an action can be constructed that completely breaks
the taste degeneracy in the continuum limit \cite{GS}.   However, this construction leads to a 
practical problems, and has not been used in practice.}   While we will not review the 
arguments supporting the ``4th-root trick,'' we can accommodate this trick in both 
WCPT and chiral perturbation theory (ChPT) by employing the {\it replica trick}.
(For a review of the rooting trick and further references, see Ref.~\cite{MGroot}.   See also
Ref.~\cite{BGS}.)

We introduce $n_r$ replicas of each staggered fermion, so that now we have
$N=4n_rN_f$ quarks in the continuum limit, and we thus expect $(4n_rN_f)^2-1$
NGBs.   Setting, at the end of any calculation (in WCPT or ChPT) $n_r=\frac{1}{4}$
thus yields effectively $N_f^2-1$ NGBs, \ie, the desired number.   We will thus develop staggered ChPT
(SChPT) for $N$ continuum quarks, and see in an example how the replica
trick works.   Of course, if we wish to consider unrooted staggered quarks, we
would set $n_r=1$.

\subsection{\label{SChPTlag} Construction of the SChPT lagrangian}
The chiral lagrangian is, like the SET, a continuum effective theory, with lattice
spacing dependence taken into account in terms of an expansion in powers of $a^2$.
It is formulated in terms of a non-linear field
\begin{equation}
\label{Sigma}
\S=e^{\frac{i}{f}\p}\ ,
\end{equation}
where $\p$ is a traceless, hermitian $4n_rN_f\times 4n_rN_f$ matrix describing 
the $(4n_rN_f)^2-1$ NGBs.   The field $\S$ transforms as
\begin{equation}
\label{Strans}
\S_{ia,jb}\sim q_{Lia}\bq_{Rjb}\ ,
\end{equation}
with $i,\ j=1,\dots,n_rN_f$ replica-flavor indices, and $a,\ b=1,\dots,4$ taste indices.
In this chapter, the left-handed and right-handed projections are defined as in continuum
QCD:
\begin{eqnarray}
\label{LR}
q_L \!\!\! &=& \!\!\! \half(1-\g_5)q\ ,\qquad q_R = \half(1+\g_5)q\ ,\\
\bq_L \!\!\! &=& \!\!\! \bq\half(1+\g_5)\ ,\qquad \bq_R = \bq\half(1-\g_5)\ .
\nonumber
\end{eqnarray}
We embed the taste matrices $\x_A$ into $4n_rN_f\times 4n_rN_f$ matrices
by simply repeating the $4\times 4$ matrices $\x_A$ $n_rN_f$ times along the 
main diagonal, thus constructing a block-diagonal matrix $\x_A^{(n_rN_F)}$. 
However, we will drop the superscript $(n_rN_F)$ below, when no
confusion is possible.   For instance, since $\S$ is a $4n_rN_f\times 4n_rN_f$
matrix, we have that
\begin{equation}
\label{xiS}
\x_A\S\equiv\x_A^{(n_rN_F)}\S\ .
\end{equation}
For reviews of continuum ChPT and applications to various discretizations of QCD,
see Refs.~\cite{SNara,MGLH}.

At leading order (LO) the chiral lagrangian is given by
\begin{equation}
\label{LOchlag}
\cl_{LO}=\frac{1}{8}f^2\,\tr(\partial_\m\S^\dagger\partial_\m\S)-\frac{1}{4}Bf^2\,\tr(\cm^\dagger\S+\cm\S^\dagger)\ .
\end{equation}
The trace is over replica/flavor and taste indices ({\it cf.} Eq.~(\ref{Strans})).
This lagrangian is invariant under
\begin{equation}
\label{LOinv}
\S\to V_L\S V^\dagger_R\ ,\qquad \cm\to V_L\cm V_R^\dagger\ ,
\end{equation}
with now $V_L\in SU(N)_L$ and $V_R\in SU(N)_R$, with $N=4n_rN_f$.   
As before, $\cm$ is a 
mass spurion, to be set equal to the quark mass matrix $M$ after the construction
of the chiral lagrangian.  $f$ and $B$ are low-energy constants (LECs), to be determined
by matching with the underlying theory, QCD.\footnote{$f$, of course, is the pion
decay constant, normalized such that at LO, $f_\p=f=130.4$~MeV.}
We will choose the quark mass matrix $M$ to be diagonal and positive.   The first four diagonal 
elements will be equal to the up quark mass $m_u$, the next four will be equal to
the down quark mass $m_d$, \etc\ (As before, the four tastes produced by each
staggered fermion are degenerate in the continuum limit.)

To incorporate terms of order $a^2$ into ChPT, we use a spurion trick just like for the
mass term in Eq.~(\ref{LOchlag}).   It is easiest to explain this through some examples.
Our first example starts from the following operator appearing in the SET lagrangian:
\begin{equation}
\label{ex1}
a^2(\bq\x_{5\n}q)(\bq\x_{5\n}q)=a^2(\bq_R\x_{5\n}q_L+\bq_L\x_{5\n}q_R)^2\to (\bq_R X_R q_L+\bq_L X_L q_R)^2\ ,
\end{equation}
where $\x_{5\n}=i\x_5\x_\n$ and $X_{L,R}$ are new spurions transforming as
\begin{equation}
\label{ex1spur}
X_L\to V_LX_L V^\dagger_R\ ,\qquad X_R\to V_RX_R V^\dagger_L\ .
\end{equation}
We can take $X_R=X_L^\dagger$, but this is not necessary.   To recover the 4-quark operator
we started from, we have to set
\begin{equation}
\label{ex1spurvev}
X_L=X_R=a\x_{5\n}\ .
\end{equation}
The next step is simply to write down all operators involving the non-linear field $\S$
and two spurions $X_L$ and/or $X_R$, such that these operators are invariant under
Eqs.~(\ref{LOinv}) and~(\ref{ex1spur}).   The possibilities are
\begin{eqnarray}
\label{ex1chpt}
&&\tr(X_L\S^\dagger)\,\tr(X_R\S)\ ,\\
&&\left(\tr(X_L\S^\dagger)\right)^2+\left(\tr(X_R\S)\right)^2\ ,\nonumber\\
&&\tr(X_L\S^\dagger X_L\S^\dagger)+\tr(X_R\S X_R\S)\ .\nonumber
\end{eqnarray}
The plus signs follow from parity symmetry (Eq.~(\ref{ex1spurvev}) requires
parity to exchange $X_L$ and $X_R$).\footnote{We note that parity $q(x)\to
\g_4\x_4 q(I_s x)$ is a symmetry of the SET.}
Setting the spurions equal to their
physical values, Eq.~(\ref{ex1spurvev}), yields the operators
\begin{eqnarray}
\label{ex1chpt2}
&&a^2\tr(\x_{5\n}\S^\dagger)\,\tr(\x_{5\n}\S)\ ,\\
&&a^2\left(\tr(\x_{5\n}\S^\dagger)\right)^2+a^2\left(\tr(\x_{5\n}\S)\right)^2\ ,\nonumber\\
&&a^2\tr(\x_{5\n}\S^\dagger \x_{5\n}\S^\dagger)+a^2\tr(\x_{5\n}\S \x_{5\n}\S)\ .\nonumber
\end{eqnarray}
Each of these operators should be added to Eq.~(\ref{LOchlag}) with an arbitary new coefficient,
thus introducing new LECs into the chiral lagrangian.

Before we go on to our next example, we note that we could proceed slightly differently.
Instead of the spurions $X_{L,R}$ we could first write
\begin{equation}
\label{spuralt}
a^2(\bq\x_{5\n}q)(\bq\x_{5\n}q)=a^2\left((\bq_R\x_{5\n}q_L)^2+(\bq_R\x_{5\n}q_L)(\bq_L\x_{5\n}q_R)+(\bq_L\x_{5\n}q_R)^2\right)\ ,
\end{equation}
and introduce $\co(a^2)$ spurions 
\begin{eqnarray}
\label{spurquad}
X_{LL}&\to& a^2\x_{5\n}\times \x_{5\n}\ ,\\
X_{LR}&\to& a^2\x_{5\n}\times \x_{5\n}\ ,\nonumber\\
X_{RR}&\to& a^2\x_{5\n}\times \x_{5\n}\ ,
\nonumber
\end{eqnarray}
for each of the three terms on the right-hand side of Eq.~(\ref{spuralt}), thus working only
with spurions of order $a^2$.   The resulting terms in the chiral lagrangian will be the same,
so instead we will use the simpler $\co(a)$ spurions above, with the extra rule that all terms
need to contain two of the spurions $X_L$ and $X_R$.

Our second example starts from 
\begin{equation}
\label{ex2}
a^2(\bq\g_\m\x_5q)(\bq\g_\m\x_5q)=a^2(\bq_L\g_\m\x_5 q_L+\bq_R\g_\m\x_5 q_R)^2\to
((\bq_L\g_\m Y_L q_L+\bq_R\g_\m Y_R q_R)^2\ ,
\end{equation}
where we introduced two new spurions $Y_L$ and $Y_R$ transforming as
\begin{equation}
\label{Yspur}
Y_L\to V_LY_LV_L^\dagger\ ,\qquad Y_R\to V_RY_RV_R^\dagger\ .
\end{equation}
Their physical values are 
\begin{equation}
\label{Yphys}
Y_L=Y_R=a\x_5\ .
\end{equation}
Again, we need the rule that any term in the chiral lagrangian should contain two $Y$
spurions, conforming with Eq.~(\ref{ex2}).
The only operator that can be constructed from $\S$ and two $Y$-spurions is
\begin{equation}
\label{Ychlag}
\tr(Y_L\S Y_R\S^\dagger)\to a^2\tr(\x_5\S\x_5\S^\dagger)\ ,
\end{equation}
which thus introduces one new LEC into the chiral lagrangian.

Performing this procedure for all dimension-6 operators we constructed in Sec.~\ref{dim6}
leads to an order-$a^2$ ``potential'' 
\begin{equation}
\label{V}
a^2\cv=a^2(\cu+\cu')
\end{equation}
to be added to Eq.~(\ref{LOchlag}), with \cite{LS,AB}
\begin{eqnarray}
\label{UUp}
-\cu&=&C_1\,\tr(\x_5\S\x_5\S^\dagger)+\half C_3\sum_\n\left(\tr(\x_\n\S\x_\n\S)+\mbox{h.c.}\right)\\
&&
+\half C_4\sum_\n\left(\tr(\x_{5\n}\S\x_{5\n}\S)+\mbox{h.c.}\right)
+C_6\sum_{\m<\n}\tr(\x_{\m\n}\S\x_{\m\n}\S^\dagger)\ ,\nonumber\\
-\cu'&=&\frac{1}{4}C_{2V}\sum_\n\left((\tr(\x_\n\S))^2+\mbox{h.c.}\right)
+\frac{1}{4}C_{2A}\sum_\n\left((\tr(\x_{5\n}\S))^2++\mbox{h.c.}\right)\nonumber\\
&&+\half C_{5V}\sum_\n\tr(\x_\n\S)\,\tr(\x_\n\S^\dagger)+\half C_{5A}\sum_\n\tr(\x_{5\n}\S)\,\tr(\x_\n\S^\dagger)\ ,
\nonumber
\end{eqnarray}
in which $\x_{\m\n}=i\x_\m\x_\n$.   We find that the operators $S\times V$, $S\times A$,
$P\times V$, $P\times A$, $T\times V$, and $T\times A$ are like example 1 above, 
and contribute to the LECs $C_3$, $C_4$, $C_{2V,A}$ and $C_{5V,A}$.   
The operators $V\times P$, $A\times P$, $V\times T$, and $A\times T$ are like
example 2 above, and contribute to $C_1$ and $C_6$.   The operators $V\times S$
and $A\times S$ do not break taste symmetry, and thus do not contribute any new
operators to the chiral lagrangian.   However, they do lead to order-$a^2$ corrections
to LECs already present in the continuum lagrangian.   The lagrangian obtained by
adding $\cv$ to Eq.~(\ref{LOchlag}) defines staggered ChPT (SChPT) at lowest-order.

Operators like $V_\m\times T_\m$, \ie, operators of type B, do contribute at order $a^2$
to the SET, but not at this order to SChPT.    The reason is that
operators that break $SO(4)$ (while preserving hypercubic rotations) can occur only
at higher orders in SChPT.  For example
 one can make a hypercubic invariant $\sim\sum_\m\partial_\m^4$
which is not an $SO(4)$ invariant.   Such an operator is $O(a^2p^4)$ in ChPT power
counting (the $a^2$ has to be present because the continuum theory respects $SO(4)$
rotational invariance), and thus always of higher order than the LO lagrangian.   This has an 
important consequence:   To order $a^2$, the physics of NGBs is invariant under 
the $SO(4)_{\rm taste}$ subgroup of $SU(4)_{\rm taste}$, and not just under the smaller staggered
symmetry group \cite{LS}!
Other examples are operators  of the schematic form $\sim\sum_\m\partial_\m^2\x_\m\x_\m$ or $\sum_\m\x_\m\x_\m\x_\m\x_\m$, which also respect only the staggered hypercubic 
invariance.   We will discuss such examples toward the end of this subsection; such operators are also 
higher order than the LO lagrangian.

This point brings us to the issue of power counting.   SChPT has three small
parameters, the typical momentum of physical low-energy pions, the quark masses
in $M_{ia,jb}=m_i\d_{ij}\d_{ab}$, and the parameter $a^2$.   Since for physical momenta, $p^2=-m_\p^2$, 
and, as we will see shortly, $m_\p^2\sim m_i$, the first two expansion parameters
are related.\footnote{However, for an alternative scenario, which appears not to be
realized in nature, see Ref.~\cite{altChPT}.}   But $a^2$ and the $m_i$ are in principle 
independent, and it depends on the actual numerical computations carried out
in lattice QCD which power counting is most appropriate.   Most staggered
computations have NGB (squared-mass) taste splittings comparable in size to the NGB 
(squared) masses.   We will thus assume a power counting
\begin{equation}
\label{power}
a^2\L_{\rm QCD}^2\sim \frac{m_i}{\L_{\rm QCD}}\sim \frac{p^2}{\L_{\rm QCD}^2}\ .
\end{equation}
Of course, if the lattice spacing is small enough, it may be possible to ignore 
lattice spacing artifacts, in which case one can use continuum ChPT.   On the other
hand, if the lattice spacing is too large, SChPT may break down, even if the masses
$m_i$ are small enough.

We can now look at the NGB masses predicted by LO SChPT.   We write the 
pion matrix in Eq.~(\ref{Sigma}) as
\begin{equation}
\label{pion}
\p_{ij}=\sum_A\x_A\p^A_{ij}\ ,\qquad \x_A\in\{\x_5,\ \x_{5\m},\ \x_{\m\n},\ \x_\m,\ 1\}
\end{equation}
and expand the LO lagrangian to second order in the pion fields.   This allows us to read
off the masses at lowest order.   For $i\ne j$ we find
\begin{equation}
\label{offdiagmasses}
m_{Aij}^2=B(m_i+m_j)+a^2\D(\x_A)\ ,
\end{equation}
with
\begin{eqnarray}
\label{Deltas}
\D(\x_5)&=&0\ ,\\
\D(\x_{5\m})&=&\frac{16}{f^2}(C_1+3C_3+C_4+3C_6)\ ,\nonumber\\
\D(\x_{\m\n})&=&\frac{16}{f^2}(2C_3+2C_4+4C_6)\ ,\nonumber\\
\D(\x_{\m})&=&\frac{16}{f^2}(C_1+C_3+3C_4+3C_6)\ ,\nonumber\\
\D(1)&=&\frac{16}{f^2}(4C_3+4C_4)\ .\nonumber
\end{eqnarray}
We see that there is a degeneracy predicted by the accidental $SO(4)$
symmetry that is present in the chiral lagrangian at LO, \ie, to order $a^2$.
In the next section, we will see that the flavor non-diagonal NGBs
form eight different multiplets (instead of the five seen in Eq.~(\ref{Deltas})) if taste symmetry is fully broken.
We also note that the $\x_5$ NGBs are massless when all quark masses vanish.   This is the NGB
associated with the exactly conserved axial currents for the $U(N_f)_L\times U(N_f)_R$ symmetry 
group of Eqs.~(\ref{UNeps}) and~(\ref{VLVR}).

The LECs in the double-trace part of the potential, $\cu'$, contribute only to
flavor-diagonal NGB masses.   To quadratic order,
\begin{equation}
\label{cupquad}
\cu'_{\rm quad}=\frac{32}{f^2}(C_{2V}-C_{5V})\sum_{ij}\p^V_{ii}\p^V_{jj}
+\frac{32}{f^2}(C_{2A}-C_{5A})\sum_{ij}\p^A_{ii}\p^A_{jj}\ ,
\end{equation}
leading to corresponding contributions to  (the squares of) their masses.
It is important to note that, even if one restricts oneself to external pions
that are flavor off-diagonal, flavor diagonal pions still contribute to loops.
For examples, see Ref.~\cite{AB}.

If any of the $\D(\x_A)$ would be negative, this could induce flavor/taste
symmetry breaking when the quark masses $m_i$ are small enough.
However, QCD inequalities (see \eg\ Ref.~\cite{NL}) prevent this from happening.
There is, however, no argument that the combinations $C_{2V,A}-C_{5V,A}$
cannot be negative.   For further discussion, see Ref.~\cite{AW}.   In practice
no spontaneous breaking of flavor/taste symmetries has been observed.

At the next order in SChPT, $SO(4)_{\rm taste}$ breaks down to $\G_4$, and rotational
symmetry breaks down to the hypercubic group.   We give a 
few examples \cite{SvdW}.   A single insertion of a 4-quark operator can lead to a 
ChPT operator of the form
\begin{equation}
\label{ex1NLO}
a^2\sum_\m\tr(\S\partial_\m\S^\dagger\x_\m\S^\dagger\partial_\m\S\x_\m)=
\frac{4a^2}{f^2}\sum_\m\tr(\partial_\m\p\x_\m\partial_\m\p\x_\m)+\dots\ ,
\end{equation}
in which the first $\x_\m$ comes from a spurion like $X_L$ and the second 
from a spurion like $X_R$.   This operator leads to direction dependence of
the dispersion relation.   This contribution appears at order $a^2p^2$,
\ie, at next-to-leading order (NLO).

Another example comes from a double insertion of two 4-quark operators,
and takes the form 
\begin{equation}
\label{ex2NLO}
a^4\sum_\m\sum_{\n\ne\m}\tr(\x_{\m\n}\S\x_\m\S\x_{\m\n}\S^\dagger\x_\m\S^\dagger)\ ,
\end{equation}
with the taste matrices coming from spurions $Y_L$, $X_R$, \etc\ Because of the four
spurions, this contribution appears at order $a^4$, which in our power counting
is NLO.   At tree level, it only contributes
to processes with four or more pions.   This is easy to see:  if we set the last
$\S^\dagger$ equal to one, we can use $\x_\m\x_{\m\n}=i\x_\n$ to write the 
operator as
\begin{equation}
\label{ex2NLOSone}
a^4\sum_\m\sum_{\n\ne\m}i\,\tr(\x_\n\S\x_\m\S\x_{\m\n}\S^\dagger)\ ,
\end{equation}
which has $SO(4)$ taste symmetry.   Similar simplifications happen when 
one of the other non-linear fields is set equal to one.   At next-to-next-to-leading
order (NNLO) this operator does contribute to two-point functions through loops.
For a complete construction of the NLO SChPT lagrangian, see Ref.~\cite{SvdW}.

Finally, in this section, we discuss whether the accidental $SO(4)$ symmetry at
order $a^2$ also occurs for other hadrons than pions.   As an example, we 
consider taste breaking for the $\r$ meson \cite{LS}, introducing a ``heavy''
$\r$ field $\r_\m$ \cite{JMW}.   In the infinite-mass limit, its 4-velocity $v_\m$ is fixed.  
We first choose $v_\m$ to be a spurion, transforming like a vector under
hypercubic rotations, and once the effective lagrangian for the $\r$ mass
has been constructed, we set $v_4=1$ and $\vec v=0$, choosing the $\r$'s
restframe.   Using this and staggered symmetries, we can write down the
most general quadratic term in the field $\r_{\m ij}=\sum_A\x_A\r^A_{\m ij}$:
\begin{eqnarray}
\label{rhomass}
\cl_{\r\rm-mass}&=&R_1\sum_\m\tr(\r_\m^\dagger\r_\m)+R_2\sum_\m\tr(\r_\m^\dagger\x_5\r_\m\x_5)
+R_3\sum_{\m\n}\tr(\r_\m^\dagger\x_\n\r_\m\x_\n)\\
&+&R_4\sum_{\m\n}\tr(\r_\m^\dagger\x_{5\n}\r_\m\x_{5\n})+R_5\sum_{\m\ \n<\k}\tr(\r_\m^\dagger\x_{\n\k}\r_\m\x_{\n\k})+R_6\sum_{\m}\tr(\r_\m^\dagger\x_\m\r_\m\x_\m)\nonumber\\
&+&R_7\sum_{\m}\tr(\r_\m^\dagger\x_{5\m}\r_\m\x_{5\m})+R_8\sum_{\m\ne\n}\tr(\r_\m^\dagger\x_{\m\n}\r_\m\x_{\m\n})+R_9\sum_\m\tr(\r_\m^\dagger\x_4\r_\m\x_4)\nonumber\\
&+&R_{10}\sum_\m\tr(\r_\m^\dagger\x_{45}\r_\m\x_{45})+R_{11}\sum_{\m\n}\tr(\r_\m^\dagger\x_{4\n}\r_\m\x_{4\n})+R_{12}\sum_\m\tr(\r_\m^\dagger\x_{4\m}\r_\m\x_{4\m})\ .
\nonumber
\end{eqnarray}
For example, the term with $R_9$ is obtained from
\begin{equation}
\label{exrho}
\sum_{\m\n}\tr(\r_\m^\dagger\x_\n\r_\m\x_\n)v_\n v_\n\big|_{v=(0,0,0,1)}=
\sum_\m\tr(\r_\m^\dagger\x_4\r_\m\x_4)\ .
\end{equation}
There are 11 different quark bilinear operators that create a $\r$ with the
form
\begin{eqnarray}
\label{rhoop}
&&\bq\g_k\x_A q\ ,\\
&&\x_A=(\x_5,\x_m,\x_{4k},\x_{m5},\x_{45},\x_{k5},\x_{m4},\x_{k4},\x_{km},\x_{\ell m})\ ,
\nonumber
\end{eqnarray}
where $k\ne\ell\ne m\ne k$ \cite{MGmesons}.  Each of these transforms in an irrep of 
the relevant staggered symmetry group (see next section). Thus, the 11 operators with LECs
$R_2$ to $R_{12}$ in Eq.~(\ref{rhomass}) completely break the degeneracy (the 
operator with LEC $R_1$ gives the same mass to all tastes).\footnote{It appears that 
the term with LEC $R_{12}$  was missed in Ref.~\cite{LS}.}

\subsection{\label{EMcurrent} Example: Electromagnetic current correlator at NLO}
As an example of SChPT at work, we will calculate the electromagnetic (EM) current two-point 
function to one loop, which is the leading order for this correlation function in ChPT.   
We usually refer to this as NLO,
since it involves a loop, but there is no tree-level contribution, as we will see.

We will use SChPT with two flavors, and $n_r$ replicas.   This theory has $4n_r$ up quarks,
and $4n_r$ down quarks in the continuum limit; we will consider the isospin-symmetric case
with $m_u=m_d$.   At the end of the calculation, we will set $n_r=1/4$, thus implementing the
rooting trick.    In the continuum limit, this will yield the expected result, which, at this order,
is that one would obtain in scalar QED (where the complex scalar is the charged pion).
This validates, in this example, the rooting trick.

In two-flavor QCD, the continuum EM current for $n_r$ replicas of two light flavors is
\begin{equation}
\label{jEM}
j_\m=\sum_i\left(\frac{2}{3}\,\bu_i\g_\m u_i-\frac{1}{3}\,\bdd_i\g_\m d_i\right)\ ,
\end{equation}
where the index $i$ is now the replica index.   

To couple SChPT to electromagnetism, we introduce the covariant derivative
\begin{equation}
\label{covder}
D_\m\S=\partial_\m\S-i\ell_\m\S+i\S r_\m\ ,
\end{equation}
where $\ell_\m$ and $r_\m$ are external vector fields gauging the groups
$U(N)_L$ and $U(N)_R$, respectively, with $N=8n_r$ (since have now
specialized to $N_f=2$).     To obtain the 
coupling of the pions to the EM field, we take $\ell_\m=r_\m=Qv_\m$, with 
\begin{equation}
\label{Qdef2}
Q=\left(\begin{array}{cc}\frac{2}{3}I & 0\\ 0 & -\frac{1}{3}I \end{array}\right)\ ,
\end{equation}
with $I$ the $4n_r\times 4n_r$ unit matrix.    We thus have
\begin{eqnarray}
\label{EMcovder}
D_\m\S&=&\partial_\m\S-iv_\m[Q,\S]\ ,\\
(D_\m\S)^\dagger &=&\partial_\m\S^\dagger-iv_\m[Q,\S^\dagger]\ .\nonumber
\end{eqnarray}
The kinetic term in the chiral lagrangian becomes
\begin{equation}
\label{kingauged}
\cl_{\rm kin}=\frac{f^2}{8}\tr\left((D_\m\S)^\dagger D_\m\S\right), 
\end{equation}
and the ChPT form of the EM current (to lowest order) is obtained from the terms linear in $v_\m$:
\begin{eqnarray}
\label{EMcurrChPT}
j_\m&=&\frac{f^2}{8}\,\tr\left(-i[Q,\S^\dagger]\partial_\m\S-i\partial_\m\S^\dagger[Q,\S]\right)\\
&=&-\frac{i}{4}\,\tr\left(Q(\p\partial_\m\p-\partial_\m\p\p)\right)+\dots\ ,\nonumber
\end{eqnarray}
where in the second line we expanded $\S=1+\frac{i}{f}\p+\dots$.   Writing the pion fields as
\begin{equation}
\label{pifield}
\p=\left(\begin{array}{cc} \frac{1}{\sqrt{2}}\p^0 & \p^+ \\ \p^- & -\frac{1}{\sqrt{2}}\p^0 \end{array}\right)\ ,
\end{equation}
and using $\p^\pm_{ij}=\sum_A \x_A\p^{\pm A}_{ij}$, this can be written as
\begin{equation}
\label{jEMdetail}
j_\m=i\left(\p^{-A}_{ij}\partial_\m\p^{+A}_{ji}-\p^{+A}_{ij}\partial_\m\p^{-A}_{ji}\right)\ .
\end{equation}
It is straightforward to check that this current is conserved by using the equation of
motion $\bo\p^{\pm A}_{ij}=m_A^2\p^{\pm A}_{ij}$.   This current corresponds to the
conserved staggered current
\begin{eqnarray}
\label{jstag}
J_\m^{\rm cons}&=&\half\frac{2}{3}\left(\bc_u(x)\eta_\m(x)U_\m(x)\c_u(x+\m)+
\bc_u(x+\m)\eta_\m(x)U^\dagger_\m(x)\c_u(x)\right)\\
&&-\half\frac{1}{3}\left(\bc_d(x)\eta_\m(x)U_\m(x)\c_d(x+\m)+
\bc_d(x+\m)\eta_\m(x)U^\dagger_\m(x)\c_d(x)\right)\ .\nonumber
\end{eqnarray}
The conserved staggered current is not the only discretization of the EM current we may
consider.   Another example, which is not conserved on the lattice, is the local current
\begin{equation}
\label{jlocal}
J_\m^{\rm local}=\frac{2}{3}\,\bc_u(x)\eta_\m(x)\z_\m(x)\c_u(x)-
\frac{1}{3}\,\bc_d(x)\eta_\m(x)\z_\m(x)\c_d(x)\ .
\end{equation}
In Eqs.~(\ref{jstag}) and~(\ref{jlocal}) we left the replica index implicit.   The 
continuum limit of $J_\m^{\rm local}$ is\footnote{We use $J$ for lattice currents, and
$j$ for continuum currents.}
\begin{equation}
\label{jloccont}
j_\m^{\rm local}=\frac{2}{3}\,\bu_i\g_\m\x_\m u_i-\frac{1}{3}\,\bdd_i\g_\m\x_\m d_i
=\bq\g_\m Q\x_\m^{(2n_r)}q\ ,
\end{equation} 
where in the second step we grouped the $u_i$ and $d_i$ quarks together into the field $q$.
To translate this current into ChPT, we set $\ell_\m=r_\m=Q\x_\m^{(2n_r)}v_\m$.   In general,
we can define ``tasteful'' currents (as opposed to the conserved current, which is tasteless) by 
generalizing this to
\begin{equation}
\label{jloccontchpt}
j_\m^A=\frac{2}{3}\,\bu_i\g_\m\x_A u_i-\frac{1}{3}\,\bdd_i\g_\m\x_A d_i
=\bq\g_\m Q\x_A^{(2n_r)}q\ ,
\end{equation} 
with the local current corresponding to $\x_A=\x_\m$, and the conserved current 
corresponding to $\x_A=1$.   In SChPT, the corresponding currents are
\begin{eqnarray}
\label{ChPTAj}
j_\m^A&=&-i\,\frac{f^2}{4}\,\tr\left(Q\x_A^{(2n_r)}[\S^\dagger,\partial_\m\S]\right)\\
&=&-\frac{i}{4}\,\tr\left(Q\x_A^{(2n_r)}[\p,\partial_\m\p]\right)+\dots\ ,\nonumber
\end{eqnarray}
where in the second line we again expanded in the pion fields.
Instead of Eq.~(\ref{pifield}), we will now write
\begin{equation}
\label{pifield2}
\p=\left(\begin{array}{cc} U & \p^+ \\ \p^- & D \end{array}\right)\ ,
\end{equation}
which, as we will see, is convenient.  In terms of the fields introduced in Eq.~(\ref{pifield2}),
\begin{equation}
\label{jA}
j_\m^A=-\frac{i}{6}\,\tr\left(\x_A\left(U\dlrm U+\p^+\dlrm\p^-\right)\right)
+\frac{i}{12}\,\tr\left(\x_A\left(\p^-\dlrm\p^++D\dlrm D\right)\right)\ .
\end{equation}
We used the notation
\begin{equation}
\label{notation}
f\dlrm g=f\partial_\m g-(\partial_\m f)g\ .
\end{equation}
Note that in these expressions, we have replaced $\x_A^{(n_r)}$ by $\x_A$, for simplicity.
If $\x_A=1$, we have that
\begin{equation}
\label{simpl}
\tr(U\dlrm U)=0\ ,\qquad \tr(D\dlrm D)=0\ ,\qquad \tr(\p^-\dlrm\p^+)=-\tr(\p^+\dlrm\p^-)\ ,
\end{equation}
and we recover Eq.~(\ref{jEMdetail}).   If $\x_A\ne 1$, 
\begin{equation}
\label{xAnontr}
\tr(\x_A U\dlrm U)=\sum_{BC}\tr(\x_A[U^B\x_B,\partial_\m U^C\x_C]=\sum_{BC}\tr(\x_A[\x_B,\x_C])U^B_{ij}\partial_\m U^C_{ji}\ne 0
\end{equation}
in general, so the terms with $U$ and $D$ in Eq.~(\ref{jA}) do not vanish.  However, if
$\x_B=1$, the commutator $[\x_B,\x_C]=[1,\x_C]=0$, and thus the contribution with 
$U^B=U^S$ (with $U^S$ the coefficient of $1$ in $U=\sum_B\x_B U^B$) does in fact vanish,
and hence the neutral singlet fields $U^S$ and $D^S$ do not contribute to the
current~(\ref{jA}).  Since there is no mixing between any of the $U^A$ and any of the $D^B$ 
for $A,\ B\ne S$, we can work on the basis of Eq.~(\ref{pifield2}).   We have that
\begin{eqnarray}
\label{consjA}
\partial_\m j_\m^A&=&-\frac{i}{6}\sum_{BC}\tr(\x_A[\x_B,\x_C])U^B_{ij}m_{\p C}^2U^C_{ji}+\dots\\
&=&-\frac{i}{12}\sum_{BC}\tr(\x_A[\x_B,\x_C])U^B_{ij}(m_{\p C}^2-m_{\p B}^2)U^C_{ji}+\dots\ne 0\ ,\nonumber
\end{eqnarray}
we see that indeed the current $j_\m^A$ is also not conserved in SChPT, unless $\x_A=1$.   
Of course, in the continuum limit, where all pions become degenerate
become equal, all currents are conserved.

We note that 
\begin{equation}
\label{Qdecomp}
Q=\half\t_3+\frac{1}{6}Y=\half\left(\begin{array}{cc} 1 & 0 \\ 0 & -1\end{array}\right)
+\frac{1}{6}\left(\begin{array}{cc} 1 & 0 \\ 0 & 1\end{array}\right)\ .
\end{equation}
Since only the commutator $[Q,\S]$ appears in the conserved current, the hypercharge
part $Y$ drops out, and thus in SChPT, the conserved EM current is the isospin-1
current.   For the other currents, with a non-trivial $\x_A$, the situation is different, 
as we will discuss shortly.

First, let us return to the $\x_A=1$ current $j_\m^S=j_\m$, in the continuum, at the
quark level.   With $\langle\dots\rangle_{U_\m}$ denoting the average over the gauge fields
\begin{equation}
\label{I1}
\langle(\bq(x)\g_\m\half\t_3 q(x))(\bq(y)\g_\n\half\t_3 q(y))\rangle=-\half\langle\tr(\g_\m S(x,y;U_\m)\g_\n S(y,x;U_\m)\rangle_{U_\m}\ ,
\end{equation}
when $m_u=m_d$, because any quark-disconnected parts cancel.   
Here $S(x,y;U_\m)$ is the quark propagator (up or down) in a gauge-field background.
The trace is over spin, taste and replica indices.
We also 
have 
\begin{equation}
\label{mixed}
\langle(\bq(x)\g_\m\half\t_3 q(x))(\bq(y)\g_\n\frac{1}{6}Y q(y))\rangle=0\ ,
\end{equation}
and, finally,
\begin{eqnarray}
\label{Y}
\langle(\bq(x)\g_\m\frac{1}{6}Y q(x))(\bq(y)\g_\n\frac{1}{6}Y q(y))\rangle
&=&-\frac{1}{18}\langle\tr(\g_\m S(x,y;U_\m)\g_\n S(y,x;U_\m)\rangle_{U_\m}\\
&&
+\frac{1}{9}\langle\tr(\g_\m S(x,x;U_\m))\,\tr(\g_\n S(y,y;U_\m))\rangle_{U_\m}\ .\nonumber
\end{eqnarray}
The hypercharge current two-point function is the sum of a quark-connected
(first line) and quark-disconnected (second line) part.

It follows that the light-quark connected part of the EM current two-point function 
is equal to
\begin{equation}
\label{lqc}
-\frac{5}{9}\langle\tr(\g_\m S(x,y;U)\g_\n S(y,x;U)\rangle_U
=\frac{10}{9}\times \langle(\bq(x)\g_\m\half\t_3 q(x))(\bq(y)\g_\n\half\t_3 q(y))\rangle\ ,
\end{equation}
\ie, the light-quark connected part of the EM current two-point function is equal to
$10/9$ times the $I=1$ current two-point function.

Now consider a similar exercise for the current $j_\m^A$ with $\x_A\ne 1$.   In this
case, the relations~(\ref{I1}) and ~(\ref{Y}) hold, with a $\x_A$ inserted next to 
every $\g_\m$ and $\g_\n$.
For
this current, there is no quark-disconnected part, because $\tr(\x_A)=0$!    Therefore
the $j_\m^A$ two-point function is equal to its light-quark connected part, and, 
in the continuum limit, the light-quark connected parts of all currents, including
the one with $\x_A$, are equal.   Therefore, in the continuum limit,
\begin{equation}
\label{A12pt}
\langle j_\m^A(x)j_\n^A(y)\rangle\big|_{A\ne S}=\frac{10}{9}\langle j_\m^S(x)j_\n^S(y)\rangle\ .
\end{equation}
We thus expect to reproduce this relation in the continuum limit of SChPT as well.

Next, we carry out the actual calculation, beginning with the $U$ part of the current~(\ref{jA}).
As we will be interested in
\begin{equation}
\label{defCt}
C^A(t)=\frac{1}{3}\sum_{i=1}^3\sum_{\vec x}\langle j_i^A(\vec x,t)j_i^A(0)\rangle\ ,
\end{equation}
we calculate, using $U_{k\ell}(x)=\sum_X \x_X U^X_{k\ell}(x)$ (where $k$ and $\ell$ are replica
indices),
\begin{eqnarray}
\label{jAjAU}
\langle j_\m^A(x)j_\n^A(y)\rangle_{U\mbox{\tiny-part}}&=&
-\frac{1}{36}\sum_{XX'YY'}\langle
(U^X_{mn}(x)\dlrm U^{X'}_{nm}(x))(U^Y_{k\ell}(y)\dlrm U^{Y'}_{\ell k}(y))\rangle\\
&&
\phantom{-\frac{1}{36}\sum_{XX'YY'}}\times\,\tr(\x_A\x_X\x_{X'})\,\tr(\x_A\x_Y\x_{Y'})\nonumber\\
&=&-\frac{n_r^2}{36}\int\frac{dp_4}{2\p}\int\frac{dq_4}{2\p}\frac{1}{V}^2\sum_{\vec p,\vec q}\sum_{XX'}
\frac{e^{ip(x-y)}e^{iq(y-x)}}{(p^2+m_X^2)(q^2+m_{X'}^2)}(p_\m+q_\m)^2\nonumber\\
&&\phantom{-}\times\Bigl(\tr(\x_A\x_X\x_{X'})\,\tr(\x_A\x_X\x_{X'})-\tr(\x_A\x_X\x_{X'})\,\tr(\x_A\x_{X'}\x_X)\Bigr)\ ,
\nonumber
\end{eqnarray}
where $V$ is the spatial volume (we take the time direction infinite), and 
\begin{equation}
\label{Uprop}
\langle U^X_{mn}(x)U_{k\ell}^Y(y)\rangle=\d_{XY}\d_{m\ell}\d_{nk}\int\frac{dp_4}{2\p}\frac{1}{V}
\sum_{\vec p}\frac{e^{ip(x-y)}}{p^2+m_X^2}\ .
\end{equation}
Setting $\m=i$, 
summing over $\vec x$,  $\vec y$ and $i$, and using
\begin{eqnarray}
\label{zmprop}
\int\frac{dp_4}{2\p}\frac{e^{ip_4t}}{p^2+m_X^2}&=&\frac{1}{2E_X(\vec p)}\,e^{-E_X(\vec p)|t|}\ ,\\
E_X(\vec p)&=&\sqrt{{\vec p}^2+m_X^2}\ ,\nonumber
\end{eqnarray}
we obtain
\begin{eqnarray}
\label{Ctresult}
3C^A(t)_{U\mbox{\tiny-part}}&=&-\frac{n_r^2}{36V}\sum_{\vec p}\sum_{XX'}\frac{e^{-(E_X(\vec p)+E_{X'}(\vec p))|t|}}{E_X(\vec p)E_{X'}(\vec p)}\,{\vec p}^2\\
&&\phantom{-}\times\Bigl(\tr(\x_A\x_X\x_{X'})\,\tr(\x_A\x_X\x_{X'})-\tr(\x_A\x_X\x_{X'})\,\tr(\x_A\x_{X'}\x_X)\Bigr)\ .
\nonumber
\end{eqnarray}
We will refer to the first double trace over taste matrices as the ``unmixed'' part, and to the second 
double trace as the ``mixed'' part.
We note that for $\x_A=1$ this vanishes, consistent with the fact that there is no $U$ contribution
to the conserved current.   For the $D$ contribution, we get $1/4$ of this, and thus the sum of the
$U$ and $D$ contributions is equal to $36\times\left(1/36+1/144\right)=5/4$ times Eq.~(\ref{Ctresult}).

It remains to add the contribution from the charged-pion part of the current.   One can verify that
choosing the term $\p^+\dlrm\p^-$ in both currents yields 1 times the mixed term in Eq.~(\ref{Ctresult}),
choosing the term $\p^-\dlrm\p^+$ in both currents yields $1/4$ times the mixed term in Eq.~(\ref{Ctresult}),
while the cross terms yields $-1$ times the unmixed term in Eq.~(\ref{Ctresult}).   Adding all contributions
leads to 
\begin{eqnarray}
\label{Ctresulttotal}
3C^A(t)&=&\frac{n_r^2}{144V}\sum_{\vec p}\sum_{XX'}\frac{e^{-(E_X(\vec p)+E_{X'}(\vec p))|t|}}{E_X(\vec p)E_{X'}(\vec p)}\,{\vec p}^2\\
&&\times\Bigl(10\tr(\x_A\x_X\x_{X'})\,\tr(\x_A\x_{X'}\x_X)-\tr(\x_A\x_X\x_{X'})\,\tr(\x_A\x_X\x_{X'})\Bigr)\ .
\nonumber
\end{eqnarray}
For $\x_A=1$ the traces set $X'=X$, and setting all masses $m_X$ equal (\ie, taking the continuum limit), the factor on the 
second line of this result is equal to $9\times 4^2\times 16=144\times 16$ (a 4 for each trace; the 16
comes from the $\sum_X$).   With $n_r=1/4$, the total prefactor becomes equal to one, which is what
one would have obtained directly in the continuum.   For $a>0$, the taste pion $\p_X$ runs around the
loop, and these loops are averaged over all tastes. 

When $\x_A\ne 1$, a little more work is needed.   First, we have that
\begin{equation}
\label{mixed1}
\tr(\x_A\x_X\x_{X'})=\tr((\x_A\x_X\x_{X'})^T)=\tr(\x_{X'}^T\x_X^T\x_A^T)=\tr(\x_A\x_{X'}\x_X')^*\ ,
\end{equation}
and thus
\begin{equation}
\label{mixed2}
\tr(\x_A\x_X\x_{X'})\,\tr(\x_A\x_{X'}\x_X)=16\d_{X',X+A}\ .
\end{equation}
This takes care of the mixed trace in Eq.~(\ref{Ctresulttotal}).   We still get the factor 16, and the 
current inserts taste $A$ into the pion loop.   For the unmixed trace, we use that
\begin{equation}
\label{AX}
\x_A\x_X=(-1)^{|A||X|+A\cdot X}\x_X\x_A\ ,
\end{equation}
with $|X|=\sum_\m X_\m$ and $A\cdot X=\sum_\m A_\m X_\m$.   Thus
\begin{equation}
\label{unmixed}
(\tr(\x_A\x_X\x_{X'}))^2=16(-1)^{|A||X|+A\cdot X}\d_{X',X+A}\ .
\end{equation}
For $\x_A=1$ the right-hand side is equal to $\d_{X,X'}$, as before.
For any $\x_A\ne 1$, $(-1)^{|A||X|+A\cdot X}=+1$ for half of the tastes $X$, and $-1$ for the
other half.   Thus, when all masses become degenerate in the continuum limit,
the trace term on the second line of Eq.~(\ref{Ctresulttotal}) becomes equal to 11 when 
$(-1)^{|A||X|+A\cdot X}=-1$, and 9 when $(-1)^{|A||X|+A\cdot X}=+1$.    The total prefactor
thus becomes
\begin{equation}
\label{prefactor}
\frac{n_r^2}{144}\times 4^2\times 8\times (11+9) =\frac{10}{9}\times 16n_r^2\ ,
\end{equation}
which is $10/9$ times the continuum limit for $\x_A=1$, as we expect from Eq.~(\ref{A12pt}).
Here the factors 4 again come from the traces, and the factor $8\times (11+9)$ from the
$\sum_X$.     Again, the current $j_\m^A$ inserts a taste $A$ into the pion loop.
This verifies Eq.~(\ref{A12pt}) at this order SChPT, in the continuum limit.   For $a>0$, 
the sums in Eq.~(\ref{Ctresulttotal}) have to be carried out to obtain the complete result
in explicit form.   The Kronecker delta in Eq.~(\ref{mixed2}) reduces the sum over $X$ and
$X'$ to a single sum.

\section{\label{states} Staggered states and operators}
In this chapter, we first consider the spectrum of QCD with staggered fermions and
general properties of staggered correlation functions,
after which we discuss the construction of operators that create and annihilate states in the
staggered Hilbert space.   Most of the chapter will deal with mesons, but in the
final section we will briefly discuss aspects of baryonic states in QCD with staggered fermions.  
The material of the first three sections can
be found in Refs.~\cite{MGmesons,GSb,MGirreps}, while the material of the 
final section can be found in
Refs.~\cite{GSb,JB}.  Another useful reference is Ref.~\cite{KStoolkit}.

\subsection{\label{spectrum} Staggered spectrum: states and operators}
We will consider states in momentum space, because the irreducible representations of a 
theory with staggered fermions live in momentum space.     
Operators acting on the staggered Hilbert space will be denoted with a hat.   

Thus, $\hS_\m$ is the shift operator acting on the Hilbert space.   Defining
$\hT_\m=\hS_\m^2$, $\hT_\m$ is a normal translation  over two lattice
spacings in the $\m$
direction.   In momentum space, the eigenvalues are
\begin{equation}
\label{evT}
\hT_k:\quad e^{2ip_k}\ ,\qquad \hT_4:\quad e^{-2E}\ ,
\end{equation}
where $p_k$ with $-\p/2<p_k\le\p/2$ are the components of the physical spatial momentum of a state,
and $E$ is the energy.   The operator $\hT_4$ can be taken as the transfer matrix of
the theory.   With Eq.~(\ref{evT}) we can define the operators $\hT_\m^{-\half}$ as having
eigenvalues $e^{-ip_k}$ and $e^E$, and thus we can define
\begin{equation}
\label{Xi}
\hX_\m=\hS_\m\hT_\m^{-\half}\ ,
\end{equation}
with 
\begin{equation}
\label{Xisq}
\hX_\m^2=1\ .
\end{equation}
We will consider states with $\vec p=0$ (except in Sec.~\ref{momentum}).   Such states transform in
representations of the geometrical restframe group ($GRF$)
\begin{equation}
\label{defGRF}
GRF=G(\X_\m,R^{(k\ell)},I_s)\ ,
\end{equation}
where the spatial rotations $R^{(k\ell)}=R^{(k\ell)}(\p/2)$ generate the little group of $\vec p=0$
states, and we recall that $I_s=I_1I_2I_3=\G_4\X_4$ in the defining representation.  The notation
$G(X)$ denotes the finite group generated by the elements $X$. Since
$\X_4\in GRF$, also parity $P=I_s\X_4=\G_4\in GRF$, and, with $P$ commuting with all other
elements of the group, 
\begin{equation}
\label{GRF}
GRF=G(\X_\m,R^{(k\ell)})\times \{1,\ P\}\ .
\end{equation}
A maximally commuting set of elements of $GRF$ is the set $\{\X_4,\ \X_1\X_2,\ R^{(12)},\ P\}$.

We will label states in the Hilbert space by the eigenvalues $E$, the eigenvalue $\s$ of $P$,
and the eigenvalue $\s_t$ of $\X_4$:
\begin{eqnarray}
\label{evs}
\hS_4|Er\s;\s_t\rangle&=&\s_t e^{-E}|Er\s;\s_t\rangle\ ,\\
\X_4|Er\s;\s_t\rangle&=&\s_t|Er\s;\s_t\rangle\ ,\nonumber\\
P|Er\s;\s_t\rangle&=&\s|Er\s;\s_t\rangle\ ,\nonumber\\
I_s|Er\s;\s_t\rangle&=&\s\s_t|Er\s;\s_t\rangle\equiv\s_s|Er\s;\s_t\rangle\ .\nonumber
\end{eqnarray}
The label $r$ denotes an irrep of $G(\X_\m,R^{(k\ell)})$; 
within such an irrep, we can label the state with the eigenvalue $\s_t$ of $\X_4$.
The final equation defines the parity $\s_s$ of $I_s$.

Next, we introduce local operators $\hf$ and $\hbf$ at fixed time $t$, which thus transform
in representations of a smaller group, the geometric time-slice group
\begin{equation}
\label{GTSdef}
GTS=G(\X_m,R^{(k\ell)},I_s)\ ,
\end{equation}
which does not contain $\X_4$.   We define time-dependent operators (\ie, we use the
Heisenberg picture) through
\begin{equation}
\label{timeop}
\hf(t)=\hS_4^{-t}\hf \hS_4^t\ ,\qquad\hbf(t)=\hS_4^{-t}\hbf \hS_4^t\ .
\end{equation}
There is a $U(1)$ charge operator $\hQ$ (quark number) with
\begin{equation}
\label{Qdef}
[\hQ,\hf]=q\hf\ ,\qquad [\hQ,\hbf]=-q\hbf\ ,
\end{equation}
with $q$ the charge of $\hf$ (we can take $q=1$ for the operator $\hchi$), and a charge conjugation operator $\hC_0$
with
\begin{equation}
\label{Cdef}
\hC_0^{-1}\hf\,\hC_0=\t\hbf\ ,\qquad \hC_0^{-1}\hbf\,\hC_0=\overline{\t}\hf\ .
\end{equation}
The phases $\t$ and $\overline{\t}$ depend on convention, but the phase
$\s_c=\t\overline{\t}$ defined by
\begin{equation}
\label{Csq}
\hC_0^{-2}\hf\,\hC_0^2=\s_c\hf
\end{equation}
is not.    Using the definition of $C_0$ in Sec.~\ref{symmetries} we find that
$\s_c=-1$ for the staggered quark field $\c$, and thus that it is equal to $+1$
for mesonic operators, and $-1$ for baryonic operators.   The fields
$\hf$ and $\hbf$ transform in a representation $r$ of $GTS$, which is
isomorphic to $G(\X_\m,R^{(k\ell)})$ and can thus be taken to be the same
irrep as in Eq.~(\ref{evs}).

We note that $GRF=GTS\times \{1,\ P\}$, but the fields $\hf$ and $\hbf$ do
not have $P$ quantum numbers, since the definition of $P$ involves a translation in time.   
We can consider operators $\f$ and $\bff$ with a well-defined value $\s_s$ of $I_s$:
\begin{equation}
\label{Isphiop}
\hI_s^{-1}\hf\hI_s=\s_s\hf\ ,\qquad \hI_s^{-1}\hbf\hI_s=\s_s\hbf\ .
\end{equation}
($\hf$ and $\hbf$ have the same $I_s$ quantum number because $\hI_s$ commutes with
$\hC_0$.)
If $\hf$, $\hbf$ have the $I_s$ quantum number $\s_s$, we have that
\begin{equation}
\label{overlap}
\langle Er\s;\s_t|\hbf|0\rangle=0\qquad\mbox{unless}\ \s_s=\s\s_t\ ,
\end{equation}
because $\hI_s=\hP\hX_4$.   

An example of $\hf$ is
\begin{equation}
\label{stagop}
\sum_{\vec m}\hchi(\vec x+2\vec m)\ ,
\end{equation}
which transforms in an 8-dimensional irrep of $GTS$.   This irrep decomposes as
\begin{equation}
\label{8decomp}
{\bf 8}\to A_1^++A_1^-+F_1^++F_1^-\ ,
\end{equation}
where $A_1$ is the trivial irrep of the cubic group and $F_1$ the vector irrep.  The
superscript denotes the value of $I_s$.   These irreps of the cubic group can be projected out as
\cite{GSb}
\begin{eqnarray}
\label{irrepsin8}
&A_1^+:\qquad &\sum_{\vec x\ {\rm even}}\hchi(\vec x)\ ,\\
&A_1^-:\qquad &\sum_{\vec x\ {\rm even}}\hchi(\vec x+1+2+3)\ ,\nonumber\\
&F_1^-:\qquad &\sum_{\vec x\ {\rm even}}\left(\begin{array}{c}\hchi(\vec x+1)\\
\hchi(\vec x+2)\\ \hchi(\vec x+3)\end{array}\right)\ ,\nonumber\\
&F_1^+:\qquad &\sum_{\vec x\ {\rm even}}\left(\begin{array}{c}\hchi(\vec x+2+3)\\
-\hchi(\vec x+1+3)\\ \hchi(\vec x+1+2)\end{array}\right)\ .
\nonumber
\end{eqnarray}
The appearance of the irreps $A_1$ and $F_1$ for a fermionic field can be understood as follows.
In the rest frame, the relevant continuum
group is $SU(2)_{\rm spin}\times SU(2)_{\rm taste}$.   The group
$G(R^{(k\ell)})$ is a discrete subgroup of the diagonal $SU(2)$ subgroup of
the continuum symmetry group, and the relevant decomposition is,
for the groups $SU(2)_{\rm spin}\times SU(2)_{\rm taste}\supset SU(2)_{\rm diag}\supset G(R^{(k\ell)})$,
\begin{equation}
\label{contlatt}
\left({\scriptstyle{\half}}_{\rm spin},\ {\scriptstyle{\half}}_{\rm taste}\right)\to {\bf 0}+{\bf 1}\to A_1+F_1\ .\nonumber
\end{equation}

Now, let us consider a correlation function of the form
\begin{equation}
\label{Ctform}
C(t)=\langle 0|\hf(t_1)\hbf(t_2)|0\rangle\ ,
\end{equation}
with $t=t_1-t_2>0$.  Using Eqs.~(\ref{evs}),~(\ref{timeop}) and~(\ref{overlap}), this can be written as
(note that $\hS_4|0\rangle=|0\rangle$)
\begin{eqnarray}
\label{Ctexpl}
C(t)&=&\sum_{E\s}\langle 0|\hf(t_1)|Er\s;\s_t\rangle\langle Er\s;\s_t|\hbf(t_2)|0\rangle\\
&=&\sum_{E\s}\langle 0|\hf\hS_4^{t_1}|Er\s;\s_t\rangle\langle Er\s;\s_t|\hS_4^{-t_2}\hbf|0\rangle\nonumber\\
&=&\sum_{E\s}\s_t^t e^{-Et}\langle 0|\hf|Er\s;\s_t\rangle\langle Er\s;\s_t|\hbf|0\rangle\nonumber\\
&=&\sum_{E_+}|\langle 0|\hf|E_+r;\s_t=\s_s\rangle|^2\s_s^t e^{-E_+t}\nonumber\\
&&+\sum_{E_-}|\langle 0|\hf|E_-r;\s_t=-\s_s\rangle|^2(-\s_s)^t e^{-E_-t}\ ,
\nonumber
\end{eqnarray}
where in the first term the sum is over states with positive parity $\s=\s_t\s_s$ and energies
$E_+$, and in the 
second term the sum is over states with negative parity and energies $E_-$.   The subscript
on $E_\pm$ indicates the parity $\s$.   If, for example, the operator $\hf$ has $I_s$ parity
$\s_s=+1$, the negative $P$ parity states yield a rapidly oscillating contribution to $C(t)$.
The correlator gets contributions from both parities because $P$ is not well-defined for 
operators localized in time.

The result~(\ref{Ctexpl}) is valid when the time extent of the lattice is infinite.   If, instead, it
is $2T$ (it is convenient to take the lattice even in all directions for staggered fermions), 
then, instead of considering $\langle 0|\hf(t_1)\hbf(t_2)|0\rangle$, we should consider
$\Tr(\hS_4^{2T-t}\hf\hS_4^t\hbf)$, with $\Tr$ the trace in Hilbert space.   Let us instead
derive the correction to Eq.~(\ref{Ctexpl}) more intuitively.

If a particle can travel from $t_2$ to $t_1$, an anti-particle can travel from $t_1$ to
$t_2'=t_2+2T$.   This leads to the replacement $e^{-Et}\to e^{-E(t_2'-t_1)}=e^{-E(2T-t)}$.
With anti-periodic/periodic boundary conditions for the staggered quarks, we pick up a phase factor $\k=+1$,
respectively, $\k=-1$ for fermionic operators, while always $\k=+1$ for bosonic operators.    Furthermore, 
\begin{equation}
\label{meeq}
\langle 0|\hf|Er\s;\s_t\rangle\langle Er\s;\s_t|\hbf|0\rangle
=\s_c\langle 0|\hC_0^{-1}\hbf\hC_0|Er\s;\s_t\rangle\langle Er\s;\s_t|\hC_0^{-1}\hf\hC_0|0\rangle\ .
\end{equation}
Using invariance under charge-conjugation, Eq.~(\ref{Ctexpl}) gets replaced by
\begin{eqnarray}
\label{CtfiniteT}
C(t)&=&\sum_{E_+}|\langle 0|\hf|E_+r;\s_t=\s_s\rangle|^2\s_s^t \left(e^{-E_+t}
+\s_c \k \,e^{-(2T-t)E_+}\right)\nonumber\\
&&+\sum_{E_-}|\langle 0|\hf|E_-r;\s_t=-\s_s\rangle|^2(-\s_s)^t 
\left(e^{-E_-t}+\s_c \k \,e^{-(2T-t)E_-}\right)+\dots\ .
\nonumber
\end{eqnarray}
Contributions proportional to $|\langle n|\hf|m\rangle|^2$ with both $|n\rangle$ and
$|m\rangle$ not equal to $|0\rangle$ are not shown.
Note that if $\langle 0|\hf|0\rangle\ne 0$, we need to subtract $\langle 0|\hf|0\rangle
\langle 0|\hbf|0\rangle$ to get the connected part of $C(t)$.

\subsection{\label{mesons} Mesons}
In this subsection, we will specialize to meson operators, in particular, operators bilinear
in $\c$ and $\bc$ \cite{MGmesons}.   If $X_\m=D(\X_\m)$ is the representation of $\X_\m$ on a mesonic
operator, they commute, because
\begin{equation}
\label{DXi}
D(\X_\m)D(\X_\n)=e^{iq\p}D(\X_\n)D(\X_\m)\ ,
\end{equation}
and $q=0$ for mesonic operators.   (This can be easily checked on the explicit
operators we will construct below.)   Defining
\begin{equation}
\tX_m=X_1X_2X_3X_m\ ,
\end{equation}
the restframe symmetry group for mesonic states is
\begin{equation}
\label{RFm}
(RF)_{\rm mesons}=G(\tX_m,R^{(k\ell)})\times\{1,\ X_4\}\times\{1,\ X_1X_2X_3\}\times
\{1,\ P\}\times\{1,\ C_0\}\ .
\end{equation}
Mesonic states can thus be labeled by 
\begin{equation}
\label{mesonstates}
|E,\br^{\s_t\s_{123}},\s,\t_0\rangle\ ,
\end{equation}
with $\br$ an irrep of $G(\tX_m,R^{(k\ell)})$, and $\s_t$, $\s_{123}$, $\s$ and $\t_0$ the
$X_4$, $X_{123}$, $P$ and $C_0$ parities, respectively.

Defining\footnote{After constructing meson operators, they can be made gauge
covariant by inserting averages over the shortest Wilson lines between $\c$ and $\bc$.}
\begin{equation}
\label{D}
D_k\c(x)=\half\left(\c(x+k)+\c(x-k)\right)\ ,
\end{equation}
we define classes of bilinear operators (dropping hats on $\c$ and $\bc$) on a fixed
time slice $t$:
\begin{eqnarray}
\label{mesonops}
M_0&=&\sum_{\vec x\ {\rm even}}\bc(x)\c(x)\ ,\\
M_1&=&\sum_{\vec x\ {\rm even}}\bc(x)D_1\c(x)\ ,\nonumber\\
M_{12}&=&\sum_{\vec x\ {\rm even}}\bc(x)D_1D_2\c(x)\ ,\nonumber\\
M_{123}&=&\sum_{\vec x\ {\rm even}}\bc(x)D_1D_2D_3\c(x)\ ,\nonumber
\end{eqnarray}
where all members of each class are obtained by taking the operator explicitly shown,
and carrying out spatial shifts and rotations on them.  The $I_s$ parities of these 
operators are $+1$, $-1$, $+1$ and $-1$, respectively.   The class obtained from
$M_0$ contains 8 components, that from $M_1$ contains 24 components, from
$M_{12}$ 24 components, and from $M_{123}$ 8 components.   Indeed, with
$\c$ and $\bc$ transforming in the $\bf 8$ irrep of $GTS$, we have that
\begin{equation}
\label{8by8}
{\bf 8}\times {\bf 8}={\bf 8}^++{\bf 24}^-+{\bf 24}^++{\bf 8}^-\ ,
\end{equation}
where the superscript is the $I_s$ parity.   

The representations on the right-hand side of Eq.~(\ref{8by8}) are reducible.   We
can project onto irreps by multiplying with sign factors, and summing over all $\vec x$.
Let us see how this works through a number of examples.
\begin{itemize}
\item[1.]  The first example is
\begin{equation}
\label{example1}
\sum_{\vec x}\bc(x)\c(x)\quad\to\quad \bq q\ ,
\end{equation}
which transforms in $1^{++}$, a 1-dimensional (trivial) irrep of $G(\tX_m,R^{(k\ell)})$;
the two superscripts are the $\s_s$ and $\s_{123}$ parities.   We also show the 
corresponding continuum limit operator in the SET notation.
\item[2.]  The second example is
\begin{equation}
\label{example2}
\sum_{\vec x}\eta_4(x)\z_4(x)\bc(x)\c(x)\quad\to\quad \bq\g_4\x_4 q\ ,
\end{equation}
which transforms in $1^{+-}$.   The parity $\s_{123}$ is negative, because $\x_1\x_2\x_3$
anti-commutes with $\x_4$. 
\item[3.]  The third example is
\begin{equation}
\label{example3}
\sum_{\vec x}\eta_k(x)\z_k(x)\bc(x)\c(x)\quad\to\quad \bq\g_k\x_k q\ ,
\end{equation}
which transforms in ${3''''}^{++}$, with $3''''$ a 3-dimensional irrep of $G(\tX_m,R^{(k\ell)})$.
This 3-dimensional irrep has components of the continuum tensor $\bq\g_k\x_\ell q$, 
and under reduction to the cubic group decomposes into $A_1$ (the trace of this tensor)
and the 2-dimensional cubic irrep $E$ (the two traceless combinations of the diagonal of
this tensor).

We can take each of these three examples, and insert an extra sign factor $\e(x)$.   This 
would lead to the continuum operators
\begin{eqnarray}
\label{threeex}
&\!\!\!\!\!\!\!\!\!\!\mbox{example}\ 1:\qquad \bq\g_5\x_5 q\qquad &1^{+-}\ ,\\
&\mbox{example}\ 2:\qquad \bq\g_4\g_5\x_4\x_5 q\qquad &1^{++}\ ,\nonumber\\
&\mbox{example}\ 2:\qquad \bq\g_k\g_5\x_k\x_5 q\qquad &{3''''}^{+-}\ .
\nonumber
\end{eqnarray}
\item[4.]  The fourth example is
\begin{equation}
\label{example4}
\sum_{\vec x}\eta_k(x)\bc(x)D_k\c(x)\quad\to\quad \bq\g_k q\ ,
\end{equation}
which transforms in $3^{-+}$, with a new 3-dimensional irrep of $G(\tX_m,R^{(k\ell)})$,
clearly not equal to $3''''$, because it reduces to $F_1$ under cubic rotations.
\item[5.]  The fifth example is
\begin{equation}
\label{example5}
\sum_{\vec x}\z_k(x)\bc(x)D_k\c(x)\quad\to\quad \bq\x_k q\ ,
\end{equation}
which transforms in ${3''}^{-+}$, yet another new 3-dimensional irrep of $G(\tX_m,R^{(k\ell)})$.
It reduces to $F_1$ under cubic rotations, but it is different under the full group, as the $\tX_m$
act non-trivially on example 5, but trivially on example 4.
\item[6.]  The sixth example is
\begin{equation}
\label{example6}
\sum_{\vec x}\eta_m(x)\z_m(x)\bc(x)\eta_k(x)D_k(\z_\ell(x)D_\ell\c(x))_{k\ne\ell\ne m\ne k}\quad\to\quad \bq\g_m\g_k\x_m\x_\ell q\ ,
\end{equation}
which transforms in $6^{++}$, a 6-dimensional irrep of $G(\tX_m,R^{(k\ell)})$.   Under
cubic rotations, the 6-dimensional irrep reduces to $F_1+F_2$.   $F_1$ we have already
encountered, and defines the vector irrep of the cubic group.  The $F_1$ irrep is obtained from
Eq.~(\ref{example6}) by anti-symmetrizing in $k$ and $\ell$.   $F_2$ is the irrep obtained
from the off-diagonal elements of a rank-2 symmetric tensor.   Under cubic rotations, 
the diagonal and off-diagonal elements do not transform into each other, and thus $F_2$
is a new 3-dimensional irrep of the cubic group, not equivalent to $F_1$.
The $F_2$ irrep is obtained from Eq.~(\ref{example6}) by symmetrizing in $k$ and $\ell$.
\end{itemize}

Each of these operators, acting on the vacuum, produces pairs of states with opposite
parity, because these operators do not have a well-defined $\s_t$, and thus not a
well-defined parity $\s$.   One of those continuum states is identified by the 
continuum operators shown in Eqs.~(\ref{example1}) through~(\ref{example6}).    To find
the other state, we should multiply the direct-product matrix  $\g_A\x_B$ by a matrix 
$\g_C\x_D$, such that $\g_C\x_D$ commutes with all elements in $(RF)_{\rm mesons}$,
but anti-commutes with $\x_4$ (to obtain the opposite value of $\s_t$).   Inspection
shows that we need to take $\g_C\x_D=\g_4\g_5\x_4\x_5$.

Working this out for the six examples above, we find
\begin{eqnarray}
\label{paritypartners}
\sum_{\vec x}\bc(x)\c(x)\quad&\to\quad &\bq q\quad\mbox{and}\quad \bc\g_4\g_5\x_4\x_5 q\ ,\\
\sum_{\vec x}\eta_4(x)\z_4(x)\bc(x)\c(x)\quad&\to\quad &\bq\g_4\x_4 q\quad\mbox{and}\quad \bq\g_5\x_5 q\ ,\nonumber\\
\sum_{\vec x}\eta_k(x)\z_k(x)\bc(x)\c(x)\quad&\to\quad &\bq\g_k\x_k q\quad\mbox{and}\quad
\bq\g_k\g_4\g_5\x_k\x_4\x_5 q\ ,\nonumber\\
\sum_{\vec x}\eta_k(x)\bc(x)D_k\c(x)\quad&\to\quad &\bq\g_k q\quad\mbox{and}\quad
\bq\g_k\g_4\g_5\x_4\x_5 q\ ,\nonumber\\
\sum_{\vec x}\z_k(x)\bc(x)D_k\c(x)\quad&\to\quad &\bq\x_k q\quad\mbox{and}\quad
\bq\g_4\g_5\x_k\x_4\x_5 q\ ,\nonumber\\
\sum_{\vec x}\eta_m(x)\z_m(x)\bc(x)\eta_k(x)D_k(\z_\ell(x)D_\ell\c(x))\quad&\to\quad &\bq\g_m\g_k\x_m\x_\ell q\quad\mbox{and}\quad\bq\g_\ell\x_k q\ ,
\nonumber
\end{eqnarray}
where in the last line $k\ne\ell\ne m\ne k$.  From the structure of the continuum 
operators and the symmetries of Sec.~\ref{symmetries}, it is straightforward to 
deduce the $J^{PC}$ quantum numbers of the continuum states excited by these
lattice meson operators.   For instance, the first continuum operator on the fourth
line of Eq.~(\ref{paritypartners}) has $J^{PC}=1^{-+}$ and is a singlet of $SU(4)_{\rm taste}$
while the second operator has $J^{PC}=1^{+-}$, and transforms in the adjoint of
$SU(4)_{\rm taste}$ (because of the presence of the taste matrix $\x_4\x_5$).\footnote{Here
$C$ is the continuum charge parity, see Ref.~\cite{GS} for the relation with $C_0$.}

We can construct more meson operators by also considering ``time-split'' operators.
Under $C_0$ 
\begin{equation}
\label{timesplC0}
\bc(x)(\vec x,t)\c(\vec x,t+1)\ \xrightarrow{C_0}\ -\bc(\vec x,t+1)\c(\vec x,t)\ ,
\end{equation}
and we can thus make operators with $\t_0=\pm 1$ by combining
\begin{equation}
\label{defC0}
\bc(x)(\vec x,t)\c(\vec x,t+1)\mp \bc(\vec x,t+1)\c(\vec x,t)\ .
\end{equation}
We can use this to make operators with values of $\t_0$ not available in the collection of 
single time-slice operators.  An example is
\begin{equation}
\label{extimesplit}
\sum_{\vec x}\eta_4(x)\left(\bc(x)\c(x+4)+\bc(x+4)\c(x)\right)\quad\to\quad \bq\g_4 q\quad\mbox{and}\quad
\bq\g_5\x_4\x_5 q\ ,
\end{equation}
which have $J^{PC}=0^{+-}$ and $0^{-+}$, with the first an $SU(4)_{\rm taste}$
singlet and the second an $SU(4)_{\rm taste}$ adjoint.   The $J^{PC}=0^{+-}$ singlet does
not occur in the complete table of single time-slice operators \cite{MGmesons}.

Let us take the classical continuum limit of Eq.~(\ref{extimesplit}).   We sum over all $x$, but 
insert a momentum $k$, which we want to take physical, so $|k|\ll 1$.\footnote{Recall
that the continuum limit is most easily obtained in momentum space.}  We have that
\begin{eqnarray}
\label{momts}
&&\half\sum_x e^{ikx}\eta_4(x)(\bc(x)\c(x+4)+\bc(x+4)\c(x))\\
&&=\int_{\tp\tq}\sum_{AB}\d(\tp+k-\tq+\p_A+\p_B+\p_{\eta_4})
\half(S_4^A e^{i\tp_4}+S_4^B e^{-i\tq_4})\bff_B(\tq)\f_A(\tp)\nonumber\\
&&=e^{\frac{i}{2}k_4}\int_{\tp\tq}\d(\tp+k-\tq)
\cos(\tp_4-\mbox{$\small\half$} k_4)\bff(\tq)\G_4\f(\tp)+O(a)\nonumber\\
&&\to \int_{\tp\tq}\d(\tp+k-\tq)\,\bq(\tq)\g_4 q(\tp)+O(a)\nonumber\\
&&=\int d^4x\, e^{ikx}\bq(x)\g_4 q(x)+O(a)\ .
\nonumber
\end{eqnarray}
In the last three steps, we used that $|k|$ is small, 
and that to take the continuum limit, the support of $\f$ and $\bff$ 
may be assumed to be at small momenta $\tp$ and $\tq$, so that the $\d$ function factorizes
up to $O(a)$ effects.  In the continuum limit, we can then set the cosine equal to 1.  The fields $q$ and $\bq$ are the
continuum (SET) fields.

The multiplication $\g_A\x_B\to\g_A\x_B\g_4\g_5\x_4\x_5q$ which leads to the opposite-parity
operators in all these examples corresponds in position space with multiplying a meson
operator by $\eta_4(x)\z_4(x)\e(x)=(-1)^{x_4}$.   Repeating the calculation of Eq.~(\ref{momts})
but multiplying under the sum over $x$ by this phase factor, similar manipulations lead to
\begin{eqnarray}
\label{momts2}
&&\half\sum_x e^{ikx}\z_4(x)\e(x)(\bc(x)\c(x+4)+\bc(x+4)\c(x))\\
&&=\int_{\tp\tq}\sum_{AB}\d(\tp+k-\tq+\p_A+\p_B+\p_{\z_4}+\p_{\e})
\half(S_4^A e^{i\tp_4}+S_4^B e^{-i\tq_4})\bff_B(\tq)\f_A(\tp)\nonumber\\
&&=\int_{\tp\tq}\sum_{AB}\d(\tp+k-\tq)
\half(S_4^A (\hp_{\z_4}\hp_{\e})_{AB}e^{i\tp_4}+(\hp_{\e}\hp_{\z_4})_{AB}S_4^B e^{-i\tq_4})\bff_B(\tq)\f_A(\tp)+O(a)\nonumber\\
&&=\int_{\tp\tq}\sum_{AB}\d(\tp+k-\tq)
\half((\G_5\X_4\X_5)_{AB}e^{i\tp_4}+(\G_5\X_5\X_4)_{AB} e^{-i\tq_4})\bff_B(\tq)\f_A(\tp)+O(a)\nonumber\\
&&=ie^{\frac{i}{2}k_4}\int_{\tp\tq}\d(\tp+k-\tq)
\sin(\tp_4-\half k_4)\bff(\tq)\G_5\X_4\X_5\f(\tp)+O(a)\nonumber\\
&&\to \int_{\tp\tq}\d(\tp+k-\tq)\,\bq(\tq)i\g_4\tp_4q(\tp)+O(a)\nonumber\\
&&=\int d^4x\, e^{ikx}\bq(x)\g_4\partial_4 q(x)+O(a)\ .
\nonumber
\end{eqnarray}
Note that $\hp_{\z_4}$ and $\hp_\e$ commute, and in the third line above we can thus write
them in the order such that $\hp_{\z_4}$ always combines with the sign factor $S_4^A$ or $S_4^B$
into $\X_4$.
Because of the extra derivative, this operator is $O(a)$ suppressed relative to the continuum
operator in Eq.~(\ref{momts}).   The amplitude of the opposite-parity state thus vanishes in the
continuum limit, and this generalizes to all time-split bilinear meson operators.

\subsection{\label{momentum} Inserting momentum}
We briefly consider (single time-slice) meson operators with non-vanishing spatial 
momentum.   We start from
\begin{eqnarray}
\label{nzmom}
&&\sum_{\vec x}e^{i{\vec k}\cdot{\vec x}}\bc(\vec x)\c(\vec x)\big|_{{\vec k}={\vec\tk}+{\vec\p_C}}\\
&&=\int_{\tp\tq}\sum_{AB}\d({\vec\tp}-{\vec\tq}+{\vec\tk}+{\vec\p_A}+{\vec\p_B}+{\vec\p_C})\bff_B(\tq)\f_A(\tp)\nonumber\\
&&=\int_{\tp\tq}\d({\vec\tp}-{\vec\tq}+{\vec\tk})\bff(\tq)\hp_C\f(\tp)\ ,
\nonumber
\end{eqnarray}
where we assume that the temporal components of $\p_{A,B,C}$ are equal to zero.
We can now make choices for $\p_C$, for example (with slight abuse of notation)
\begin{equation}
\label{exmom}
e^{i{\vec\p_C}\cdot{\vec x}}=\eta_i({\vec x})\z_i({\vec x})\e({\vec x})=(-1)^{x_i}\qquad
\Rightarrow\qquad \hp_C=\G_i\G_5\X_i\X_5\ .
\end{equation}
The operator irrep is determined from the little group of $\vec k$, \ie, the subgroup of the 
cubic group (with spatial reflections) which leaves $\vec k$ invariant.   For example:
\begin{eqnarray}
\label{kex}
{\vec k}&=(0,0,k)\qquad&\Rightarrow\qquad D_4=G(R^{(12)}(\p/2),I_1)\ ,\\
&=(\p,\p,k)\qquad&\Rightarrow\qquad D_4=G(R^{(12)}(\p/2),I_1)\ ,\nonumber\\
&=(0,\p,k)\qquad&\Rightarrow\qquad Z_2\times Z_2\ ,
\nonumber
\end{eqnarray}
where $k\ne 0$ and $|k|<\p/2$.  
In this equation, we list the little group for each
choice of $\vec k$ shown.  These examples correspond to the choices
$\hp_C=1$, $\G_1\G_2\X_1\X_2$ and $\G_2\G_5\X_2\X_5$, respectively.
 The group $D_4=G(R^{(12)}(\p/2),I_1)$ is the dihedral
group, which has 8 elements and is not abelian, because
\begin{equation}
\label{Dnonab}
R^{(12)}(\p/2)I_1R^{(12)}(-\p/2)=I_2\ .
\end{equation}
It has four 1-dimensional irreps, obtained by mapping $R^{(12)}(\p/2)$ and
$I_1$ to $\pm 1$ in four different ways.   It has one 2-dimensional irrep, 
defined by what happens to a vector in the $(12)$ plane.   For this irrep,
we have
\begin{equation}
\label{2dirrepD}
D(R^{(12)}(\p/2))=\left(\begin{array}{cc}0&-1\\1&0\end{array}\right)=-i\s_2\ ,
\qquad D(I_1)=\left(\begin{array}{cc}-1&0\\0&1\end{array}\right)=-\s_3\ .
\end{equation}
From Eq.~(\ref{Dnonab}) we then have that, as an example,
\begin{equation}
\label{exD}
D(I_2)=(-i\s_2)(-\s_3)(i\s_2)=\s_3\ ,
\end{equation}
as one would expect.   For more information on general irreps of the staggered
symmetry group, see Ref.~\cite{MGirreps}.\footnote{The examples in Eq.~(\ref{kex})
correspond to the 8th and last entries in Table 5 of Ref.~\cite{MGirreps}.}

\subsection{\label{baryons} Baryons}
We will review the theory of staggered baryons only briefly, referring to Refs.~\cite{GSb,JB} for 
a much more extensive treatment.
The group $GTS$ has, of course, both mesonic and fermionic irreps.   There are
three fermionic ones: the $\bf 8$ we already encountered in Eq.~(\ref{stagop}), 
another 8-dimensional one labeled by $\bf 8'$, and a 16-dimensional one, $\bf 16$ \cite{GSb}.
There are two 1-dimensional, one 2-dimensional, six 3-dimensional and one 6-dimensional
mesonic irreps of the mesonic subgroup $G(\tX_m,R^{(k\ell)})$, further dressed by
parities $\s_s$ and $\s_{123}$ to make them mesonic irreps of $GTS$.   Some of these we already encountered in Sec.~\ref{mesons}.
We thus have that the sum
\begin{equation}
\label{sumdsq}
\sum_r d^2(r)=2\times 8^2+16^2+4\times(2\times 1^2+2^2+6\times 3^2+6^2)=768\ ,
\end{equation}
where $d(r)$ is the dimension of irrep $r$.   This is indeed the number of elements of
the group.   Rotations over multiples of $\p/2$ and reflections constitute 48 elements
(6 for all permutations of three objects, times 8 for $\pm$ signs in each direction).  This
gets multiplied by the number of elements in the group $\G_3=G(\X_m)$, which is 16.

The decomposition of the fermionic irreps to irreps of the cubic group are
\begin{eqnarray}
\label{fermdecomp}
{\bf 8}&\to & A_1^++A_1^-+F_1^++F_1^-\ ,\\
{\bf 8'}&\to & A_2^++A_2^-+F_2^++F_2^-\ ,\nonumber\\
{\bf 16}&\to & E^++E^-+F_1^++F_1^-+F_2^++F_2^-\ .\nonumber
\end{eqnarray}
We already encountered the first of these in Eq.~(\ref{8decomp}).  An example of an
operator transforming in $\bf 8'$ is \cite{GSb}
\begin{equation}
\label{8prime}
\e_{abc}\sum_{{\vec x}\ \rm even}(D_1(U)\c(x))^a (D_2(U)\c(x))^b (D_3(U)\c(x))^c\ ,
\end{equation}
where we made the color indices explicit, and we used
\begin{equation}
\label{covD}
D_k(U)\c(x) = \half(U_k(x)\c(x+k)+U^\dagger_k(x-k)\c(x-k))\ .
\end{equation}

Let us now compare baryonic states in the continuum with those on the lattice
\cite{JB}.
We will consider QCD with three flavors and isospin symmetry, \ie, we will take
\begin{equation}
\label{quarkmasses}
m_\ell\equiv m_u=m_d\ll m_s\ .
\end{equation}
While we can use the rooting trick to remove the 4-fold taste multiplicity for 
sea quarks, this degeneracy still exists in the valence sector.   We thus have
four valence up quarks, four valence down quarks and four valence strange
quarks, with the relevant flavor symmetry group thus being $SU(12)$.  
Taking also the baryon spin $SU(2)$ into account, the relevant continuum irreps are thus
\begin{eqnarray}
\label{birreps}
&\mbox{``octet''}:\qquad &\left(\mbox{\small$\half$},\ 572_M\right)\ ,\\
&\mbox{``decuplet''}:\qquad &\left(\mbox{\small$\frac{3}{2}$},\ 364_S\right)\ .\nonumber
\end{eqnarray}
Here the labels $S$, $M$ and $A$ denote the 3-index irreps fully symmetrized in 
the three indices, with mixed symmetry, and fully anti-symmetrized, respectively.   For $SU(N)$,
the dimensions of these irreps are
\begin{equation}
\label{irrepsizes}
S:\quad \left(\begin{array}{c}N+2\\3\end{array}\right)\ ,\qquad
M:\quad 2\left(\begin{array}{c}N+1\\3\end{array}\right)\ ,\qquad
A:\quad\left(\begin{array}{c}N\\3\end{array}\right)\ .
\end{equation}
Note that, because of the many valence quarks, there are many more states than in QCD
with three flavors!   This is why we put quotes around ``octet'' and ``decuplet.''

Since we are considering the valence sector, we will encounter two different cases:
baryons for which all three valence quark masses have the same mass, such as the nucleon,
the $\D$ and the $\O$, and baryons for which one valence mass is different from the other
two, for all other members of the octet and decuplet.  In the latter case, the valence flavor symmetry
group gets reduced to $SU(8)\times SU(4)$.  

The relevant continuum irreps for the first case are those of Eq.~(\ref{birreps}).   To decompose
these irreps with respect to the lattice symmetry group, we consider the chain of
decompositions 
\begin{eqnarray}
\label{groupdec}
SU(2)_{\rm spin}\times SU(12)&\to& SU(2)_{\rm spin}\times SU(3)_{\rm flavor}\times SU(4)_{\rm taste}\\
&\to& SU(3)_{\rm flavor}\times GTS\ .\nonumber
\end{eqnarray}
with \cite{JB}
\begin{eqnarray}
\label{decomp1}
\left(\mbox{\small$\half$},572_M\right)&\to&
\left(\mbox{\small$\half$},10_S,20_M\right)+\left(\mbox{\small$\half$},8_M,20_S\right)+\left(\mbox{\small$\half$},8_M,20_M\right)\\
&&+\left(\mbox{\small$\half$},8_M,{\overline{4}}_A\right)
+\left(\mbox{\small$\half$},1_A,20_M\right)\nonumber\\
&\to&3(10_S,{\bf 8})+(10_S,{\bf 16})+5(8_M,{\bf 8})+3(8_M,{\bf 16})+3(1_A,{\bf 8})+(1_A,{\bf 16})\ ,
\nonumber\\
\left(\mbox{\small$\frac{3}{2}$},364_S\right)&\to&
\left(\mbox{\small$\frac{3}{2}$},10_S,20_S\right)+\left(\mbox{\small$\frac{3}{2}$},8_M,20_M\right)+\left(\mbox{\small$\frac{3}{2}$},1_A,{\overline{4}}_A\right)\nonumber\\
&\to&2(10_S,{\bf 8})+2(10_S,{\bf 8'})+3(10_S,{\bf 16})+(8_M,{\bf 8})\nonumber\\
&&+(8_M,{\bf 8'})+
4(8_M,{\bf 16})+(1_A,{\bf 16})\ .
\nonumber
\end{eqnarray}
We see that many of the lattice irreps appear in both decompositions, often multiple times, so
these irreps couple to, for instance, both the nucleon and the $\D$ (if we take all quark masses
to be $m_\ell$).   Moreover,
all different lattice states in one decomposition are expected to become degenerate in the continuum
limit, so the mass splittings between these lattice states are expected to be small, thus creating a 
severe excited-state contamination problem.   Furthermore, there is a mixing problem:  for example
the irrep $(10_S,{\bf 8})$ appears five times on the right-hand side of Eq.~(\ref{decomp1}), and
corresponding operators can mix \cite{JB}.

However, there are useful irreps on the lattice side of these decompositions.   The lattice irrep
$(8_M,{\bf 8'})$ appears only once, and does not appear in the first decomposition; it thus only couples to the $\D$.
Therefore, the operator~(\ref{8prime}) is suitable if one wishes to compute the $\D$ mass
from the ground state contribution \cite{GSb,JB} (instead of trying to obtain its mass from an 
excited-state contribution).  Likewise, if we take all valence masses
equal to $m_s$, this operator gives access to the $\O$.
 The irrep $(1_A,{\bf 16})$ appears only 
once in each decomposition in Eq.~(\ref{decomp1}), and thus couples to the nucleon and the
$\D$, without mixing to other nearly-degenerate lattice states.    This operator is thus
favorable for extracting the nucleon mass, as the $\D$ is about 300~MeV heavier.   Near the 
continuum limit, one expects all different lattice states in the first decomposition of
Eq.~(\ref{decomp1}) to be much closer in mass.   

We can also consider operators made out of two staggered quarks with mass $m_\ell$,
and one with mass $m_s$.   The relevant group decomposition is now
\begin{eqnarray}
SU(2)_{\rm spin}\times SU(12)&\to&SU(2)_{\rm spin}\times SU(8)\times SU(4)\\
&\to& SU(2)_{\rm spin}\times SU(2)_{\rm isospin}\times U(1)_{\rm strange}\times SU(4)_{\rm taste}\nonumber\\
&\to&SU(2)_{\rm isospin}\times U(1)_{\rm strange}\times GTS\ .\nonumber
\end{eqnarray}
Considering the decomposition of irreps of $SU(2)\times SU(8)\times SU(4)$ with respect to
the subgroup $SU(2)_{\rm isospin}\times U(1)_{\rm strange}\times GTS$, one finds
\begin{eqnarray}
\label{decomp2}
\left(\mbox{\small$\frac{3}{2}$},120_S,1\right)&\to&\left(\mbox{\small$\half$},{\bf 8'}\right)_0+\dots\ ,\\
\left(\mbox{\small$\frac{3}{2}$},36_S,4\right)&\to&\left(0,{\bf 8'}\right)_{-1}+\dots\ ,\nonumber
\end{eqnarray}
where the complete decompositions can be found in Ref.~\cite{JB}.   On the left-hand side the
labels are for spin, $SU(8)$ and $SU(4)$, while on the right-hand side the labels are for
isospin and $GTS$, with the subscript denoting strangeness.   The lattice irreps shown on the
right-hand side do not appear anywhere else in these decompositions, and are thus again
practical for extracting baryon masses.   The state $\left(\mbox{\small$\half$},{\bf 8'}\right)_0$
is again our lattice $\D$, since it has strangeness 0; it is the operator found in the
$SU(3)_{\rm flavor}\to SU(2)_{\rm isospin}\times U(1)_{\rm strange}$ decomposition
\begin{equation}
\label{SU3dec}
8_M\to \left(\mbox{\small$\half$}\right)_0+(1)_{-1}+
\left(\mbox{\small$\half$}\right)_{-2}\ ,
\end{equation}
where on the right-hand side the subscript denotes the $U(1)_{\rm strange}$ charge.
   However, the state $\left(0,{\bf 8'}\right)_{-1}$
has strangeness $-1$, and thus gives access to a new decuplet member, the $\S^*$.

We can also consider operators with two strange quarks, and only one light quark, 
by reinterpreting the group $SU(8)$ as the valence symmetry group for four times two strange
valence
quarks, and the group $SU(4)$ as the valence symmetry group for the four light valence quarks.
Likewise, $SU(2)_{\rm isospin}$ is now ``isospin'' for the two strange valence quarks, and
$U(1)$ is the light valence quark number.   Therefore, if we ``flip'' the quark masses in the
two lattice irreps shown in Eq.~(\ref{decomp2}), we now find that the irrep
$\left(\mbox{\small$\half$},{\bf 8'}\right)_0$ again yields the $\O$, but now the 
irrep $\left(0,{\bf 8'}\right)_{-1}$ yields the $\X^*$.    

We note that, because of the large number of valence quarks available, many more states
exist on the lattice than in three-flavor QCD, and a number of ``unphysical'' states are thus
accessible.   For instance, one can make a $N_s$, which is a nucleon with only strange
quarks \cite{JB}.   Such a state should be considered to be a ``partially quenched'' state, as it 
only appears because of the many valence quarks available.  To
``unquench'' this state, one should not use the rooting trick, so that all 12 quarks are
also present in the sea.\footnote{For an introduction to quenched and partially quenched
QCD, see Ref.~\cite{MGLH}.}   But, of course, in the real world of QCD, there do not exist
eight light quarks and four strange quarks, so this unquenched theory is not QCD.

\section{\label{anomaly} The anomaly}
In this section, we consider the (abelian) axial anomaly with staggered fermions. 
There exists an exactly conserved staggered current,  the Noether
current for $U(1)_\e$ symmetry.   Clearly, this current is not anomalous.   This is 
consistent with what we expect in the continuum limit, since this current has taste
$\x_5$ and is thus not equal to the singlet axial current.

In Sec.~\ref{axial} we construct a singlet axial current, which is not conserved on
the lattice.  This is as it should be; otherwise it cannot be anomalous in the continuum limit
\cite{KS}.   Then, in Sec.~\ref{divergence}, we calculate the 
divergence of this singlet current in a background (abelian) gauge field in
two dimensions, following Ref.~\cite{STW}, where this calculation was first
carried out.   In Sec.~\ref{direct}, we verify the result of Sec.~\ref{divergence} by a 
direct calculation of the anomalous diagrams.   We limit ourselves to the theory
in two dimensions, for simplicity, and refer to Ref.~\cite{STW} for the four-dimensional case.   
For the non-abelian anomaly, see Ref.~\cite{CKN}.

\subsection{\label{axial} Singlet axial current}
Let us return to the naive fermion theory, Eq.~(\ref{naive}).   This theory has a conserved axial 
current,
\begin{equation}
\label{consaxial}
\frac{i}{2}\left(\bj(x)\g_5\g_\m U_\m(x)\j(x+\m)+\bj(x+\m)\g_5\g_\m U^\dagger_\m(x)\j(x)\right)\ .
\end{equation}
However, if we spin-diagonalize this current, we find the conserved staggered current
associated with $U(1)_\e$ symmetry, which is not a singlet axial current.  Indeed, with or
without spin diagonalization, the anomaly cancels precisely because of the species doublers
\cite{KS}.

Instead, we consider
\begin{equation}
\label{singletax}
j_\m^5(x)=\frac{i}{2}\left(\bj(x)\g_5\g_\m U_{\m 5}(x)\j(x+\m+5)+\bj(x+\m+5)\g_5\g_\m U^\dagger_{\m 5}(x)\j(x)\right)\ ,
\end{equation}
in which 
\begin{equation}
\label{psi5}
x+\m+5=x+\m+1+2+3+4\ ,
\end{equation}
and $U_{\m 5}(x)=U_\m(x)U_5(x+\m)$, with $U_5(x+\m)$ a Wilson line from $x+\m$ to $x+\m+5=x+\m+1+2+3+4$.
Using Eq.~(\ref{spindiag}), we have that
\begin{eqnarray}
\label{psimu5}
\j(x+\m+5)&=&\eta_\m(x+5)\g_\m\g_1^{x_1+1}\g_2^{x_2+1}\g_3^{x_3+1}\g_4^{x_4+1}\c(x+\m+5)\\
&=&\eta_\m(x+5)\eta_1(x)\eta_2(x)\eta_3(x)\eta_4(x)\g_\m\g_5
\g_1^{x_1}\g_2^{x_2}\g_3^{x_3}\g_4^{x_4}\c(x+\m+5)\ ,\nonumber\\
\bj(x+\m+5)&=&\bc(x+\m+5)\eta_\m(x+5)\eta_1(x)\eta_2(x)\eta_3(x)\eta_4(x)\g_4^{x_4}\g_3^{x_3}
\g_2^{x_2}\g_1^{x_1}\g_5\g_\m\ ,\nonumber
\end{eqnarray}
and thus
\begin{equation}
\label{axdiag}
j_\m^5(x)=\frac{i}{2}\left(\prod_\n\eta_\n(x)\right)\eta_\m(x+5)\left(\bc(x) U_{\m 5}(x)\c(x+\m+5)-\bc(x+\m+5) U^\dagger_{\m 5}(x)\c(x)\right)\ .
\end{equation}
Note the minus sign in this equation, which is needed to make the current hermitian.
We also note that
\begin{equation}
\label{prodeta}
\prod_\n\eta_\n(x)=\eta_1(x)\eta_2(x+1)\eta_3(x+1+2)\eta_4(x+1+2+3)\ .
\end{equation}
This implies that 
\begin{equation}
\label{interpret}
\prod_\n\eta_\n(x)\eta_\m(x+5)\c(x+\m+5)=T_1T_2T_3T_4T_\m\c(x)\ ,
\end{equation}
with $T_\m$ defined in Eq.~(\ref{Tmu}).  With
\begin{eqnarray}
\label{momax}
T_1T_2T_3T_4T_\m:\ \c(x)\quad&\to&\quad \G_5\G_\m\f(\tp)\ ,\\
T_1T_2T_3T_4T_\m:\ \bc(x)\quad&\to&\quad \bff(\tp)\G_\m\G_5\ ,\nonumber
\end{eqnarray}
where we jumped to
momentum space, one can verify that Eq.~(\ref{axdiag}) reproduces the singlet axial
current in the classical continuum limit.

\begin{boldmath}
\subsection{\label{divergence} Divergence of singlet axial current in $d=2$}
\end{boldmath}
In this subsection, we will outline the calculation of Ref.~\cite{STW} of the divergence of the singlet axial
current in an abelian background gauge field, in $d=2$ dimensions.   As was done in Ref.~\cite{STW},
we will use naive instead of staggered fermions.   Since in this calculation the fermions appear in a 
loop, the result is basis independent, and it does not matter whether one uses the fields $\j$
and $\bj$, or $\c$ and $\bc$ of Eq.~(\ref{spindiag}).   Of course, dropping the spin index on $\c$
reduces the number of degrees of freedom by a factor $2^{d/2}$, so the staggered axial-current
divergence is a factor $2^{d/2}$ smaller than the naive axial-current divergence.   

It turns out to be convenient to use, instead of Eq.~(\ref{singletax}), the current
\begin{equation}
\label{sax}
j^A_\m(x)=\frac{1}{2^d}\sum_\x\frac{i}{2}\left(\bj(x)\g_5\g_\m U_{\m\x}(x)\j(x+\m+\x)+\bj(x+\m+\x)\g_5\g_\m U^\dagger_{\m\x}(x)\j(x)\right)\ .
\end{equation}
Here the vectors $\x$ have components $\pm 1$, and there are thus $2^d$ different $\x$ vectors.
As indicated, the average over all $2^d$ $\x$ vectors is taken. 
The current~(\ref{singletax}) takes all components of $\x$ equal to $+1$, and (thus) does not 
average.  The classical continuum limit of both currents is the
same.
 $U_{\m\x}(x)=U_\m(x)U_\x(x+\m)$ with
 $U_\x(x+\m)$ a Wilson line from $x+\m$ to $x+\m+\x$, replacing $U_5(x+\m)$.\footnote{The ordering of the links on this Wilson
line will turn out not to matter.}   Below, we will need the properties
\begin{equation}
\label{xiprop}
\frac{1}{2^d}\sum_\x\x_\m=0\ ,\qquad \frac{1}{2^d}\sum_\x\x_\m\x_\n=\d_{\m\n}\ ,\qquad
\frac{1}{2^d}\sum_\x\x_\m\x_\n\x_\k=0\ .
\end{equation}
Next, we turn to the calculation, in $d=2$, of the expectation value of the lattice divergence of the axial current in an (abelian) gauge-field background,
\begin{equation}
\label{divax}
\langle\sum_\m\partial^-_\m j_\m^A(x)\rangle\equiv \langle\sum_\m\left(j_\m^A(x)-j_\m^A(x-\m)\right)
\rangle\ .
\end{equation}
Using Eq.~(\ref{sax}), this can be written as
\begin{eqnarray}
\label{divax2}
\langle\sum_\m\partial^-_\m j_\m^A(x)\rangle
&=&\frac{i}{2}\frac{1}{4}\sum_{\x\m}\langle\bj(x)\g_5\g_\m\left(U_\m(x)U_\x(x+\m)-U_\x(x)U_\m(x+\x)\right)\j(x+\m+\x)\nonumber\\
&&\hspace{2cm}+\hc\rangle+\dots\ .
\end{eqnarray}
It was shown that the terms not shown give rise to a contribution that vanishes in the
continuum limit (in $d=2$) \cite{STW}.\footnote{We do not show this latter point here, but we will
calculate the anomaly in a different way in the next subsection.}
We thus need, using $U_\m(x)=1+iagA_\m(x)+\dots$,
\begin{eqnarray}
\label{gfpart}
&&U_\m(x)U_\x(x+\m)-U_\x(x)U_\m(x+\x)\\
&&=
iagA_\m(x)+iag\sum_\k\x_\k A_\k(x+\m)-iag\sum_\k\x_\k A_\k(x)-iagA_\m(x+\x)+\dots\nonumber\\
&&=ia^2g\partial_\m\sum_\k\x_\k A_\k(x)-ia^2g\sum_\k\x_\k\partial_\k A_\m(x)+\dots
=ia^2g\sum_\k\x_\k F_{\m\k}(x)+\dots\ ,\nonumber
\end{eqnarray}
where we restored the lattice spacing $a$, and the ellipsis denotes terms of higher order 
than quadratic in $a$.   We also need
\begin{equation}
\label{fermloop}
-\tr\left(\g_5\g_\m S(x+a\m+a\x,x)\right)\ ,
\end{equation}
where $S$ is the fermion propagator defined by 
\begin{equation}
\label{fermpropinv}
S^{-1}(x,y)=\sum_\n\frac{1}{2a}\,\g_\n\left(\d_{y,x-a\n}-\d_{y,x+a\n}\right)\ ,
\end{equation}
and thus
\begin{eqnarray}
\label{fermprop}
S(x,y)&=&\int_p e^{-ip(x-y)}S(p)\ ,\\
S(p)&=&\frac{-\frac{i}{a}\sum_\n\g_\n\sin(ap_\n)}{\frac{1}{a^2}\sum_\k\sin^2(ap_\k)}\ .
\nonumber
\end{eqnarray}
(We set the quark mass $m$ equal to zero, as the anomaly does not depend on it.)

Using $\g_5=\g_1\g_2$ (in $d=2$), 
\begin{equation}
\label{trg}
\tr(\g_5\g_\m\g_\n)=-2\e_{\m\n}\ ,
\end{equation}
where\footnote{Do not confuse $\e_{\m\n}$ with $\varepsilon_{\m\n}$ of Sec.~\ref{action}.} $\e_{\m\n}=-\e_{\n\m}$ and $\e_{12}=+1$, we find that
\begin{equation}
\label{tragain}
-\tr\left(\g_5\g_\m S(x+a\m+a\x,x)\right)=2\sum_\n\e_{\m\n}\int_p
\frac{-\frac{i}{a}\sum_\n\g_\n\sin(ap_\n)}{\frac{1}{a^2}\sum_\k\sin^2(ap_\k)}\,e^{-iap_\m-ia\sum_\r\x_\r p_\r}\ .
\end{equation}
We can now write $p_\m+\sum_\r\x_\r p_\r=p_\m(1+\x_\m)+p_\n\x_\n$, with $\m\ne\n$ because of the 
factor $\e_{\m\n}$ and use the fact that only the part of the integrand even in $p_\n$ yields a non-vanishing
integral to simplify this to
\begin{equation}
\label{tragain2}
-\tr\left(\g_5\g_\m S(x+a\m+a\x,x)\right)=
-\frac{2}{a}\sum_\n\e_{\m\n}\x_\n\,4\int_{\tp}\frac{\sin^2(\tp_\n)}{\sum_\k\sin^2(\tp_\k)}\left(\cos^2(\tp_\m)-\x_\m\sin^2(\tp_\m)\right)\ .
\end{equation}
Here we wrote the integral over $p$ as $2^d=4$ times the integral over $\tp$, the momentum in the
reduced BZ, while at the same time rescaling the integration momentum to be dimensionless.  
Combining this with Eq.~(\ref{gfpart}), we find
\begin{eqnarray}
\label{resultdivax}
\langle\sum_\m\partial^-_\m j_\m^A(x)\rangle
&=&-\frac{i}{2}\frac{1}{4}\sum_{\x\m\n\k}\frac{8}{a^2}\e_{\m\n}\x_\n (ia^2g\x_\k F_{\m\k}(x))
\int_{\tp}\frac{\sin^2(\tp_\n)}{\sum_\l\sin^2(\tp_\l)}\cos^2(\tp_\m)\nonumber\\
&=&4g\sum_{\m\n}\e_{\m\n}F_{\m\n}(x)\int_{\tp}\frac{\sin^2(\tp_1)\cos^2(\tp_2)}{\sum_\l\sin^2(\tp_\l)}\ ,
\end{eqnarray}
where we took the continuum limit, and used Eq.~(\ref{xiprop}).   The integral in this result can be
calculated analytically as follows \cite{STW}.   We first observe that
\begin{equation}
\label{1ststep}
\int_{\tp}\frac{\sin^2(\tp_1)\cos^2(\tp_2)}{\sum_\l\sin^2(\tp_\l)}
=-\int_{-\p/2}^{\p/2}\frac{dp_2}{(2\p)^2}\int_{-\p/2}^{\p/2}dp_1\,\cos^2(p_2)\frac{\partial}{\partial p_1}
\frac{\sin(p_1)\cos(p_1)}{\sum_\k\sin^2(p_\k)}\ ,
\end{equation}
because
\begin{equation}
\label{partial2}
\frac{\partial}{\partial p_1}
\frac{\sin(p_1)\cos(p_1)}{\sum_\k\sin^2(p_\k)}=-\frac{\sin^2(p_1)}{\sum_\k\sin^2(p_\k)}
+\frac{\cos^2(p_1)(\sin^2(p_2)-\sin^2(p_1))}{(\sum_\k\sin^2(p_\k))^2}\ ,
\end{equation}
and the second of the terms on the right gives rise to a vanishing integral.  We can now carry out
the integral over $p_1$, but have to be careful to regulate the integral near $p=0$, which
we do by inserting $\theta(p^2-\e^2)$, where we will take $\e\to 0$ at the end of the calculation.
The integral over $p_1$ is equal to twice the integral from 
$p_1=\sqrt{\e^2-p_2^2}$ to $\p/2$, which implies that $p_2^2\le\e^2$.  The 
integral thus reduces to 
\begin{equation}
\label{2ndstep}
2\int_{-\p/2}^{\p/2}\frac{dp_2}{4\p^2}\,\cos^2(p_2)\,
\frac{\sin\sqrt{\e^2-p_2^2}\cos\sqrt{\e^2-p_2^2}}{\sin^2\sqrt{\e^2-p_2^2}+\sin^2(p_2)}
\,\theta(\e^2-p_2^2)\ .
\end{equation}
Taking $\e$ small, this simplifies to
\begin{equation}
\label{3rdstep}
\frac{1}{\p^2}\int_0^{\p/2}dp_2\,\theta(\e^2-p_2^2)\sqrt{\e^2-p_2^2}\,\frac{1}{\e^2}
=\frac{1}{\p^2}\int_0^1 dp_2\sqrt{1-p_2^2}=\frac{1}{4\p}\ ,
\end{equation}
where we rescaled $p_2\to\e p_2$, 
and we thus find
\begin{equation}
\label{axanres}
\langle\sum_\m\partial^-_\m j_\m^A(x)\rangle=\frac{2g}{\p}\sum_{\m\n}\e_{\m\n}F_{\m\n}(x)
+\co(a)\ ,
\end{equation}
where the extra factor of 2 comes from the ``$\hc$'' in Eq.~(\ref{divax2}).   This is indeed four
times the anomaly one obtains using a continuum regulator, as one expects for naive
fermions in two dimensions with a four-fold species doubling.   To obtain the staggered
result, we simply divide the result by two.

\vskip0.8cm
\begin{boldmath}
\subsection{\label{direct} Direct calculation in $d=2$}
\end{boldmath}
Instead of completing the calculation of the previous subsection, in $d=2$ it is not difficult to
calculate the two diagrams of Fig.~\ref{anomalyd2} directly.   Somewhat surprisingly, the anomaly
comes from the right-hand panel in this figure, while the left-hand panel does not contribute, as
we will see.   This is true even though the right-hand panel is a ``lattice artifact'' diagram, as it only
exists because of the link variables in the current~(\ref{sax}).

\begin{figure}[!t]
\begin{center}
\vskip 3ex
\begin{picture}(50,100)(5,0)
   \put(-130,0){\includegraphics*[height=1.in]{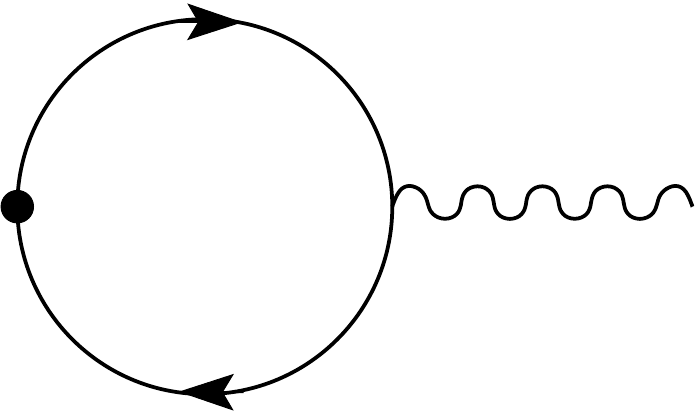}}
   \put(60,0){\includegraphics*[height=1.in]{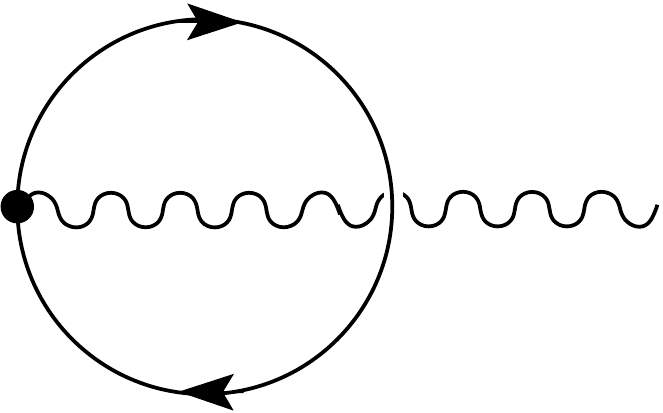}}
   \put(-140,33){$\m$}
   \put(-5,33){$\n$}
   \put(50,33){$\m$}
   \put(180,33){$\n$}
\end{picture}
\vskip 3ex
\floatcaption{anomalyd2}{The anomaly in two dimensions.   The black dot denotes an insertion
of $\g_5\g_\m$, the wiggly line is a photon.}
\end{center}
\end{figure}

Let us thus start with the panel on the right.    The vertex, obtained from Eq.~(\ref{sax}), is
\begin{eqnarray}
\label{anvertex}
&&-\frac{g}{2}\,\frac{1}{4}\sum_{\x}\Biggl(\bj(x)\g_5\g_\m(A_\m(x)+\sum_\k\x_\k A_\k(x+\m))\j(x+\m+\x)\\
&&\phantom{-\frac{g}{2}\,\frac{1}{4}\sum_{\x\m}\Biggl(}-\bj(x+\m+\x)\g_5\g_\m(A_\m(x)+\sum_\k\x_\k A_\k(x+\m))\j(x)\Biggr)\ .\nonumber
\end{eqnarray}
The expectation value (treating the gauge field as an external field) is thus equal to
\begin{eqnarray}
\label{tadpolean}
&&\frac{g}{2}\,\frac{1}{4}\sum_{\x}(A_\m(x)+\sum_\k\x_\k A_\k(x+\m))\,\tr(\g_5\g_\m(S(x+\m+\x,x)-S(x,x+\m+\x)))\nonumber\\
&=&4g\,\frac{1}{4}\sum_{\x\n}(A_\m(x)+\sum_\k\x_\k A_\k(x+\m))\,\e_{\m\n}\int_p\,\frac{\sin(p_\n)}{\sum_\l\sin^2(p_\l)}\,\sin(p_\m+\x p)\ ,
\end{eqnarray}
where $\x p=\sum_\r\x_\r p_\r$ and we followed steps similar to those used in the previous subsection.
We expand ($\m\ne\n$)
\begin{equation}
\label{sinexp}
\sin(p_\m+\x p)=\sin(p_\m)\cos(p_\m)\cos(p_\n)(1+\x_\m)+\x_\n\cos^2(p_\m)\sin(p_\n)+\x_\m\x_\n\sin^2(p_\m)\sin(p_\n)\ ,
\end{equation}
and substitute this into Eq.~(\ref{tadpolean}).   Since $\m\ne\n$ the first term yields a vanishing
integral upon symmetric integration.   The last term also contributes zero, using Eq.~(\ref{xiprop})
and the presence of $\e_{\m\n}$.   Only the middle term
contributes, and, using Eq.~(\ref{xiprop}), yields
\begin{equation}
\label{RHpanel}
4g\sum_{\n} \e_{\m\n}A_\n(x+\m)\int_p\,\frac{\sin^2(p_\n)\cos^2(p_\m)}{\sum_\l\sin^2(p_\l)}
=\frac{4g}{\p}\sum_{\n} \e_{\m\n}A_\n(x)+\co(a)\ ,
\end{equation}
where  the integral is equal to four times the reduced-BZ integral, and thus equal to $1/\p$.
In the last step we took the continuum limit, $A_\m(x+\m)\to A_\n(x)$.   Applying $\partial_\m$
to this result recovers Eq.~(\ref{axanres}).

This implies that the left-hand panel of Fig.~\ref{anomalyd2} had better contribute zero, and
we will now demonstrate that this is indeed the case.   As the anomaly (in two dimensions)
is independent of momentum, we can set the external momentum in this diagram equal to zero.
Labeling the axial-current vertex by $x$, the other vertex by $y$, using the Feynman rules, and summing over $x$ to set the external momentum equal to zero,
the left-panel diagram is
\begin{eqnarray}
\label{LHpanel}
&&-\frac{g}{2}\sum_{xz\n}\frac{1}{4}\sum_\x\tr\biggl(S(x+\m+\x,y)\half(\d_{x,z+\n}+\d_{x,z-\n})\g_\n S(z,x)\\
&&\phantom{-\frac{g}{2}\sum_{xz\n}\frac{1}{4}\sum_\x\tr}
+S(x,y)\half(\d_{x,z+\n}+\d_{x,z-\n})\g_\n S(z,x+\m+\x)
\biggr)\nonumber\\
&=&g\,\frac{1}{4}\sum_{\x\r\s}\tr(\g_5\g_\m\g_\r\g_\n\g_\s)\int_p\,\frac{\sin(p_\r)\sin(p_\s)}{(\sum_\k\sin^2(p_\k))^2}\,\cos(p_\n)\cos(p_\m+\x p)\ ,
\nonumber
\end{eqnarray}
where we used Eqs.~(\ref{sax}) (setting all gauge links equal to one) and~(\ref{fermprop}).

Writing $p_\m+\x p=p_\m(1+\x_\m)+\x_\l p_\l$ with $\l\ne \m$, 
\begin{eqnarray}
\label{cosexp}
&&\frac{1}{4}\sum_\x\cos(p_\m+\x p)\\
&&=\frac{1}{4}\sum_\x\left(\cos^2(p_\m)\cos(p_\l)-\x_\m\sin^2(p_\m)\cos(p_\l)
-\x_\l\sin(p_\m)\cos(p_\m)\sin(p_\l)(1+\x_\m)\right)\nonumber\\
&&= \cos^2(p_\m)\cos(p_\l)\ ,
\nonumber
\end{eqnarray}
because $\sum_\x\x_\m\x_\l=0$ when $\m\ne \l$.
If, in the trace, we set $\m=\n$, we have that $\r\ne\s$, and it is straightforward to check
that for this choice the integral in Eq.~(\ref{LHpanel}) vanishes.   Hence, we have that 
$\m\ne\n$, and thus (in two dimensions!) $\r=\s$, with Eq.~(\ref{LHpanel}) simplifying to
\begin{equation}
\label{LHsimple}
g\sum_\r\tr(\g_5\g_\m\g_\r\g_\n\g_\r)\int_p\,\frac{\sin^2(p_\r)}{(\sum_\k\sin^2(p_\k))^2}\,\cos^2(p_\m)\cos^2(p_\n)\ ,
\end{equation}
because $\l=\n$ if both $\l$ and $\n$ are not equal to $\m$.   We now use that
\begin{eqnarray}
\label{temp}
\sum_\r\tr(\g_5\g_\m\g_\r\g_\n\g_\r)\sin^2(p_\r)
&=&2\e_{\m\n}(\sin^2(p_1)+\sin^2(p_2)-2\sin^2(p_\n))\\
&=&2|\e_{\m\n}|(\sin^2(p_1)-\sin^2(p_2))\ ,\nonumber
\end{eqnarray}
and we find that also this contribution vanishes.

\begin{figure}[!t]
\begin{center}
\begin{picture}(50,100)(5,0)
   \put(-130,0){\includegraphics*[height=1.in]{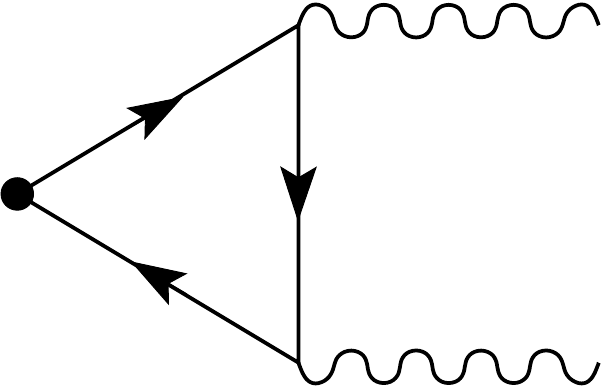}}
   \put(60,0){\includegraphics*[height=1.in]{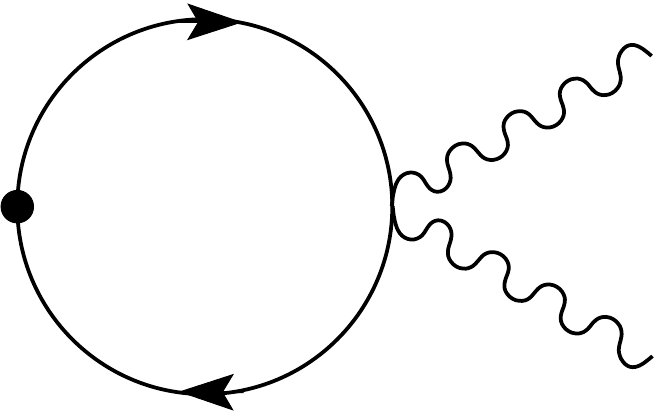}}
\put(-130,-110){\includegraphics*[height=1.in]{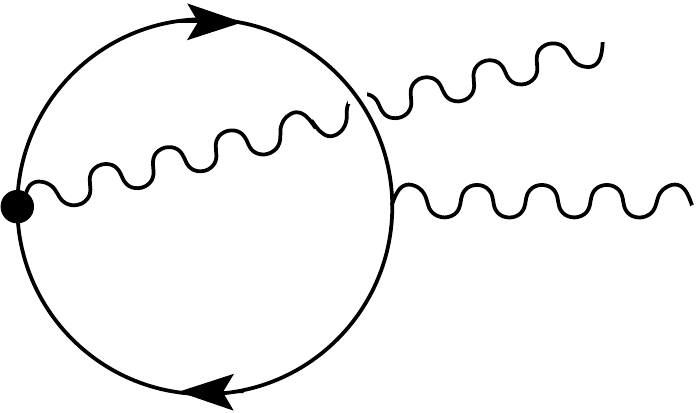}}
   \put(60,-110){\includegraphics*[height=1.in]{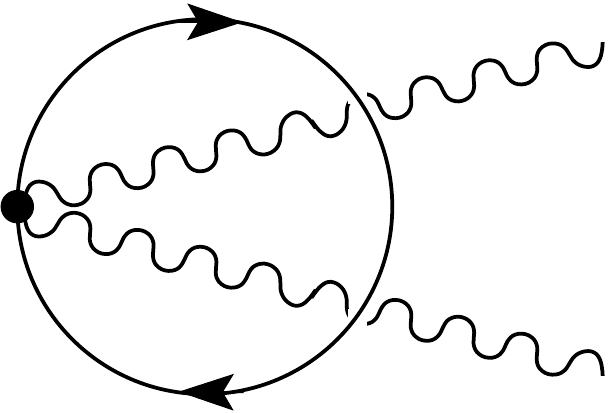}}
\end{picture}
\vskip 23ex
\floatcaption{anomalyd4}{The anomaly in four dimensions.   The black dot denotes an insertion
of $\g_5\g_\m$, wiggly lines are photons or gluons.}
\end{center}
\end{figure}

We end this section with a comment on the anomaly in four dimensions, for which the diagrams are
shown in Fig.~\ref{anomalyd4}.   Starting from the upper left, we see that the last three diagrams
vanish.   For the fourth diagram, this is obvious, because $\tr(\g_5\g_\m\g_\n)=0$ in $d=4$.
Then, if we take the divergence of the axial current, we expect a result proportional to
$\e_{\m\n\r\s}k_{1\r}k_{2\s}$, where $k_1$ and $k_2$ are the two photon (or gluon) momenta.
But the second and third diagrams have only one momentum flowing through them, so no such
contribution can come from these two diagrams.   Thus, the anomaly has to come from
the triangle diagram on the upper left.   For a detailed calculation of the divergence of the
axial current in four dimensions following the strategy of Sec.~\ref{divergence}, see Ref.~\cite{STW}.

\vskip0.8cm
\section{\label{improvement} Improvement}
In this section, we specifically discuss ``asqtad'' and ``HISQ'' improvement.   The goal of
these improvement schemes is {\em not} to systematically remove all $\co(a^2)$ effects from 
the theory.   As we have seen in Sec.~\ref{SET}, there are very many dimension-6 contributions
to the SET,
and thus very many lattice dimension-6 operators would have to be added to the staggered
action in order to cancel their appearance in the SET.   This would require all their
coefficients to be tuned, a demanding task
especially if one would want to do this non-perturbatively.

Instead, the idea is to improve the theory only with regard to taste-breaking effects,
which are generally believed to be a major cause of scaling violations in 
QCD with unimproved staggered fermions.   Again, this could be done by adding and tuning 
many taste-breaking $\co(a^2)$ operators to the lattice action, which is again not
practical.   However, as we have seen in Sec.~\ref{tastebrsec}, at least to leading order in the 
gauge coupling, taste breaking is caused by the exchange of gluons with momenta
of order $\p/a$ between staggered quarks.   By suppressing the corresponding
vertices, one may thus hope to suppress taste-breaking effects.   Doing this at tree level
will not completely remove $\co(a^2)$ affects; instead, taste-breaking effects will be suppressed
to order $\a_s a^2$ (with $\a_s$ the strong coupling).  Suppressing taste-breaking effects
should reduce the mass slittings between pions ({\it c.f.} Sec.~\ref{SChPT}), which, because
all pions are light, removes an important source of scaling violations from the infrared
physics of staggered QCD.
 It is this idea that
underlies the ``a-squared, tadpole improved'' (asqtad) and "highly improved 
staggered quark'' (HISQ) improvement schemes.   For an extensive review,
see Ref.~\cite{MILC}.   In this section, we will not set the lattice spacing $a=1$, but
instead explicitly keep track of it.

\subsection{\label{asqtad} Asqtad}
First, we have seen that the vertex~(\ref{1gvertex}) already suppresses interactions
with gluons with momentum $k_\m=\p/a$, when $\m$ is equal to the index of that vertex.   
Our first goal is thus to modify this vertex 
such that interactions with gluons with momenta $k$ in which at least one other component
of $k$ equal to $\p/a$ are suppressed, \ie, not only when $k_\m=\p/a$ and the other
components of $k$ equal to zero.   Of course, this needs to be done such that the classical
continuum limit of the vertex remains the same.

In the asqtad scheme,\footnote{For references, see Ref.~\cite{MILC}.} this is accomplished by smearing the gauge field in several
steps.   The first step is to replace the link variables in the staggered action by
the symmetrized ``staple''
\begin{equation}
\label{F3}
U_\m(x)\to\cf^{\rm f3}U_\m(x)=U_\m(x)+\frac{1}{4}a^2\sum_{\n\ne\m}\D^\ell_\n U_\m(x)\ ,
\end{equation}
in which 
\begin{eqnarray}
\label{Deltaell}
\D^\ell_\n U_\m(x)&=&\frac{1}{a^2}\Bigl(U_\n(x)U_\m(x+a\n)U^\dagger_\n(x+a\m)
\\&&\phantom{\frac{1}{a^2}\Bigl(}+U^\dagger_\n(x-a\n)U_\m(x-a\n)U_\n(x-a\n+a\m)
-2U_\m(x)\Bigr)\ .\nonumber
\end{eqnarray}
Here we are showing the lattice spacing explicitly.   Let us work out how this modifies
Eq.~(\ref{1gvertex}).   We expand to linear order in the gauge field $A_\m$ and go to 
momentum space by writing\footnote{In this section, we use a different convention
for the definition of the gauge field $A_\m$ from the link variables $U_\m(x)$ from 
the one used in previous sections.}
\begin{eqnarray}
\label{linmom}
U_\m(x)&=&1+iagA_\m(x+\mbox{\small$\half$} a\m)+\co(a^2g^2)\ ,\\
A_\m(x)&=&\int_k e^{ikx}A_\m(k)\ .\nonumber
\end{eqnarray}
This replaces Eq.~(\ref{F3}) by
\begin{equation}
\label{F3mom}
A_\m(k)\to A_\m(k)+\frac{1}{4}\sum_{\n\ne\m}\left(2(\cos(ak_\n)-1)A_\m(k)
+4\sin(\mbox{\small$\half$} ak_\m)\sin(\mbox{\small$\half$} ak_\n)A_\n(k)\right)\ .
\end{equation}
If we choose $k=\bp$ with $\bp_\n=\p/a$ and the other components physical
(\ie, such that $a\bp_\r$ is small, for $\r\ne\n$), the right-hand side of Eq.~(\ref{F3mom}) becomes
\begin{equation}
\label{F3momspec}
\sum_{\n\ne\m}2a\bp_\m A_\n(\bp)\ ,
\end{equation}
and we see that indeed the coupling of a gluon with momentum $\bp$ is suppressed.

In order to also suppress momenta with two components equal to $\p/a$, we add to
Eq.~(\ref{F3}) the 3-dimensional 5-link staple
\begin{equation}
\label{F5}
\frac{a^4}{32}\sum_{\n\ne\r,\n\ne\m,\r\ne\m}\D^\ell_\r\D^\ell_\n U_\m(x)\ .
\end{equation}
At the one gluon level, in momentum space, this means adding to Eq.~(\ref{F3mom}) the
terms
\begin{eqnarray}
\label{F5mom}
&&\frac{1}{32}\sum_{\n\ne\r,\n\ne\m,\r\ne\m}\Bigl(4(\cos(ak_\n)-1)(\cos(ak_\r)-1)A_\m(k)\\
&&\phantom{\frac{1}{32}\sum_{\n\ne\r,\n\ne\m,\r\ne\m}\Bigl(}+8((\cos(ak_\r)-1)\sin(\mbox{\small$\half$} ap_\n)\sin(\mbox{\small$\half$}ap_\m)A_\n(p)\Bigr)\ .\nonumber
\end{eqnarray}
Taking $k=\bp$ with $\bp_\n=\bp_\r=\p/a$ and the other components zero, this expression
vanishes, and we thus suppress the coupling of gluons with momenta close to $\bp$.
For this momentum, the coefficient of $A_\m$ becomes
\begin{equation}
\label{coeffAmu}
1+\frac{1}{4}\times 2\times (-4) +\frac{1}{32}\times 2\times 16=0\ .
\end{equation}
Here the 1 comes from the first term in Eq.~(\ref{F3mom}), and the second term from the
cosine term in Eq.~(\ref{F3mom}), where now both components $\n$ and $\r$ contribute
$-4$ to the sum over $\n$.   The final term comes from Eq.~(\ref{F5mom}), where we took into
account that the pair $\n\ne\r$ appears twice in the sum.
We also note that for momenta with only one component $\bp_\n=\p/a$, with all three other
components vanishing, Eq.~(\ref{F5mom}) also vanishes, so the improvement effect of adding the 3-link
staple in Eq.~(\ref{F3}) remains intact.

There is one more step to be done, which is to add a 4-dimensional 7-link staple, 
\begin{equation}
\label{F7}
\frac{a^6}{384}\sum_{\n\ne\r\ne\s\ne\n,\n\ne\m,\r\ne\m,\s\ne\m}\D^\ell_\s\D^\ell_\r\D^\ell_\n U_\m(x)\ .
\end{equation}
This leads to the 1-gluon-vertex contribution\footnote{I guessed the second term, which I leave as
an exercise to the reader.   The key point is that for $k_\m=\bp_\m$ small, it is $\co(a)$.}
\begin{eqnarray}
\label{F7mom}
&&\frac{1}{384}\sum_{\n\ne\r\ne\s\ne\n,\n\ne\m,\r\ne\m,\s\ne\m}\Bigl(
8(\cos(ak_\n)-1)(\cos(ak_\r)-1)(\cos(ak_\s)-1)A_\m(k)\\
&&\hspace{3.5cm}+16(\cos(ak_\n)-1)(\cos(ak_\r)-1)\sin(\mbox{\small$\half$} ak_\n)\sin(\mbox{\small$\half$} ak_\m)A_\n(k)\Bigr)\ .\nonumber
\end{eqnarray}
For $k=\bp$ with $\bp_\n=\bp_\r=\bp_\s=\p/a$ and $\bp_\m=0$ (all indices different), the 
coefficient of $A_\m$ becomes
\begin{equation}
\label{coeffAmuagain}
1+\frac{1}{4}\times 3\times (-4)+\frac{1}{32}\times 6\times 16+\frac{1}{384}\times 6\times(-64)=0\ .
\end{equation}
The 3 in the second term comes from the sum over $\n$ in the 3-link term; the 6 in the third term from the sum over $\n\ne\r$ in the 
5-link term, and the 6 in the fourth term from the sum over all permutations of the indices
$\n$, $\r$ and $\s$ in the 7-link term.   Again, this new addition does not spoil the effect of the
3-link and 5-link staples.  We conclude that the asqtad scheme suppresses the coupling of
gluons with at least one component of the gluon momentum close to $\p/a$, which is what
causes taste breaking to leading order in the gauge coupling $g$.

There is a subtlety.   Consider again Eq.~(\ref{F3mom}), rewriting this as
\begin{equation}
\label{F3momagain}
A_\m(k)+\sum_{\n\ne\m}\sin(\mbox{\small$\half$} ak_\n)\left(\sin(\mbox{\small$\half$} ak_\m)A_\n(k)-\sin(\mbox{\small$\half$} ak_\n)A_\m(k)\right)\ .
\end{equation}
For $\bp_\n=\p/a$ and $\bp_\m$
small, the first term in the sum is $\co(a)$, and not $\co(a^2)$, as one would naively expect for staggered
fermions ({\it c.f.} Sec.~\ref{SET}).   However, as we will see, 
this $\co(a)$ term can be removed by a gauge
transformation, and should thus not affect the physics.   

A gauge transformation
\begin{equation}
\label{gaugetr}
U_\m(x)\to e^{ia\l(x)}U_\m(x)e^{-ia\l(x+a\m)}
\end{equation}
can be written to leading order and in momentum space as
\begin{equation}
\label{gaugemom}
A_\m(k)\to A_\m(k)-\frac{1}{g}\,2i\sin(\mbox{\small$\half$} ak_\m)\l(k)\ .
\end{equation}
We now choose 
\begin{equation}
\label{choice}
\l(k)=-\half\,ig\sum_\n\sin(\mbox{\small$\half$} ak_\n)A_\n(k)\ ,
\end{equation}
which transforms Eq.~(\ref{F3momagain}) into
\begin{equation}
\label{F3momgtr}
A_\m(k)+\frac{1}{4}\sum_{\n\ne\m}2(\cos(ak_\n)-1)A_\m(k)-\sin^2(\mbox{\small$\half$} ak_\m)A_\n(k)\ ,
\end{equation}
and we now see that the last term is $\co(a^2)$, thus removing the uncomfortable $\co(a)$
term of Eq.~(\ref{F3mom}).   We expect that similar considerations apply to the 5- and 7-link
additions~(\ref{F5}) and~(\ref{F7}).   Using all multilink terms discussed above, \ie, 
adding Eqs.~(\ref{F5}) and~(\ref{F7}) to Eq.~(\ref{F3}) is usually referred to as ``fat7 smearing.'' 

Let us again return to Eq.~(\ref{F3mom}).   If we take all momentum components small,
this takes the form
\begin{equation}
\label{F3momexp}
A_\m(k)+\frac{1}{4}\,a^2\sum_{\n\ne\m}k_\n(k_\m A_\n(k)-k_\n A_\m(k))\ ,
\end{equation}
and this vertex modification thus adds a new $\co(a^2)$ contribution to the staggered
action, which one may want to cancel.   (We do not have to worry about this for the
5-link and 7-link additions, since they are of higher order in $a$ for small momenta.)
This can be done by adding the ``Lepage'' term \cite{Lepage}
\begin{equation}
\label{lepterm}
-\frac{a^2}{4}\sum_{\n\ne\m}\D^{2\ell}_\n U_\m(x)\ ,
\end{equation}
in which
\begin{eqnarray}
\label{Delta2ell}
\D^{2\ell}_\n U_\m(x)&=&\frac{1}{4a^2}\Bigl(U_\n(x)U_\n(x+a\n)U_\m(x+2a\n)U^\dagger(x+a\n+a\m)U^\dagger_\n(x+a\m)
\\&&\hspace{-1cm}+U^\dagger_\n(x-a\n)U^\dagger(x-2a\n)U_\m(x-2a\n)U_\n(x-2a\n+a\m)U_\n(x-a\n+a\m)\nonumber\\
&&
\hspace{-1cm}-2U_\m(x)\Bigr)\ .\nonumber
\end{eqnarray}
To linear order and in momentum space Eq.~(\ref{lepterm}) adds
\begin{eqnarray}
\label{lepmom}
&&-\frac{1}{16}\Biggl(2(\cos(2ak_\n)-1)A_\m(k)+4\sin(\mbox{\small$\half$} ak_\m)\left(\sin(\mbox{\small$\half$} ak_\n)
+\sin(\mbox{\small$\frac{3}{2}$} ak_\n)\right)A_\n(k)\Biggr)\nonumber\\
&&=\frac{1}{4}\,a^2\sum_{\n\ne\m}k_\n(k_\n A_\m(k)-k_\m A_\n(k))+\dots
\end{eqnarray}
to Eq.~(\ref{F3mom}).
For small momenta, shown in the second line of Eq.~(\ref{lepmom}), Eq.~(\ref{lepterm}) cancels Eq.~(\ref{F3momexp}).   At the same time,
Eq.~(\ref{lepmom}) vanishes for $k=\bp$ with (only) $\bp_\n=\p/a$, and it thus does not affect
asqtad improvement.

\subsection{\label{Naik} Naik term}
Also the standard discretization of the derivative on the fermion fields can be
improved.   We can do so by replacing \cite{SN}
\begin{eqnarray}
\label{Naikterm}
\nabla_\m\c(x)&\to& \nabla_\m\c(x)-\frac{a^2}{6}\nabla^3_\m\c(x)\\
&=&\frac{9}{16a}\left(U_\m(x)\c(x+a\m)-U^\dagger_\m(x-a\m)\c(x-a\m)\right)\nonumber\\
&&-\frac{1}{48a}\bigl(U_\m(x)U_\m(x+a\m)U_\m(x+2a\m)\c(x+3a\m)\nonumber\\
&&\phantom{-\frac{1}{48a}\bigl(}-U^\dagger_\m(x-a\m)U^\dagger_\m(x-2a\m)U^\dagger_\m(x-3a\m)\c(x-3a\m)\bigr)\ ,\nonumber
\end{eqnarray}
using
\begin{equation}
\label{nabla}
\nabla_\m\c(x)=\frac{1}{2a}\left(U_\m(x)\c(x+a\m)-U^\dagger_\m(x-a\m)\c(x-a\m)\right)\ .
\end{equation}
In the free theory (\ie, setting $U_\m(x)=1$), in momentum space, this improved
derivative becomes
\begin{equation}
\label{Naikmom}
\frac{i}{a}\left(\sin(ap_\m)+\frac{1}{6}\sin^3(ap_\m)\right)\tc(p)=ip_\m\left(1+\co((ap)^4)\right)\tc(p)\ ,
\end{equation}
thus improving the fermion derivative to $\co(a^2)$ at tree level.\footnote{This improvement also works
for momenta $p=\p_A/a+\tp$ with $\tp$ physical.}  Using Eq.~(\ref{linmom})
(instead of Eq.~(\ref{link}); note our change of convention for expressing $U_\m$ in terms of $A_\m$
in this section), we find for the
one-gluon vertex including the Naik term:
\begin{eqnarray}
\label{1gluonNaik}
&&-ig\d(p+k-q+\p_{\eta_\m})\Bigl(\frac{9}{8}\cos(ap_\m+\mbox{\small{$\half$}}ak_\m)\\
&&\hspace{2cm}-\frac{1}{24}\left(\cos(3ap_\m+\mbox{\small{$\half$}}ak_\m)+\cos(3ap_\m+\mbox{\small{$\frac{3}{2}$}}ak_\m)+\cos(3ap_\m+\mbox{\small{$\frac{5}{2}$}}ak_\m)\right)\Bigr)\ .\nonumber
\end{eqnarray}
It is straightforward to check that this vertex is still suppressed when we take $k_\m=\p/a$.
Of course, we can still use the results of the previous subsection, by substituting the smeared
link instead of the straight link into Eq.~(\ref{Naikterm}).  Using asqtad plus the Naik term leads
to an action with taste breaking suppressed to order $\a_s a^2$ (where $\a_s=g^2/(4\p)$).
One thus expects taste splittings to be of order $\a_s^2 a^2$, where, since taste splitting is a 
short-distance effect, one should take the strong coupling $\a_s$ at the lattice scale, $\sim 1/a$.
Because $\a_s$ is small at the lattice scale, one thus expects a significant reduction of the
taste breaking, as borne out by numerical simulations \cite{MILC2004}.
As in all other discretizations of QCD, one can also apply ``tadpole improvement'' \cite{tadpole},
which generally improves the behavior of the theory in WCPT.   As this is not specific to 
the use of staggered fermions, we will not discuss this here.

\subsection{\label{HISQ} HISQ}
Finally, we define the HISQ improvement scheme \cite{HISQref} as implemented by the
MILC collaboration \cite{HISQMILC}.   Denoting the operation of fat7 smearing by $\cf^{\rm f7}$, 
the HISQ prescription is to replace the link variables $U_\m(x)$ by
\begin{equation}
\label{HISQdef}
\cf^{\rm f7L}\cu\cf^{\rm f7}U_\m(x)\ .
\end{equation}
Here $\cu$ denotes a unitarization back to $U(3)$, and the label f7L denotes fat7 smearing
with the addition of a Lepage term~(\ref{lepterm}).  Reference~\cite{HISQref} proposed
to unitarize back to $SU(3)$ between the two fat7 smearings.   The advantage of unitarizing to $U(3)$ is that it can be
done analytically, as described in detail in Appendices B and C of Ref.~\cite{HISQMILC}.  
The complete HISQ action, as used by the MILC collaboration in the generation of HISQ
ensembles and including adjustments for the treatment of the charm quark, is described
in detail in Appendices A through D of Ref.~\cite{HISQMILC}.

It was found in Ref.~\cite{HISQMILC} that taste splittings in the pion multiplet are significantly
suppressed with HISQ, in comparison with asqtad, see for example Fig. 3 in Ref.~\cite{HISQMILC}
where taste splittings for lattice spacings 0.06, 0.09, 0.12, and 0.15 fm are shown.  
Since taste splittings for these lattice spacings are small, one would expect that they can be
fit using SChPT, and an attempt was made to do so in Ref.~\cite{ABGP22}.     In general, one
can try semi-phenomenological fits of the form
\begin{equation}
\label{fittbr}
\D(\x)=A_\x\a_s^{n_2}(1/a)a^2+B_\x\a_s^{n_4}(1/a)a^4+C_\x\a_s^{n_6}(1/a)a^6\ ,
\end{equation}
where $\D(\x)$ is one of the taste splittings of Eq.~(\ref{Deltas}), and $n_{2,4,6}$ are non-negative
integers.\footnote{In any given fit, one chooses values for $n_{2,4,6}$, as it is not possible,
with current data, to treat these integers as free parameters.}   In Ref.~\cite{ABGP22}, it was found that, for lattice spacings from 0.06 fm to 0.15 fm, the $a^6$ 
term appears to be indispensible.    While in Ref.~\cite{ABGP22} only the case $n_4=n_6=0$ was
studied, one finds in general that, for HISQ taste splittings with lattice spacings 0.06, 0.09, 0.12, 
and 0.15 fm the $a^6$ term is needed, while the fits are not sensitive to the $a^2$ term (so the
value of $n_2$ does not matter).   The optimal powers of $\a_s$ in the $a^4$ and $a^6$ terms
appear to be $n_4=3$ and $n_6=0$.   Since taste splittings should be suppressed to $\a_s$, 
the $a^6$ term in Eq.~(\ref{fittbr}) plays a purely phenomenological in these fits.
An important caveat is that these fits were attempted
without knowing the correlations between different taste splittings at the same lattice spacing,
since these correlations were not available.

These results are puzzling \cite{ABGP22}.  The partial degeneracy predicted in Eq.~(\ref{Deltas}) is clearly
seen in the HISQ taste splittings, but this degeneracy is predicted at order $a^2$ in SChPT,
while it does not persist at higher orders.  From $\co(a^4)$ on, \ie, NLO and higher orders, SChPT predicts that the degeneracy is
completely broken down to the restframe symmetry group (with splittings between 
$\x_{5i}$ and $\x_{54}$, between $\x_{ij}$ and $\x_{i4}$ and between $\x_i$ and $\x_4$
breaking the ``accidental'' $SO(4)$ symmetry of LO SChPT).   This would lead one to expect that 
the $a^2$ terms in Eq.~(\ref{fittbr}) should dominate, which, however, appears not to be the case.
It would be interesting to understand this better, with a dedicated study of taste splittings on
HISQ ensembles.

\section{\label{reading} Further reading}
Many topics were not discussed in these lectures; instead, we collect a few relevant references
here.   Examples are operator mixing \cite{SP}, topology with staggered fermions \cite{SV},
topological freezing \cite{BT}, ChPT for mixed actions with Ginsparg--Wilson valence and 
staggered sea fermions \cite{BBRS,Chenetal},
SChPT for heavy-light mesons \cite{ABheavy}, a study of
taste breaking to higher order in SChPT \cite{BL}, and other improvement schemes
\cite{finT,BMW}.   For more information on the 4th root trick, see the reviews in Ref.~\cite{MGroot,SSroot,AKroot}, and references therein; for issues with applying the rooting trick
at non-zero chemical potential, see Refs.~\cite{GSS,BFetal}.   
It was shown that the tastes of staggered fermions can be
used as physical flavors by adding terms to the staggered action that break the mass degeneracy
completely.   These mass term break the staggered symmetry group only softly, and
renormalization of all masses is still multiplicative \cite{GS}.
Employing tastes as physical flavors would avoid the need to use the 4th root trick.
However, this leads to a non-positive fermion determinant, and to a serious
fine-tuning problem if one wants to interpret the four tastes as the physical up, down, strange
and charm flavor \cite{GS2}.

We emphasize again that no attempt has
been made for the list of references of these lecture notes to be complete.

\vspace{3ex}
\noindent {\bf Acknowledgments}
\vspace{2ex}

I would like to thank Laurent Lellouch for the invitation to give these lectures
at the Centre de Physique Th\'eorique in Luminy, Marseille, and I would like to thank Laurent and
the lattice group at the CPT for the warm hospitality.   I also would like to thank
Claude Bernard and Steve Sharpe for answering
many questions about staggered fermions, and Yigal Shamir for help with the figures.     
This material is based upon work supported by the U.S. Department of
Energy, Office of Science, Office of High Energy Physics,
Office of Basic Energy Sciences Energy Frontier Research Centers program
under Award Number DE-SC0013682.

\vspace{5ex}
\def\refttl#1{\medskip\noindent \underline{#1}}

\end{document}